\pdfoutput=1

\documentclass[twocolumn]{aastex62}
\usepackage{balance}
\usepackage{enumerate}
\usepackage{mathrsfs}
\usepackage{graphicx}
\usepackage{subfigure}
\hypersetup{backref,colorlinks=true,linkcolor=blue,citecolor=blue}
\usepackage{amsmath,bm}
\usepackage{diagbox}
\usepackage{slashbox}
\usepackage[figuresright]{rotating}

\def\nh{\textit{$N_{\rm H}$}}
\def\lognh{\textit{${\rm log}\,N_{\rm H}$}}

\def\xspec{\textit{\rm XSPEC}}

\def\cm{\textit{$\rm 10^{23}\ cm^{-2}$}}
\def\ct{\textit{$\rm 10^{24}\ cm^{-2}$}}

\def\ergs{\textit{$\rm erg\ s^{-1}$}}
\def\ergcms{\textit{$\rm erg\ cm^{-2}\ s^{-1}$}}
\def\nhu{\textit{$\rm cm^{-2}$}}

\def\>{\textit{$\textgreater$}}
\def\<{\textit{$\textless$}}

\def\lx{\textit{$L_{\rm X}$}}

\def\fs{\textit{$f_{\rm exs}$}}
\def\fr{\textit{$f_{\rm ref}$}}
\def\gm{\textit{$\Gamma$}}
\def\nxv{\textit{$\sigma^2_{\rm nxv}$}}
\def\cf{\textit{$\rm CF_{tor}$}}
\def\aic{\textit{$\rm \Delta AIC$}}
\def\lb{\textit{$L_{\rm bol}$}}

\def\loglx{\textit{${\rm log}\,L_{\rm X}$}}
\def\lognh{\textit{${\rm log}\,N_{\rm H}$}}
\def\edd{\textit{$\rm \lambda_{Edd}$}}
\def\loglb{\textit{${\rm log}\,L_{\rm bol}$}}
\def\logedd{\textit{${\rm log\,} \rm \lambda_{Edd}$}}

\def\chandra{\emph{Chandra}}

\shorttitle{Highly Obscured and Compton-thick AGNs in CDFs}
\shortauthors{Li, Xue, Sun et al.}

\begin{document}

\title{Piercing Through Highly Obscured and Compton-thick AGNs in the \textit{Chandra} Deep Fields: I. X-ray Spectral and Long-term Variability Analyses}

\author{Junyao Li}
\affiliation{CAS Key Laboratory for Research in Galaxies and Cosmology, Department of Astronomy, University of Science and Technology of China, Hefei 230026, China; lijunyao@mail.ustc.edu.cn, xuey@ustc.edu.cn}
\affiliation{School of Astronomy and Space Science, University of Science and Technology of China, Hefei 230026, China}

\author{Yongquan Xue}
\affiliation{CAS Key Laboratory for Research in Galaxies and Cosmology, Department of Astronomy, University of Science and Technology of China, Hefei 230026, China; lijunyao@mail.ustc.edu.cn, xuey@ustc.edu.cn}
\affiliation{School of Astronomy and Space Science, University of Science and Technology of China, Hefei 230026, China}

\author{Mouyuan Sun}
\affiliation{CAS Key Laboratory for Research in Galaxies and Cosmology, Department of Astronomy, University of Science and Technology of China, Hefei 230026, China; lijunyao@mail.ustc.edu.cn, xuey@ustc.edu.cn}
\affiliation{School of Astronomy and Space Science, University of Science and Technology of China, Hefei 230026, China}

\author{Teng Liu}
\affiliation{Max-Planck-Institut f\"{u}r extraterrestrische Physik, Giessenbachstrasse 1, D-85748 Garching bei M\"{u} nchen, Germany}

\author{Fabio Vito}
\affiliation{Instituto de Astrofisica and Centro de Astroingenieria, Facultad de Fisica, Pontificia Universidad Catolica de Chile, Casilla 306, Santiago 22, Chile}
\affiliation{Chinese Academy of Sciences South America Center for Astronomy, National Astronomical Observatories, CAS, Beijing 100012, China}

\author{William N. Brandt}
\affiliation{Department of Astronomy  \& Astrophysics, 525 Davey Lab, The Pennsylvania State University, University Park, PA 16802, USA}
\affiliation{Institute for Gravitation and the Cosmos, The Pennsylvania State University, University Park, PA 16802, USA}
\affiliation{Department of Physics, The Pennsylvania State University, University Park, PA 16802, USA}

\author{Thomas M. Hughes}
\affiliation{CAS Key Laboratory for Research in Galaxies and Cosmology, Department of Astronomy, University of Science and Technology of China, Hefei 230026, China; lijunyao@mail.ustc.edu.cn, xuey@ustc.edu.cn}
\affiliation{School of Astronomy and Space Science, University of Science and Technology of China, Hefei 230026, China}
\affiliation{CAS South America Center for Astronomy, China-Chile Joint Center for Astronomy, Camino El Observatorio \#1515, Las Condes, Santiago, Chile}
\affiliation{Instituto de F\'{i}sica y Astronom\'{i}a, Universidad de Valpara\'{i}so, Avda. Gran Breta\~{n}a 1111, Valpara\'{i}so, Chile}

\author{Guang Yang}
\affiliation{Department of Astronomy  \& Astrophysics, 525 Davey Lab, The Pennsylvania State University, University Park, PA 16802, USA}
\affiliation{Institute for Gravitation and the Cosmos, The Pennsylvania State University, University Park, PA 16802, USA}

\author{Paolo Tozzi}
\affiliation{Istituto Nazionale di Astrofisica (INAF) -- Osservatorio Astrofisico di Firenze, Largo Enrico Fermi 5, I-50125 Firenze Italy}

\author{Shifu Zhu}
\affiliation{Department of Astronomy  \& Astrophysics, 525 Davey Lab, The Pennsylvania State University, University Park, PA 16802, USA}
\affiliation{Institute for Gravitation and the Cosmos, The Pennsylvania State University, University Park, PA 16802, USA}

\author{Xuechen Zheng}
\affiliation{CAS Key Laboratory for Research in Galaxies and Cosmology, Department of Astronomy, University of Science and Technology of China, Hefei 230026, China; lijunyao@mail.ustc.edu.cn, xuey@ustc.edu.cn}
\affiliation{School of Astronomy and Space Science, University of Science and Technology of China, Hefei 230026, China}
\affiliation{Leiden Observatory, Leiden University, PQ Box 9513, NL-2300 RA Leiden, the Netherlands}

\author{Bin Luo}
\affiliation{School of Astronomy and Space Science, Nanjing University, Nanjing 210093, China}
\affiliation{Key Laboratory of Modern Astronomy and Astrophysics (Nanjing University), Ministry of Education, Nanjing, Jiangsu 210093, China}
\affiliation{Collaborative Innovation Center of Modern Astronomy and Space Exploration, Nanjing 210093, China}

\author{Chien-Ting Chen}
\affiliation{Astrophysics Office, NASA Marshall Space Flight Center, ZP12, Huntsville, AL 35812}

\author{Cristian Vignali}
\affiliation{Dipartimento di Fisica e Astronomia, Alma Mater Studiorum, Universit\`a degli Studi di Bologna, Via Gobetti 93/2, I-40129 Bologna, 
Italy}
\affiliation{INAF -- Osservatorio di Astrofisica e Scienza dello Spazio di Bologna, 
Via Gobetti 93/3, I-40129 Bologna, Italy}

\author{Roberto Gilli}
\affiliation{INAF -- Osservatorio di Astrofisica e Scienza dello Spazio di Bologna, 
Via Gobetti 93/3, I-40129 Bologna, Italy}

\author{Xinwen Shu}
\affiliation{Department of Physics, Anhui Normal University, Wuhu, Anhui, 241000, China}

\begin{abstract}
\noindent
We present a detailed X-ray spectral analysis of 1152 AGNs selected in the \chandra~Deep Fields (CDFs), in order to identify highly obscured AGNs (\nh~\>~\cm). By fitting spectra with physical models, 436 (38\%) sources with $\lx > 10^{42}\ \ergs$ are confirmed to be highly obscured, including 102 Compton-thick (CT) candidates. 
We propose a new hardness-ratio measure of the obscuration level which can be used to select highly obscured AGN candidates. The completeness and accuracy of applying this method to our AGNs are 88\% and 80\%, respectively.
The observed log$\,N-$ log$\,S$ relation favors cosmic X-ray background models that predict moderate (i.e., between optimistic and pessimistic) CT number counts.
19\% (6/31) of our highly obscured AGNs that have optical classifications are labeled as broad-line AGNs, suggesting that, at least for part of the AGN population, the heavy X-ray obscuration is largely a line-of-sight effect, i.e., some high-column-density clouds on various scales (but not necessarily a dust-enshrouded torus) along our sightline may obscure the compact X-ray emitter.
After correcting for several observational biases, we obtain the intrinsic \nh~distribution and its evolution. The CT-to-highly-obscured fraction is roughly 52\% and is consistent with no evident redshift evolution.  We also perform long-term ($\approx$ 17 years in the observed frame) variability analyses for 31 sources with the largest number of counts available. Among them, 17 sources show flux variabilities: 31\% (5/17) are caused by the change of \nh, 53\% (9/17) are caused by the intrinsic luminosity variability, 6\% (1/17) are driven by both effects, and 2 are not classified due to large spectral fitting errors.
\end{abstract}


\section{introduction} \label{sec:intro}
Highly obscured active galactic nuclei (AGNs), which are defined as AGNs with hydrogen column density (\nh) larger than $\rm 10^{23}\ cm^{-2}$, are believed to represent a crucial phase of active galaxies. According to our knowledge of co-evolution of supermassive black holes (SMBHs) and their host galaxies \cite[for reviews, see, e.g.,][]{Alexander2012,KH2013} and the hierarchical galaxy formation model, AGN activity may be triggered in a dust enshrouded environment, into which gas inflows due to either internal \cite[e.g., disk instabilities; e.g.,][]{Hopkins2006} or external \citep[e.g., major mergers; e.g.,][]{Dimatteo2005} processes both fuel and obscure the SMBH accretion, resulting in short-lived heavily obscured AGNs \citep[e.g.,][]{Fiore2012, Morganti2017}. The subsequent AGN feedback process may blow out the obscuring material and leave out an unobscured optically bright quasar. Compared with unobscured AGNs, AGNs in the highly obscured phase tend to have smaller BH masses, higher Eddington ratios \cite[\edd; e.g.,][]{Lanzuisi2015} and larger merger fractions \citep[e.g.,][]{Kocevski2015, Ricci2017a}, which may indicate a fast growth state of central SMBHs \citep[e.g.,][]{Goulding2011}.
Moreover, the cosmic X-ray background (CXB) synthesis models also require a sizable population of highly obscured AGNs, or even Compton-thick (CT) AGNs (\nh~$\rm \gtrsim10^{24}\ cm^{-2}$; see, e.g., \citealt{Comastri2004,Xue2017,Hickox2018} for reviews), to reproduce the peak of CXB at 20$-$30 keV 
\citep[e.g.,][but see \citealt{Treister2009}]{Gilli2007}. Therefore, the study of highly obscured AGNs across cosmic epochs is vital for our understanding of the AGN triggering mechanism, SMBH growth, AGN environment and the origin of CXB.

Thanks to the powerful penetrability of high energy X-ray photons, X-ray observations provide a great window to uncover the mysterious veil of these heavily obscured sources that are likely missed in optical surveys. In the past twenty years or so, the deep X-ray surveys conducted by \chandra~\citep[e.g.,][]{Alexander2003, Xue2011, Xue2016, Luo2017},  \emph{XMM-Newton} \citep[e.g.,][]{Ranalli2013}, \emph{Swift}/BAT \citep[e.g.,][]{Baumgartner2013} and \emph{NuSTAR} \citep[e.g.,][]{Lansbury2017} have provided relatively unbiased AGN samples thanks to their unprecedented depths and sensitivities, which allow us to identify a significant population of heavily obscured AGNs \citep[e.g.,][]{Risaliti1999, Brightman2012, Ricci2015} using either X-ray color, spectral analysis and/or stacking technique \citep[e.g.,][]{Alexander2011, Iwasawa2012, Georgantopoulos2013, Brightman2014, Corral2014, DelMoro2016, Koss2016}. Moreover, the combination of mid-infrared (MIR), optical and X-ray data provides additional methods to select heavily obscured systems, such as MIR excess \citep[e.g.,][]{Daddi2007, Alexander2008, Luo2011}, high 24~$\mu m$ to optical flux ratio \citep[e.g.,][]{Fiore2009} and high X-ray to optical flux ratio \citep[e.g.,][]{Fiore2003}. 

Among various methods, X-ray spectroscopy provides the most direct and unambiguous way to measure the column density of the obscuring materials. Several previous studies have focused on deriving the intrinsic \nh~distribution corrected for the survey biases. \cite{Tozzi2006} presented a \nh~distribution that has an approximately log-normal shape peaking at $\sim$\cm~and with an excess at $\sim$\ct. \cite{Liu2017} (hereafter L17) reported a similar \nh~distribution that peaks in a higher \nh~range, due to the inclusion of more sources with low X-ray luminosities (\lx) and high redshifts than that of \cite{Tozzi2006}, which are expected to have relatively high \nh~values. However, both works only focused on bright AGNs, and neither was dedicated to or optimized for investigating highly obscured sources. In particular, L17 excluded CT AGNs in their work and only focused on the Compton-thin population. Hence the absorption distribution and evolution of the most deeply buried AGNs are still unclear, especially at high redshifts \citep{Vito2018}. Therefore, unveiling the apparently faint, CT regime using the deepest X-ray survey data is indispensable for fully understanding the entire AGN population.

There have also been several attempts to constrain the obscured AGN fraction and CT fraction on the basis of modeling the CXB \citep[e.g.,][]{Gilli2007, Akylas2012}, the X-ray luminosity function \citep[e.g.,][]{Aird2015, Buchner2015} or X-ray spectral analysis \citep[e.g.,][]{Burlon2011}.  Although most of the studies support the picture of evolved absorption with redshift and luminosity \citep[e.g.,][]{Hasinger2008, Liu2017}, the value of the CT fraction varies from study to study, ranging from a few percent to $\sim$50\%, largely due to the limited sample sizes, the use of different X-ray spectral models, the unknown contribution from Compton reflection \citep{Treister2009, Akylas2016} and the relatively poor quality of X-ray spectra at high \nh~circumstances. Therefore, deeper X-ray observations as well as the physically appropriate spectral models for highly obscured AGNs are needed to robustly characterize the obscuration properties.

Among the X-ray surveys, the \chandra~Deep Fields (CDFs) surveys (see, e.g., \citealt{Brandt2015} and \citealt{Xue2017} for reviews), which consist of the $\rm 7\ Ms$ \chandra~Deep Field-South survey \cite[CDF-S;][]{Luo2017} and the $\rm 2\ Ms$ \chandra~Deep Field-North survey \cite[CDF-N;][]{Xue2016} along with the $\rm 250\ ks$ Extended \chandra~Deep Field-South survey \cite[E-CDF-S;][]{Xue2016}, provide us the most promising data to study highly obscured AGNs.  In particular, the $\rm 7\ Ms$ CDF-S, which is the deepest X-ray survey to date, significantly improves the count statistics that allow us to extract high-quality X-ray spectra, detect more faint, highly obscured sources and perform more robust spectral analyses compared with previous 4~Ms analyses \citep[e.g.,][]{Brightman2014}. Furthermore, recent works suggest that the power spectral density (PSD) break frequency of AGN light curves might be related to \nh~variability \citep{Zhang2017, Gonzales2018}, which makes obscuration a very important factor in investigating the driving mechanism of AGN variability. Benefiting from the very long timespan of the $\rm 7\ Ms$ CDF-S data (16.4 years in the observed frame), we are able to, for the first time, quantify the detailed variability behavior for a large, dedicated sample of highly obscured AGNs, in order to better understand the location of the obscuring materials and their contribution to AGN variability \citep[also see][Y16 hereafter]{Yang2016}. 

In this study, we construct the largest dedicated highly obscured AGN sample in the deepest \chandra~surveys which enables us to extend the studies of deeply buried sources to lower luminosities and higher redshifts in great details, and present systematic X-ray spectral and long-term variability analyses to study their evolution and physical properties. This paper is organized as follows. We describe our data reduction procedure and sample selection in Section \ref{sec:data}. In Sections \ref{sec:spectra} and \ref{sec:spec_result}, we present detailed X-ray spectral analyses of our sample, focusing on the column density and luminosity distributions of highly obscured AGNs as well as their relations; the number counts of CT AGNs and the constraint on CXB models. The reprocessed components, the covering factor of the obscuring materials and their indications to AGN structures are discussed in Section \ref{sec:ref}. In Section \ref{sec:intrin_nh}, by correcting for several observational biases, we constrain the intrinsic \nh~distribution representative for the highly obscured AGN population and study its evolution across cosmic time. In Section \ref{sec:var},  we select a subsample of highly obscured AGNs that have largest counts available and perform detailed long-term X-ray variability analyses, in order to find out the variable fraction as well as the main driven mechanism of variability. 

Throughout this paper, we adopt a Galactic column density of $N \rm_H = 8.8 \times 10^{19}\ cm^{-2}$ for the CDF-S, and $N \rm_H = 1.6 \times 10^{20}\ cm^{-2}$ for the CDF-N, respectively \citep{Stark1992}. We adopt cosmological parameters of $\rm H_0 = 70.0\ km\ s^{-1}\ Mpc^{-1}$, $\rm \Omega_M = 0.30$, and $\rm \Omega_\Lambda = 0.70$. All given errors are at 1$\sigma$ confidence level unless otherwise stated. 

\section{data reduction and sample selection} \label{sec:data}
\begin{figure}
\includegraphics[width=\linewidth]{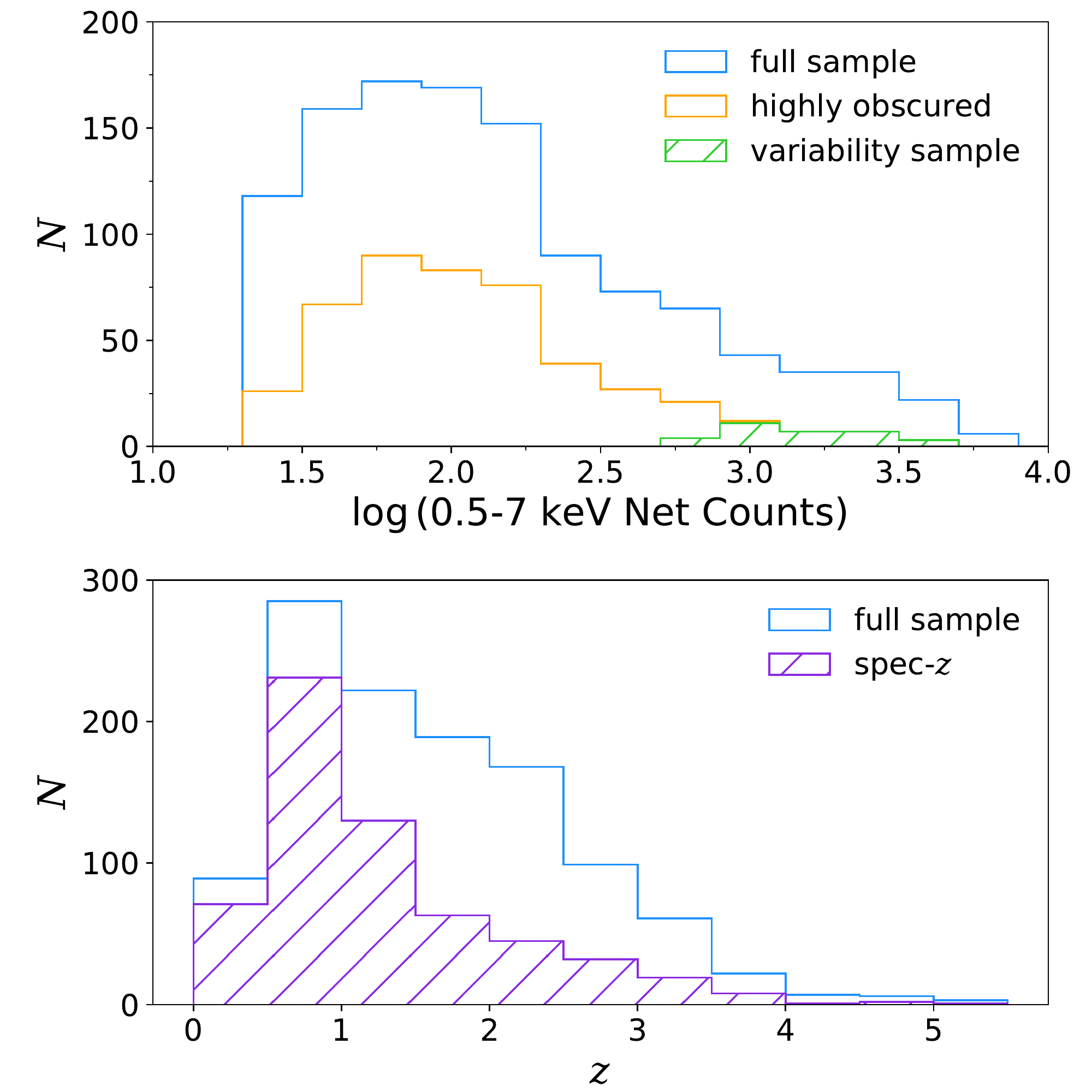}
\caption{Top: {\it Chandra} net counts distributions of the full sample of 1152 AGNs (blue solid histogram), the highly obscured sources confirmed with X-ray spectral fitting in Section~\ref{sec:spec_result} (orange dashed histogram), and the subsample selected to perform variability analyses in Section~\ref{sec:var} (solid green histogram), respectively. 
Bottom: Redshift distributions for the full sample (blue histogram) and the 603 sources with spectroscopic redshifts (purple histogram), respectively.
}
\label{fig:cnts}
\end{figure} 

This work is based on the \chandra~data in the $\rm 7\ Ms$ CDF-S main-source catalog \citep{Luo2017} and the $\rm 2\ Ms$ CDF-N main-source catalog \citep{Xue2016}. Since the exposure in the E-CDF-S is $\sim$8--28 times shallower than in the CDF-N and CDF-S, we exclude the E-CDF-S data in this work. 
The redshift information for each source is adopted from the two catalogs that presented a preferred redshift value given by the ZFINAL keyword. The ZFINAL redshift values were collected and selected based on various published catalogs and followed a general preference order of spectroscopic redshift over photometric redshift (see Section~4.3 of \citealt{Luo2017} and Section~2.3.4 of \citealt{Xue2016} for details). The redshift (spectroscopic redshift) completenesses in the CDF-S and the CDF-N main source catalogs are 99.4\% (64.8\%) and 95.2\% (52.4\%), respectively; and the corresponding mean $1\sigma$ photometric redshift errors are about 0.21 and 0.17, respectively.

The source spectra from individual observations were extracted using the ACIS Extract (AE) software package \citep{Broos2010}. AE generates the point-spread function (PSF) model based on the MARX ray-tracing simulator and constructs a polygonal extraction region that corresponds to an encircled-energy fraction of $\sim$90\%. For crowded sources, AE adopts smaller extraction regions to avoid overlapping polygonal regions. The background spectra were extracted using the {\tt{BETTER\_BACKGROUNDS}} algorithm. The most significant aspect of the above photometry and spectral extraction procedure, compared to the widely used circular-aperture extraction, is that it can obtain photometry and spectra as accurate as possible and remove the contamination from neighboring sources to faint sources to the greatest extent (see Section~3.2 of \citealt{Xue2011} for details). This is extremely important for our work since we are dealing with the highly obscured AGNs that are generally fainter and expected to have limited counts due to significant obscuration.

The spectra eventually used in this work are the merged spectra for which all the individual observations were matched to a common $K_s$-band astrometric frame (see Section 2.2.1 in \citealt{Xue2016} ) and stacked using the {\tt{MERGE\_OBSERVATIONS}} algorithm in AE. The corresponding response matrix files (rmf) and ancillary response files (arf) were also generated and combined during this stage. 

We construct our sample by selecting sources which (1) are classified as AGN (TYPE=AGN; see Section 4.5 in \citealt{Luo2017} for details); (2) have 0.5--7~keV net counts $> 20$ (FB$\_$COUNTS $>$ 20) to allow basic X-ray spectral fitting; and (3) have redshift measurements (ZFINAL $!= -1$) in the main catalogs. 
The resulting full sample consists of 1152 sources, with 660 from the CDF-S and 492 from the CDF-N, respectively. 
The counts and redshift distributions for the full sample are shown in Figure \ref{fig:cnts}. The median counts are 112, and 53\% (33\%) of the sources have counts larger than 100 (200).
The median redshift is 1.45, and 603 (52\%) sources have spectroscopic redshifts.

\begin{figure*}
\centering
\includegraphics[width=0.8\linewidth]{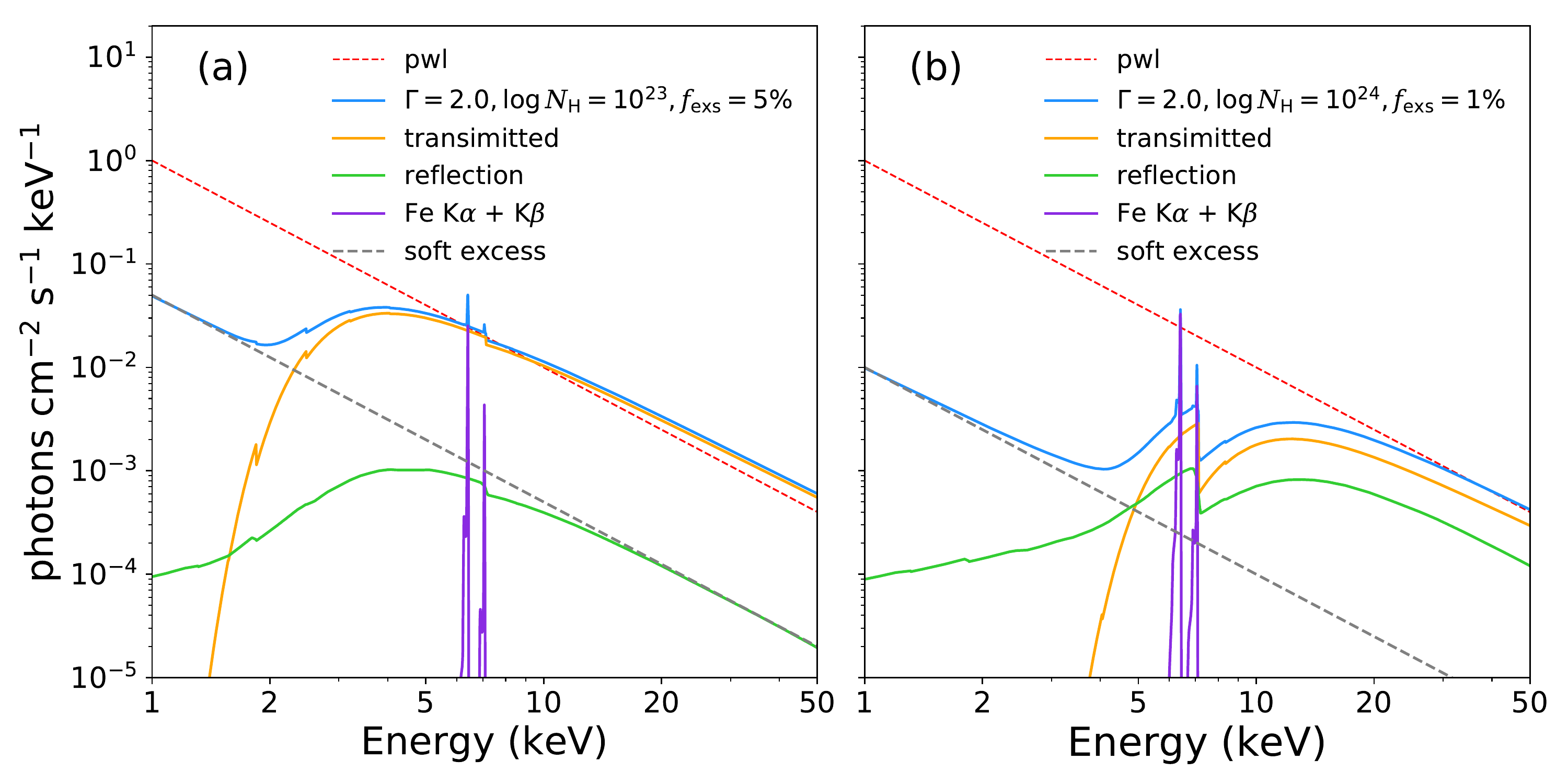}
\caption{Examples of MYTorus-based models with different parameters used in spectral fitting.  The unattenuated continuum (pwl), absorbed continuum (blue curve), transmitted zeroth-order continuum, reflection component, iron emission lines and the soft excess component are shown in different colors. (a) Model used to generate curve A in Figure \ref{fig:HR} with \nh, \gm~and \fs~fixed at \cm, 2.0 and 5\%, respectively. (b) Model used to generate curve B in Figure \ref{fig:HR} with \nh, \gm~and \fs~fixed at \ct, 2.0 and 1\%, respectively.}
\label{fig:curve}
\end{figure*} 

\section{X-ray Spectral Analysis}\label{sec:spectra}
\subsection {Spectral Fitting Models} \label{subsec:model}
\begin{figure}
\includegraphics[width=\linewidth]{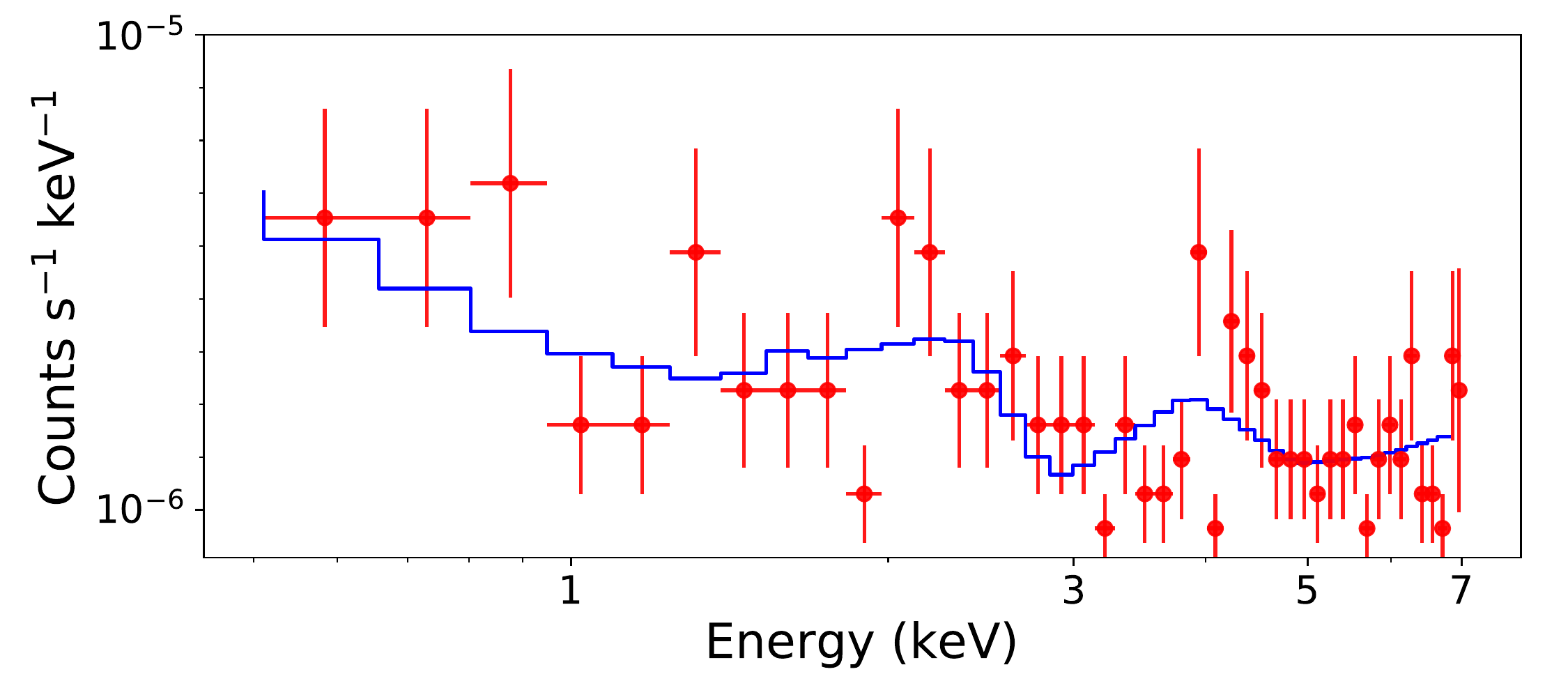}
\caption{An example of fitting the \chandra~background spectrum that has 332 counts using the \textit{cplinear} model (binned only for illustration purpose).}
\label{fig:bkg}
\end{figure}

We use MYTorus-based models \citep{Murphy2009a} to fit the observed-frame 0.5--7~keV spectra
of the full sample in order to identify heavily obscured sources. 
Due to limited counts, we do not bin the spectra because it may lose some key information of the sources. We use the Cash statistic \citep[Cstat in \xspec;][]{Cash1979} as our spectral fitting statistic. Cstat has a similar probability distribution to $\chi ^2$ statistics and has been proved to be more appropriate in the low-counts regime. Since Cstat is not appropriate for the background-subtracted spectra, we simultaneously fit the source and background spectra, with the latter (with 1642 median counts) being fit with the $cplinear$ model \citep{Broos2010} that properly describes the observed {\it Chandra} background by a continuous piecewise-linear function in ten energy segments (an example is shown in Figure \ref{fig:bkg}). In this way, we are able to maximize the usage of information relevant to the sources. 

We adopt two models to fit the source spectra with different components and degrees of freedom (d.o.f) in order to find a statistically robust best-fit one:
\begin{enumerate}[1.]

\item The MYTorus baseline model:
$phabs \times {\rm (MYTZ} \times \\zpow + f_{\rm ref} \times {\rm MYTS} + f_{\rm ref} \times gsmooth {\rm (MYTL)})$.

\item The soft-excess model:
$phabs \times {\rm (MYTZ} \times \\zpow + f_{\rm ref} \times {\rm MYTS} + f_{\rm ref} \times gsmooth {\rm (MYTL)} + \fs \times zpow)$.

\end{enumerate}

These models include all the typical spectral features found in highly obscured AGNs. The $phabs$ component models the Galactic photoelectric absorption. The MYTZ$\times zpow$ term represents the zeroth-order transmitted power-law continuum across the torus that takes into account both photoelectric absorption and Compton scattering processes. MYTS and $gsmooth\,$(MYTL) stand for the reflection component and the broadened Fe $\rm K{\footnotesize \alpha}$, $\rm K{\footnotesize \beta}$ fluorescent emission lines, respectively. A second power law is added to represent the soft excess component often found in AGN spectra, possibly originating from the zeroth-order continuum being scattered by the extended Compton-thin (CN) materials in obscured AGNs \citep{Guainazzi2005, Bianchi2006, Corral2011}. During spectral fitting, the inclination angle $\theta$ is fixed at $75^\circ$, which is the average value within a range of $60^\circ$ to $90^\circ$ where the torus intercepts our line of sight (l.o.s). The nominal normalizations of MYTS, MYTL and the second power law are set to be the same as that of the first power law, and we use the constants \fr~and \fs~to represent the real normalizations. \fs~is allowed to vary between $0-0.1$, and we assume a 200 keV high-energy cutoff throughout. One thing to mention is that the best-fit \nh~values in MYTorus represent the equatorial values, and we always use the l.o.s values $N_{{\rm H,l.o.s}}=(1-4\,\cos^2\theta)^{\frac{1}{2}} N \rm _H$ in this work.

\subsection {Model Selection} \label{subsec:select}
Several previous works suggested that the inclusion of the soft excess and reflection components in the spectral models is crucial for us to correctly estimate \nh~of highly obscured sources \citep[e.g.,][]{Brightman2014, Lanzuisi2015}. However, the low counts of many sources do not allow us to apply complex models with free parameters since it could lead to large degeneracies. Therefore, we choose the model components and the parameter spaces according to the following criteria:

\begin{enumerate}[1.]
\item We fix \gm~at 1.8 \citep[e.g.,][]{Tozzi2006, Marchesi2016} for the 851 sources with counts less than 300. 
For the 301 sources with counts larger than 300, we set \gm~free to find a best-fit value. 

\item For all sources, we first fit the spectra with a free $f_{\rm ref}$. If \fr~is less than $10^{-5}$, we then fix it to $10^{-8}$ that is an arbitrary value set as the lower limit for \fr~(i.e., indicating a negligible reflection component); otherwise, we fix \fr~at 1 that is the default value adopted in MYTorus.

\item To determine whether we should add a soft excess component, we compare the Cstat between models with and without considering soft excess emission. The best-fit model is chosen to be the one that has a statistically robust low Cstat value. More specifically, adding a new soft excess component should improve the Cstat at least for $\Delta C=3.84$ with 1 more d.o.f. This criterion is based on the fact that $\Delta C$ approximately follows the $\chi^2$ distribution, thus $\Delta C = 3.84$ is roughly consistent with a \>~95\% confidence level \citep{Tozzi2006, Liu2017}.

\item In order to avoid extremely untypical \gm~caused by the degeneracy between \gm~and \nh, we re-fit the spectra of those sources with \gm~pegged at 1.4 (i.e., the lowest value permitted by MYTorus) or \gm~\>~2.4 \citep[i.e., the typical maximum photon index for high-\edd~AGNs; e.g.,][]{WangJM2004, Fanali2013} by fixing their \gm~at 1.8. If $\Delta C=C_{\rm fix} - C_{\rm free}$ \>~3.84, we adopt the free \gm~value; otherwise, we adopt \gm=1.8.
\end{enumerate}

\begin{table}
\centering
\caption{Model definitions used in this work}
\begin{tabular}{c c c c}
\hline
\hline
model & \fr~& \fs\\
\hline
A & $10^{-8}$ & 0\\
B & $10^{-8}$ & $0 < \fs~ < 0.1$\\
C & $1.0$ & 0\\
D & $1.0$ & $0 < \fs~ < 0.1$\\
\hline
\end{tabular}
\label{table:modeldef}
\end{table}

From now on, for convenience, we refer model 1 (2) with negligible \fr~as model A (B) and model 1 (2) with \fr~fixed at 1.0 as model C (D), as summarized in Table \ref{table:modeldef}. In order to ensure the consistency of the models used for subsequent bias corrections (see Section  \ref{sec:intrin_nh}) as much as possible, we have to assume fixed parameters in case of low counts.
We justify our models in Appendix \ref{app:model} by validating that the usage of fixed parameters does not significantly affect our results. We also compare our fitting results with several previous works (see Section \ref{subsec:cmp_pre}) and those obtained from the Borus model (\citealt{Balokovic2018}; see Section \ref{subsec:ref}) and find good consistency.

We note that the simple absorbed power-law model (e.g., $phabs \times zwabs \times zpow$; hereafter model $z$) has been widely used in the literature to obtain rough estimates of AGN parameters. However, such a model is not appropriate for highly obscured AGNs since $zwabs$ only models photoelectric absorption and does not take into account the Compton scattering process, which is particularly important in the high \nh~regime. Therefore, we also fit the spectra using model $z$, aiming at directly testing how accurately such a simple model reproduces the main spectral parameters.

\begin{table*}
\centering
\caption{X-ray spectral fitting results for highly obscured AGNs}
\begin{tabular}{c c c c c c c c c c c c}
\hline
\hline
XID & field & RA & DEC & $z$ & ztype & HR & \gm & \nh & \lx &  counts & model\\
(1) & (2) & (3) & (4) & (5) & (6) & (7) & (8) & (9) & (10) & (11) & (12)\\
\hline
   8 & CDF-N & 188.841072 & 62.250256 & 2.794 &  zphot  &  0.17 & 1.80f & $24.42$ & 45.30 & 39 & A\\
  10 & CDF-N & 188.846392 & 62.292835 & 1.173 &  zphot & 0.07 & 1.80f & $23.88$ & 43.15 & 41 & C\\
  11 & CDF-N & 188.847062 & 62.217590 & 1.498 &  zphot & -0.23 & 1.80f & $24.63$ & 44.90 & 51 & D\\
  14 & CDF-N & 188.853852 & 62.256847 & 1.652 &  zphot & 0.59 & 1.80f & $23.81$ & 44.22 & 105 & A\\
  16 & CDF-N & 188.869802 & 62.240976 & 1.732 &  zphot & 0.25 & 1.80f & $23.37$ & 43.97 & 197 & A\\
\hline
\end{tabular}

\vspace{5.0 pt}
{\sc \bf{Notes.}}
Column 1: source ID in the \cite{Luo2017} and \cite{Xue2016} catalogs. Column 2: field. Columns 3 and 4: right ascension (RA) and declination (DEC) for the X-ray source position. Column 5: ZFINAL redshift. Column 6: redshift type (``zspec'': spectroscopic; ``zphot'': photometric). Column 7: hardness ratio. Column 8: photon index (``f'': fixed value). Column 9: line-of-sight column density in units of \ct. Column 10: logarithm of absorption-corrected rest-frame 2--10~keV luminosity in units of erg~s$^{-1}$. Column 11: 0.5--7~keV background-subtracted net counts. Column 12: best-fit model (see Section~\ref{subsec:model}). The full version of this table is available online.
\label{table:fitting_result}
\end{table*}

\begin{table*}
\centering
\caption{The best-fit models for highly obscured sources with 0.5--7~keV net counts \<~100, $\rm 100 \leq counts < 300$, counts $\geq$ 300, counts $\geq$ 1000, \cm~\<~$\nh < \ct$, and \nh~$\geq$ \ct, respectively}
\begin{tabular}{c c c c c c c c c c c c c c c c c}
\hline
\hline  
cases & $N$ & A & B & C & D & $\frac{\rm C+D}{N}$ & $N_s$ & A$_s$ & B$_s$ & C$_s$ & D$_s$ & $\frac{{\rm C}_s+{\rm D}_s}{N_s}$\\
(1)   & (2) & (3)&(4)&(5)&(6)&(7)&(8)&(9)&(10)&(11)&(12)&(13)\\
\hline
$<$ 100 counts & 210 & 80 & 21 & 89 & 20 & 51.9\% & 79 & 14 & 11 & 38 & 16 & 68.4\%\\
$100-300$ counts & 146 & 51 & 7 & 69 & 19 & 60.3\% & 67 & 17 & 5 & 29 & 16 & 67.2\%\\
$\geq 300$ counts & 80 & 14 & 5 & 52 & 9 & 76.2\% & 45 & 4 & 2 & 31 & 8 & 86.7\%\\
$\geq 1000$ counts & 18 & 3 & 2 & 9 & 4 & 72.2\% & 14 & 3 & 1 & 6 & 4 & 71.4\%\\
\cm~\<~\nh~\<~\ct & 328 & 120 &  17 & 166 & 24 & 57.9\% & 136 & 27 & 10 & 80 & 19 & 72.8\%\\
\nh~$\geq$ \ct & 108 & 25 & 16 & 44 & 23 & 62.0\% & 55 & 8 & 8 & 18 & 21 & 70.9\%\\
\nh~$\geq$ \ct \ and \ count $\geq$ 150 &16 & 3 & 1 & 6 & 6 & 75.0\% & 9 & 1 & 1 & 1 & 6 & 77.8\%\\
total & 436 & 145 & 33 & 210 & 48 & 58.9\% & 191 & 35 & 18 & 98 & 40 & 72.3\%\\
\hline
\end{tabular}

\vspace{5.0 pt}
{\sc \bf{Notes.}}
Column 1: cases of four count bins and two \nh~ranges corresponding to CN and CT AGNs. Column 2: total source number in each case. Columns 3$-$6: numbers of sources that are best-fitted by models A$-$D, respectively. 
Note that models C and D have \fr~fixed at 1.0 while models A and B have negligible \fr~thus negligible reflection component and Fe K lines, and we show the fraction of the sources that have reflection and iron emissions lines in Column 7.
Columns 8--13: same as Columns 2--7, but for the spectroscopic-redshift subsample.
\label{table:model}
\end{table*}

\section{Spectral Fitting Results}\label{sec:spec_result}

On the basis of the spectral fitting results, we find that 39\% (458/1152) of sources in the full sample are identified to be highly obscured AGNs (hereafter the highly obscured sample; HOS). The main fitting results for HOS using MYTorus models A--D are presented in Table \ref{table:fitting_result} and the full version is available on line. The best-fit models for sources in different count bins and \nh~ranges are summarized in Table \ref{table:model}. In the following analyses, we only consider the 436 sources that have the absorption-corrected, rest-frame 2--10~keV luminosity (calculated from the {\tt{lumin}} command after deleting all the additional components and only keeping the $zpow$ component) \lx~\>~$\rm 10^{42}\ erg\ s^{-1}$ to avoid possible contamination from star-forming galaxies.

\begin{figure*}
\includegraphics[width=\linewidth]{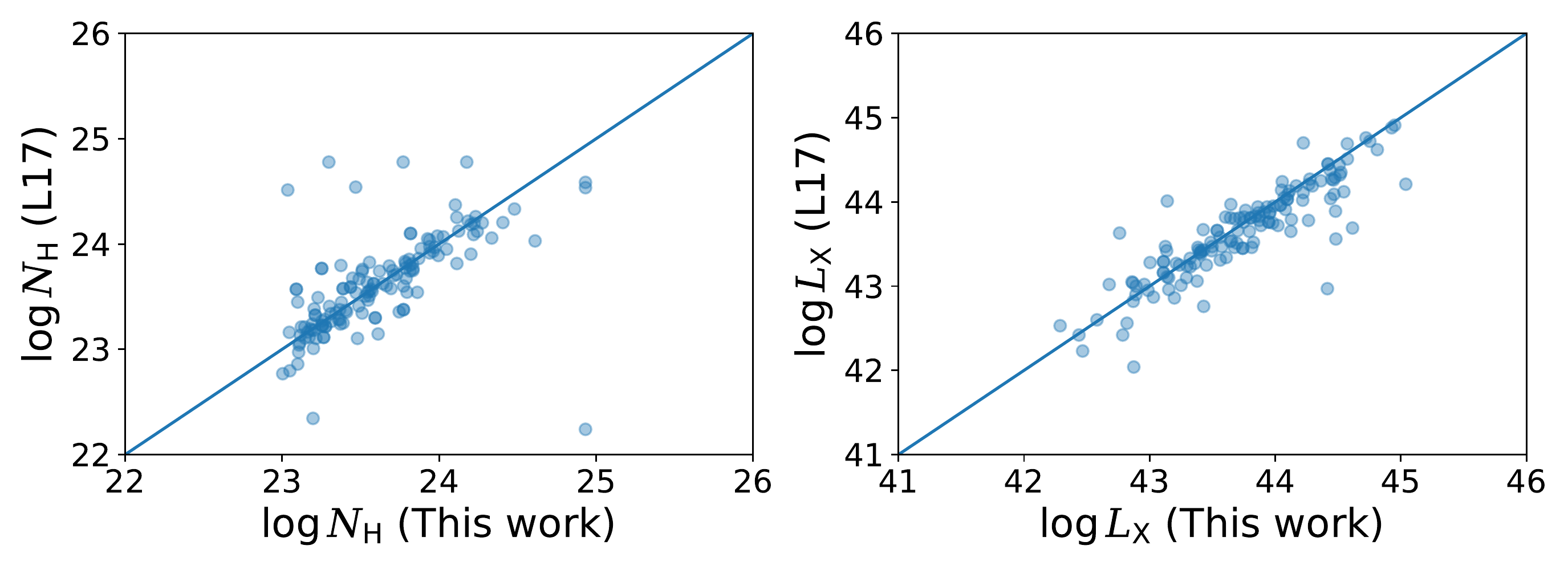}
\includegraphics[width=\linewidth]{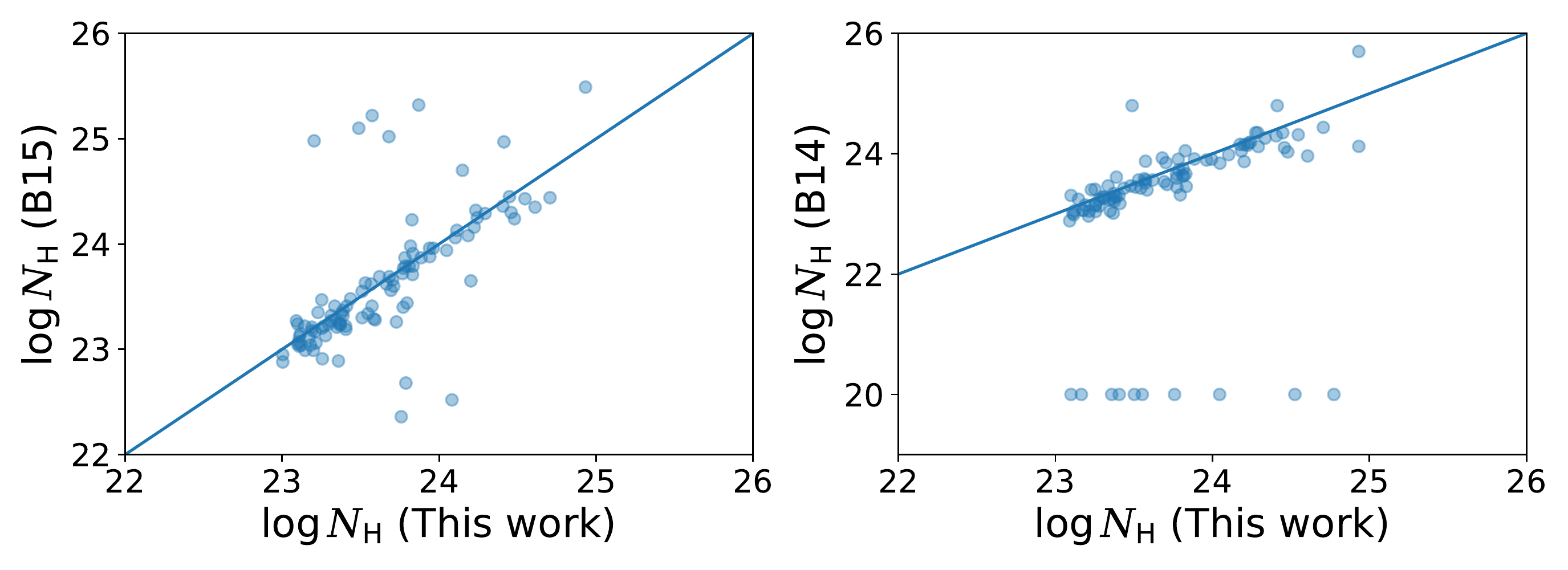}
\caption{Comparing the spectral fitting results with \cite{Liu2017}, \cite{Brightman2014} and \cite{Buchner2015} for common sources with redshift difference $\Delta z < 0.2$.}
\label{fig:cmp_pre}
\end{figure*}

\subsection {Comparisons With Previous Works and the Model of $phabs \times zwabs \times zpow$} \label{subsec:cmp_pre}
Thanks to the increased exposure time, the improved data reduction procedure (see Table 1 in \citealt{Xue2016} for a summary), the updated redshift measurements and the usage of physically appropriate spectral fitting models for highly obscured AGNs, we are able to extract higher-quality spectra and obtain more robust parameter constraints than previous works in the same field \citep[e.g.,][]{Tozzi2006}, especially for faint sources. Before performing further analyses, we first make direct comparisons with several works which presented spectral fitting parameters in the CDF-S to understand how different methods influence the spectral fitting results. The works used for comparisons are:

\begin{enumerate}
\item \cite{Liu2017} (L17): L17 performed an extensive X-ray spectral analysis for the bright sources (hard-band counts $>$ 80) in the 7 Ms CDF-S using the $wabs \times (plcabs \times power-law+ zgauss + power-law + plcabs \times pexrav \times constant)$ model. The background spectrum was also fitted with the \textit{cplinear} model. The redshift information used in the two works is the same.

\item \cite{Brightman2014} (B14): B14 presented X-ray spectral fitting results in the 4 Ms CDF-S using the BNtorus model \citep{Brightman2011}. 112 sources are found common between the two works; 24 of them have redshifts different by more than 0.2 ($\Delta z > 0.2$) compared with the values adopted in the updated 7 Ms catalog, and are neglected in the following comparison.

\item \cite{Buchner2015} (B15): The 4~Ms~CDF-S spectra were analyzed using a physically motivated torus model and a Bayesian methodology to estimate spectral parameters. 114 sources are found common between the two works and 30 of them with $\Delta z > 0.2$ are excluded.

\end{enumerate}

As shown in Figure \ref{fig:cmp_pre}, our measured \nh~and \lx~values are in general agreement with previous works, despite that for individual sources, the usage of different model configurations and data may result in large discrepancies. The largest distinction happens between B14 and our work that ten highly obscured AGNs in our sample are reported to be unobscured in B14. To understand the discrepancy, we re-fit their spectra using the same model and method as described in B14 and the results again confirm their heavily obscured nature. Therefore, the large discrepancy could be due to the adopted data of different depths as well as the different data reduction and spectral extraction methods used in the two works.

\begin{figure*}
\includegraphics[width=\linewidth]{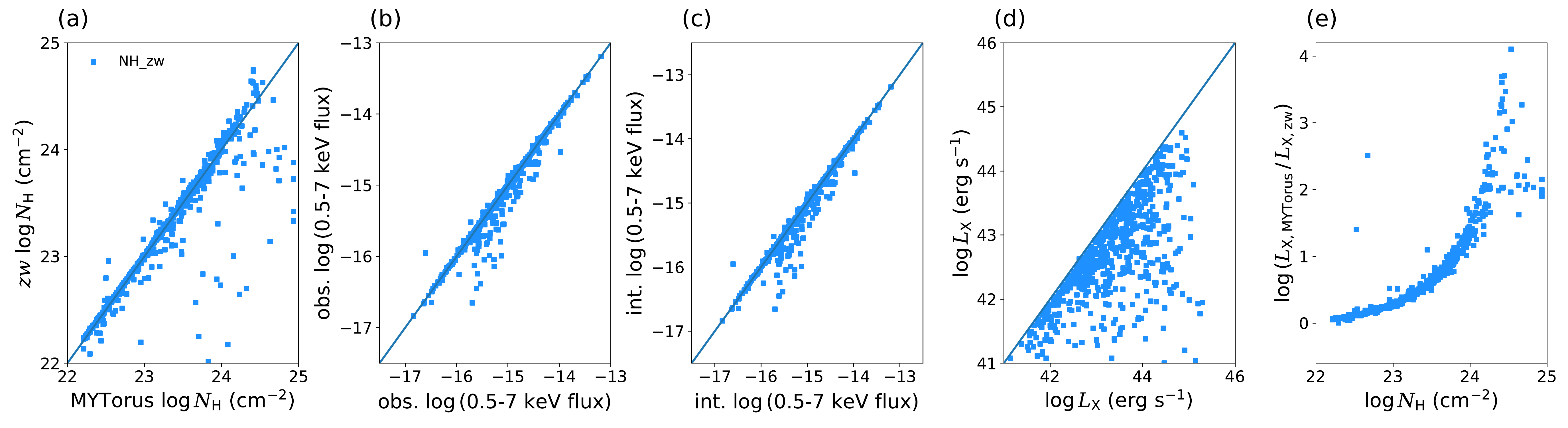}
\caption{Comparing the spectral fitting results between using the MYTorus ($x$-axes) and $phabs \times zwabs \times zpow$ ($y$-axes in all panels but $e$) models. $a$-$d$: Comparison results of \lognh, observed 0.5--7~keV flux, intrinsic 0.5--7~keV flux and intrinsic 2--10~keV luminosity, respectively. $e$. The log-ratio of the intrinsic 2--10~keV luminosity between MYTorus and model $z$ as a function of the MYTorus \nh. 
}
\label{fig:zwabs}
\end{figure*}

The comparisons between the results obtained through MYTorus and model $z$ are shown in Figure \ref{fig:zwabs}. Note that since MYTorus does not allow \nh~to vary below $10^{22}\ \nhu$, a number of unobscured sources are thus have best-fit $\nh = 10^{22}\ \nhu$. Therefore, we only consider sources with MYTorus $\lognh > 22.2\ \nhu$ in the comparison. 

The \nh~and the observed 0.5--7~keV flux derived by the two models are generally consistent. But at a given \nh, the absorption-corrected intrinsic $\rm 0.5 - 7\ keV$ flux ratio and the intrinsic \lx~ratio between the two models dramatically increase with \nh. It is obvious that model $z$ underestimates the intrinsic luminosity due to neglect of the Compton scattering process, and the discrepancy is already evident at $\nh \sim \cm$ \citep[see also][]{Burlon2011}. Therefore such models with only photoelectric absorption taken into consideration (e.g., $zwabs$, $zphabs$, $ztbabs$) must be used with caution in the highly obscured regime.

\subsection {Photon Index Distribution}\label{subsec:gamma}

\begin{figure}
\includegraphics[width=\linewidth]{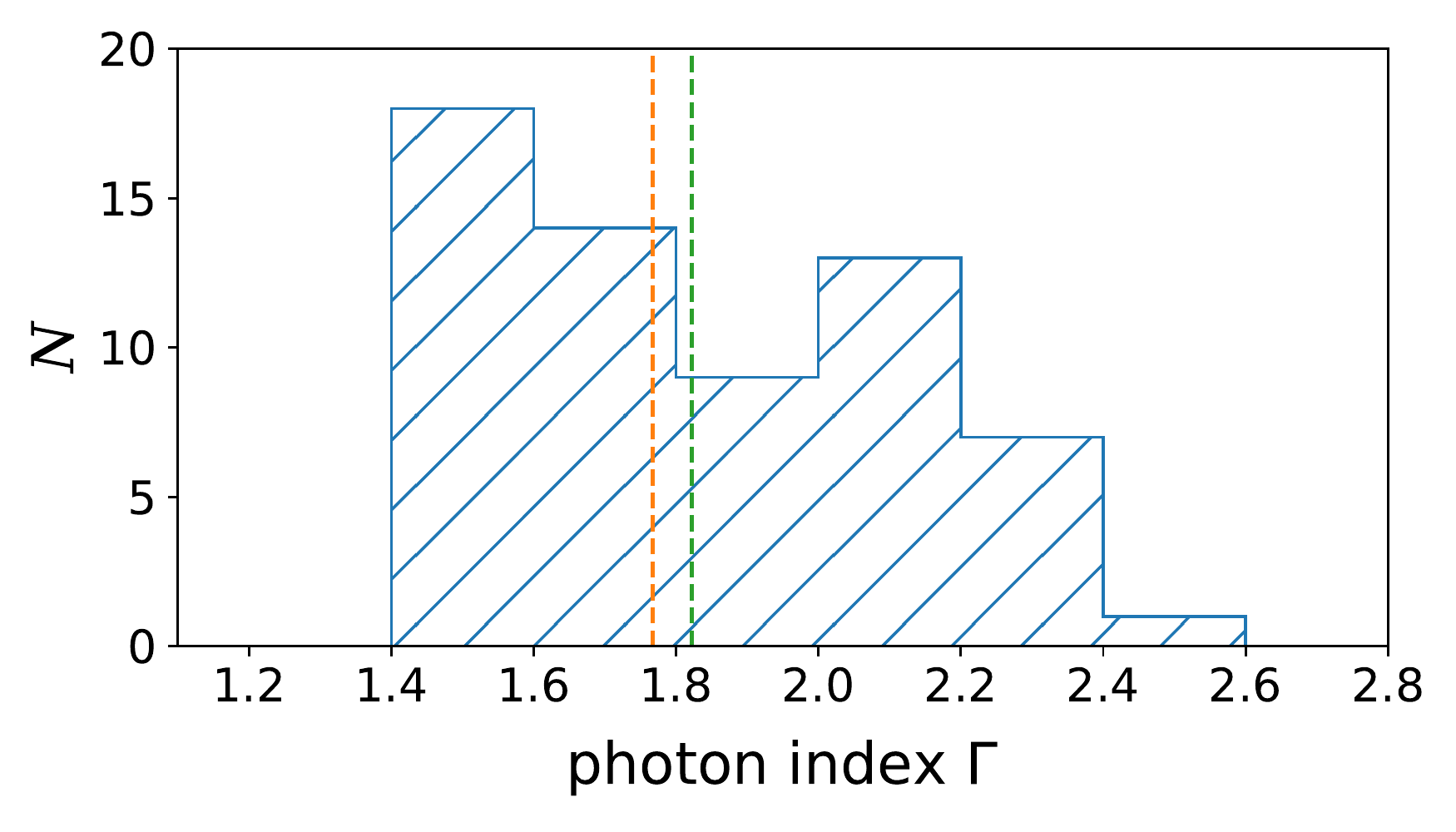}
\caption{The distribution of the best-fit photon index for the 62 sources with free \gm. The green and red vertical dashed lines show the median (\gm~= 1.77) and mean (\gm~= 1.82) values, respectively.}
\label{fig:gamma}
\end{figure}

The photon index distribution for the 62 highly obscured sources with free-\gm~during the fitting process is shown in Figure \ref{fig:gamma}. 
The mean value of the distribution is $1.82\pm0.04$, in agreement with previous X-ray spectral analyses in the CDF-S \citep[e.g.,][]{Tozzi2006, Liu2017} and COSMOS \citep[e.g.,][]{Lanzuisi2013}. 
Note that our \gm~distribution peaks at \gm~=~1.4 instead of the mean value 1.8. This is because the photon index is restricted within $1.4-2.6$ in MYTorus. 
If we use MYTZ alone which does not limit the \gm~range to fit the spectra, most sources with \gm~pegged at 1.4 will have smaller \gm, thus the distribution will appear more symmetric with a larger dispersion. 

We also try to search the potential correlations among \gm, \nh~and \lx~using the Spearman rank correlation test. No correlation between \gm~and redshift (Spearman's $\rho$ = 0.00, $p$-value = 0.99) is detected, indicating that the inner disk and corona structures have little evolution across cosmic time. The Spearman tests also suggest no significant correlation between \gm~and \nh~(\lx).

\begin{table}
\caption{Information of sources with extreme photon indexes}
\footnotesize
\begin{tabular}{c | c c c c c c}
\hline
\hline
field & ID & $\Gamma$ & $N\rm _H$ & counts & ztype & zqual\\
\hline
CDFS &   98 & 1.40 & 0.24$^{+0.02}_{-0.01}$ & 3689 & zphot & insecure\\
&  135 & 2.53 & 1.59$^{+0.21}_{-0.24}$ &  617 & zspec & insecure\\
&  172 & 1.40 & 0.33$^{+0.07}_{-0.08}$ &  534 & zphot & ...\\
&  243 & 1.40 & 0.12$^{+0.01}_{-0.01}$ &  372 & zspec & secure\\
&  249 & 1.40 & 0.18$^{+0.02}_{-0.02}$ &  830 & zphot & ...\\
&  597 & 1.40 & 0.24$^{+0.03}_{-0.04}$ &  351 & zspec & secure\\
&  760 & 1.40 & 0.72$^{+0.08}_{-0.07}$ &  351 & zspec & insecure\\
\hline
CDF-N &  546 & 1.40 & 0.11$^{+0.01}_{-0.01}$ &  456 & zspec & secure\\
\hline
\end{tabular}

\vspace{5.0 pt}
{\sc \bf{Notes.}}
Columns are the same as in Table~\ref{table:fitting_result}. The zqual column represents the quality of the spectroscopic redshift (Secure or Insecure). For sources with insecure spectroscopic redshifts, the X-ray source catalogs may choose to adopt the photometric redshifts as ZFINAL (see Section 4.4 in \citealt{Xue2011} for details).
\label{table:peculiar}
\end{table}

It is noteworthy that there are 8 sources with \gm~= 1.4 or \gm~\>~2.5 (see Table \ref{table:peculiar}).
Unlike some of the high-count sources that we artificially fix \gm~at 1.8 (see Section \ref{subsec:select}), these sources have significantly worse fits if we do not allow \gm~to vary freely (the average improvement of $\Delta C$ is 9.0 if we set \gm~free). Sources with very flat photon indexes are often considered to be reflection-dominated CT candidates and their extremely obscured nature may not be revealed by our best-fit \nh~\citep{Georgantopoulos2009, Georgantopoulos2011}. Moreover, since we are dealing with the stacked spectra here, sources which possessed large spectral variability may also exhibit untypical spectral shape, as might be the case for XID 249 (see Section \ref{subsec:srcvariab}). Additionally, we cannot rule out the possibility that the extreme photon indexes of some sources are wrongly measured due to insecure photometric redshifts.

\subsection {Hardness Ratio versus Redshift}\label{subsec:hr}

\begin{figure}
\includegraphics[width=0.95\linewidth]{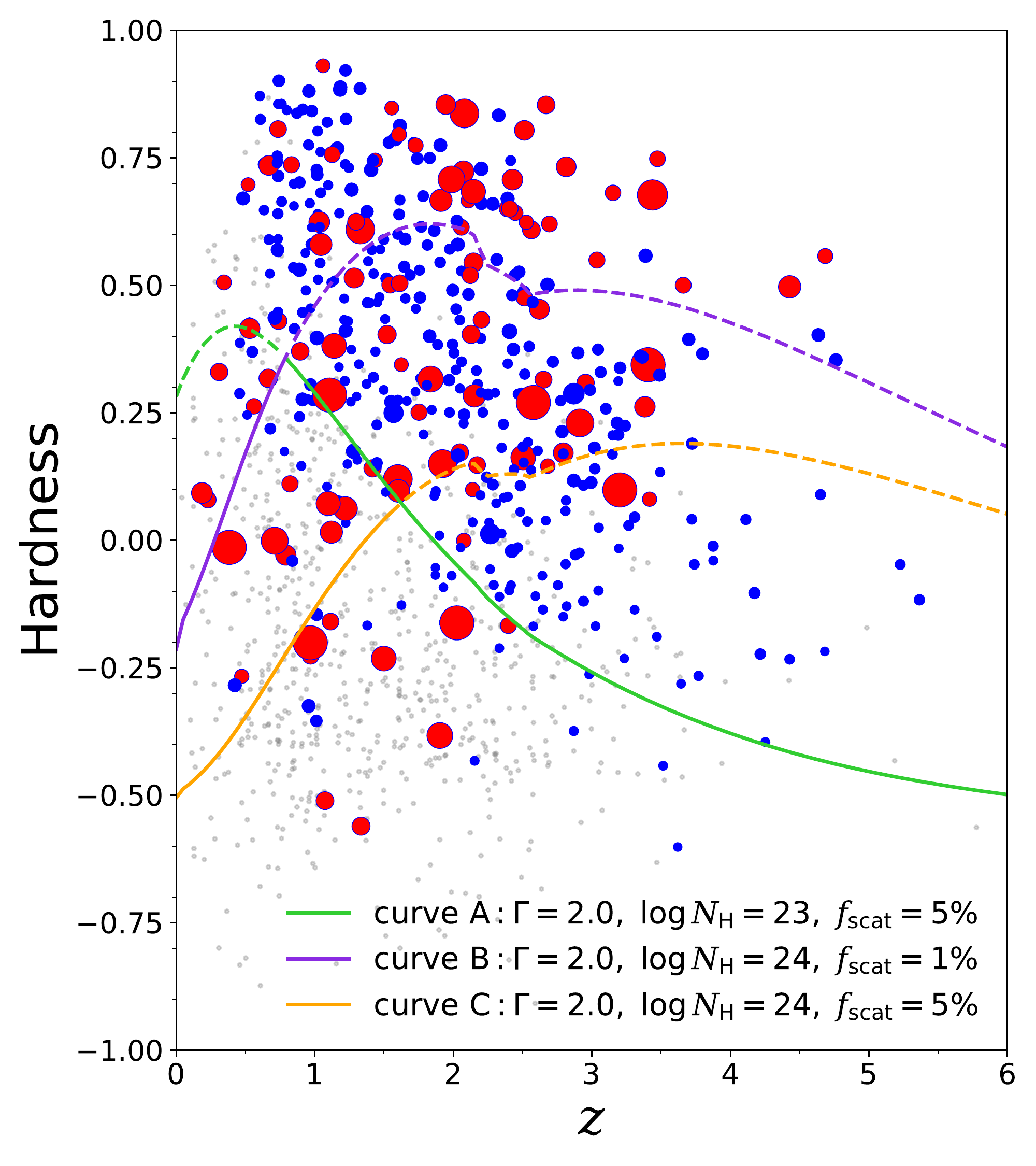}
\caption{Hardness ratio of the highly obscured sample as a function of source redshift.  CT candidates selected in Section  \ref{subsec:nh} with best-fit $N \rm_H\ \>\ 10^{24}\ \rm cm^{-2}$ and the 1$\sigma$ lower limit of \nh~\>~$\rm 5 \times 10^{23}\ cm^{-2}$ are shown in red while other highly obscured AGNs are shown in blue. The size of the symbol indicates their best-fit $N_{\rm H}$ value. Less obscured sources with best-fit $\nh < \cm$ are shown in gray. Curves in different colors represent different selection curves presented in Section \ref{subsec:hr}. }
\label{fig:HR}
\end{figure}

\begin{figure*}
\includegraphics[width=\linewidth]{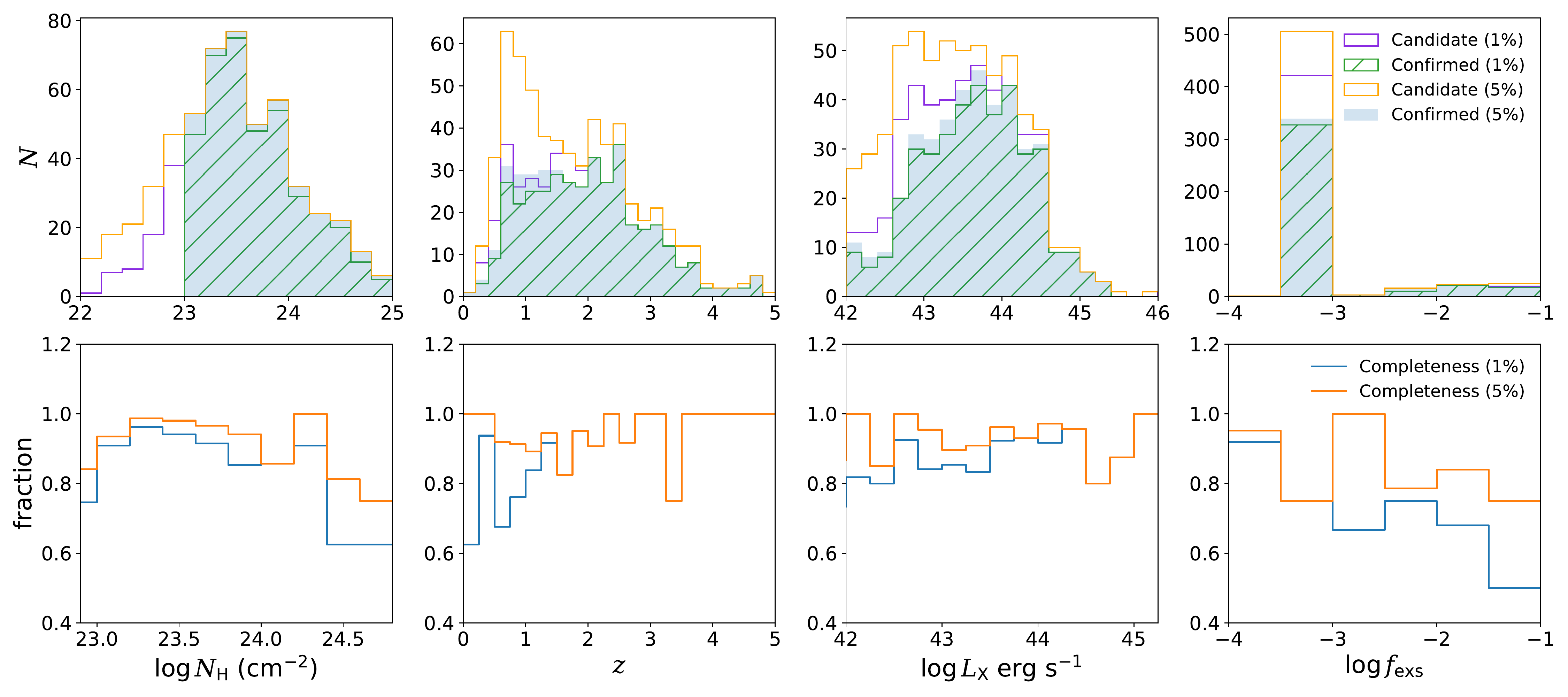}
\caption{Top: Distributions of source properties of the candidate samples selected from different HR-curves (Candidate) and those confirmed to be heavily obscured through spectral fitting (Confirmed). Bottom: Completeness fractions of the candidate samples selected by different HR-curves.   The 1\% and 5\% in the legend represent different soft excess fractions assumed while deriving the HR-curves in Figure \ref{fig:HR}. Sources without a detected soft-excess component are shown as $\rm log\,\fs = -4$ in the plot.}
\label{fig:select}
\end{figure*}

For highly obscured AGNs, the significant absorption and scattering of soft X-ray photons may lead to large hardness ratio (HR), thus the HR may be used as an indicator of obscuration level \citep[e.g.,][]{Wang2004}. 
In Figure \ref{fig:HR} we show the observed HR (here we adopt the definition of HR as (H$-$S)/(H+S), where H stands for the observed-frame 2--7~keV count rate and S stands for the 0.5--2~keV count rate) as a function of redshift for highly obscured sources and CT candidates confirmed by the spectral fitting as well as less obscured AGNs (best-fit \nh~\<~\cm) that contaminate our sample. Heavily obscured sources have significantly larger HR than less obscured sources as expected. 
We also calculate the effective photon index $\Gamma_{\rm eff}$ (obtained by fitting the spectra using XSPEC model $phabs \times zpow$ with \nh~fixed at the Galactic value) for the HOS. 90\% of them have $\Gamma_{\rm eff}$ \<~1.0, and the median $\Gamma_{\rm eff}$ is only $-0.38$ for CT AGNs, which again verify their heavily obscured nature.

The anti-correlation between HR and $z$ is due to that for high redshift sources, the observed soft band actually corresponds to a much harder band in the rest-frame and the photons may not be absorbed significantly. Note that the HRs of the most heavily obscured CT AGNs are not always the largest at a given redshift. This can be ascribed to the additional soft excess component softens the CT AGN spectra, and Compton-thin (CN) AGNs with intrinsically flat photon indexes are also plausible to have larger HRs than CT AGNs. 

To test whether a simple HR value can be used to select heavily obscured AGNs, we calculate the evolution of HR as a function of redshift for typical X-ray spectral parameters. Since our purpose is to find the critical HR for highly obscured AGNs as a function of redshift, we simply consider the threshold condition: a source with $\nh = \cm$. Additionally, we set \fr~at 1.0, the high-energy cutoff at 200 keV and fix the photon index \gm~of the two power-laws at 2.0. The constant \fs~is set to 5\% that is typical for AGNs with $\nh \sim \cm$ \citep[e.g.,][]{Brightman2014}. The reason why we adopt $\gm = 2.0$ instead of the mean value 1.8 is that $\sim 96\%$ of the sources with a well-constrained photon index in L17 have $\gm \lesssim 2.0$. Therefore, with this value we expect that our derived HR curve will be able to successfully select out most of highly obscured AGNs. The  model is shown in Figure \ref{fig:curve}a and the derived model HRs at different reshifts are shown in green in Figure \ref{fig:HR} (curve A).

It is worth noting that not all CT AGNs have larger HRs than CN AGNs. We also show the HR curve derived for a CT AGN with $\nh = \ct$. This time \fs~is chosen to be 1\% due to the fact that \fs~decreases with increasing \nh~(see Section  \ref{subsec:softexcess}). The model is shown in Figure \ref{fig:curve}b and the simulated HR curve is also shown in Figure \ref{fig:HR} (curve B). It is clear that at low redshifts, even a very small soft excess fraction can easily dominate the soft X-ray spectra that could make CT AGNs even softer than CN AGNs.
Therefore, to avoid missing a large number of high-\nh~sources at low redshifts, we use curve B at $z\le 0.8$ and curve A at $z > 0.8$ as our selection curve in the following analysis (i.e., the combined curve). 

For a source with given HR and redshift values, we consider it as a heavily obscured candidate if the observed HR lies above the combined HR curve. By applying this curve to our full sample,  480 sources will be selected as highly obscured candidates (hereafter the candidate sample).  80\% (382/480) of sources in the candidate sample are indeed highly obscured as confirmed from spectral fitting, i.e., the accuracy is 80\%. The parameter distributions for the selected candidates and confirmed sources are shown in Figure \ref{fig:select} (top panel). Only 20\% of the candidate sources are contaminated by less obscured sources with most of them having \nh~only slightly smaller than \cm~and lying close to the boundary curve, as expected. The redshift, luminosity and soft excess fraction distributions for the selected candidates are very similar to sources which are confirmed to have $\nh > \cm$. 

In addition, we define the completeness fraction as the ratio of the amount of highly obscured sources (confirmed) lie above the selection curve to the total highly obscured source amount (i.e., all red and blue points in Figure \ref{fig:HR}). The completeness fractions as a function of several parameters are shown in the second row in Figure \ref{fig:select}. The average completeness fraction for the selection curve is 88\% (382/436) and remains $>90\%$ for most high-redshift and high-luminosity bins. While at lower redshifts and larger \fs, the completeness fraction significantly drops to $< 80\%$, since the soft excess dominates the observed spectrum which makes highly obscured sources lie below the selection curve.

We also show the result after changing the soft excess fraction for the model CT AGN to 5\% (see curve C in Figure \ref{fig:HR}). This time the completeness fraction at low redshift is improved to 93\%, but the accuracy dramatically drops to 68\% since a large amount of less obscured AGNs are mistakenly included. As a trade-off, we choose to adopt the combined curve A and B as our selection curve. The functional form of this curve can be written as:
\begin{equation}
{\rm HR}=
\left\{
             \begin{array}{lr}
             -0.240z^3 + 0.230 z^2 + 0.680 z -  0.200,\\ \ (z\ \le\ 0.8); \\
	     -0.004 z^3 + 0.077 z^2 - 0.530 z + 0.740,\\ \ (z\ >\ 0.8).
             \end{array}
\right.
\label{eq:curve}
\end{equation}

We note that the conclusions in the following sections remain largely unchanged if we use the HR-selected sample instead, suggesting that a simple HR value can be used to identify highly obscured AGNs and select a representative highly obscured sample as long as the redshift and soft excess effects are properly taken into account while deriving the boundary curve.

\begin{figure*}
\centering
\includegraphics[width=0.85\linewidth]{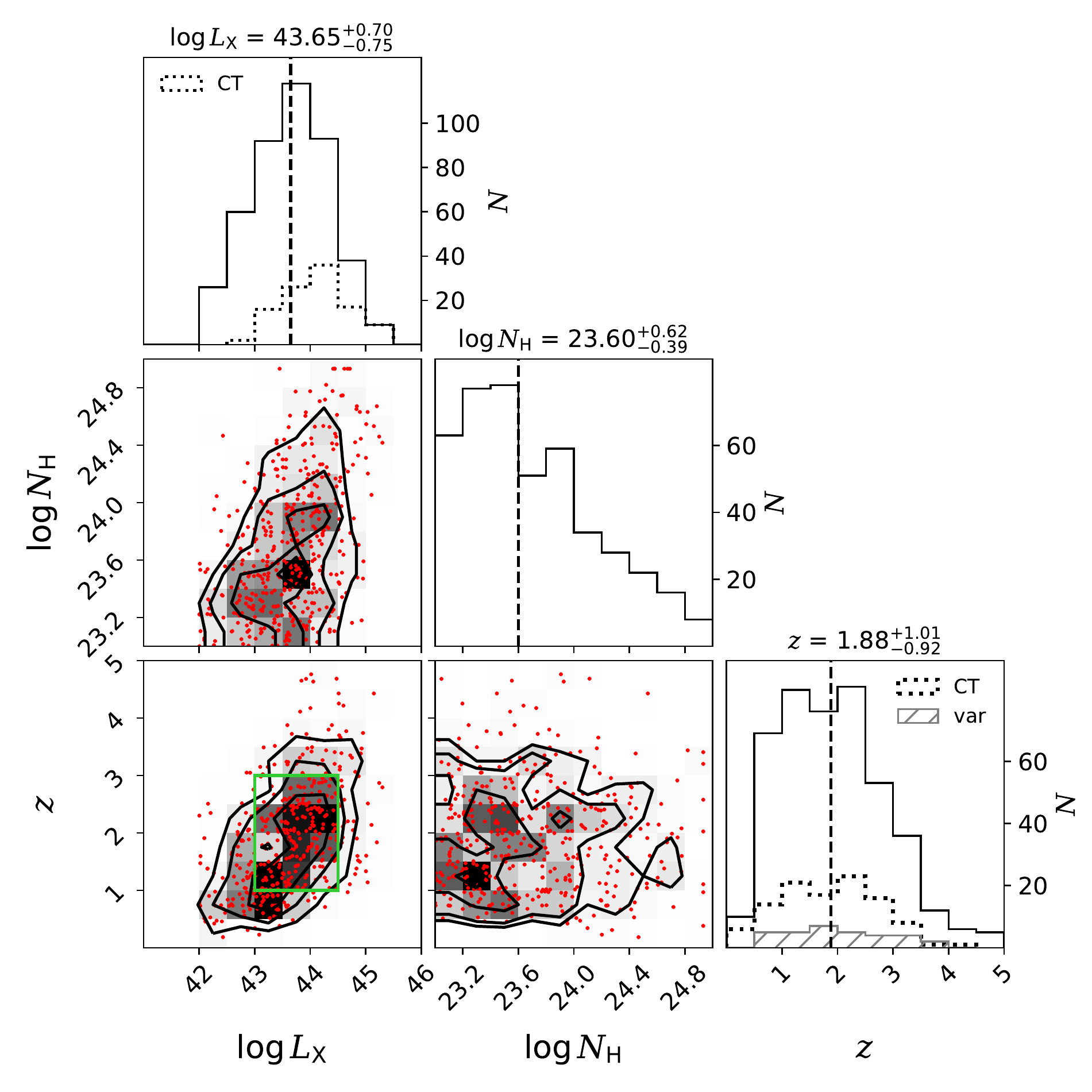}
\caption{The triangle plot of the correlations among \lx, \nh~and redshift. The outer parts of the triangle show the distributions of \lx, \nh~and redshift from the top-left to the bottom-right panels, respectively. The vertical dashed lines show the median value for each distribution. In the redshift and \lx~distribution panels, the solid histograms represent the total sample and the dotted ones show the distributions of CT candidates selected in \ref{subsec:nh}. The shaded histogram shows the redshift distribution of our variability sample in Section  \ref{sec:var}. 
The inner parts of the triangle show the scatter plots and corresponding density maps overlaid with contours among \lx, \nh~and redshift, respectively.
In the \lx~versus $z$ plane, the green rectangle shows the subsample we select in Section \ref{subsec:intrinsic_nh} to investigate the evolution of the intrinsic fraction of CT AGNs among highly obscured ones.}
\label{fig:corner}
\end{figure*}

\subsection{Redshift and Luminosity Distributions}
\label{subsec:lx}

In Figure \ref{fig:corner} we show the redshift distribution of the HOS in the right-bottom panel. 191 sources have spectroscopic redshifts and 245 sources have photometric redshifts. The peak appears at $z\sim 1-2.5$, which is known as the peak epoch of both star-formation and black hole accretion activities \citep[e.g.,][]{Aird2015}. The source amount slightly decreases down to $z \sim 0.5$, and significantly drops at $z < 0.5$.
Since \emph{Chandra} only pointed at small patches of sky in these two surveys, the small number of sources detected at $z < 0.5$ is mainly due to the small volume probed and the insufficient penetrability of the corresponding rest-frame energies at low redshifts; thus, the detection probability is severely limited in finding nearby highly obscured AGNs. 
The downward trend toward high redshifts is primarily due to the flux limits and may be partly due to the fact that photoelectric absorption in the soft band of mildly obscured ($\nh \sim \cm$) sources shifts out of the \emph{Chandra} soft bandpass at $z\sim2.5$, making it more difficult to constrain \nh~\citep{Vito2018}. 

The \lx~distribution for the HOS is displayed in the top-left panel of Figure \ref{fig:corner}. The \loglx~(\lx~in units of \ergs) peaks in the range of 43.5--44.0 with the mean value of $43.65 \pm 0.03$. The luminous sources with \loglx~\>~44.0 (44.5) account for 32\% (11\%) of the whole sample. 
CT AGNs, in particular, have even higher luminosities with the mean value of $44.10 \pm 0.06$, because of the minimum detectable \lx~for a flux-limited survey significantly increases with increasing \nh~(see Figure \ref{fig:boundary}).
The correlations among \lx, \nh~and redshift are also plotted in Figure \ref{fig:corner}. The well-known Malmquist bias can be clearly seen in the bottom-left \lx-$z$ corner and we will discuss the influence of this bias to our results in Section  \ref{subsec:malm}.

\subsection{Observed \nh~Distribution}\label{subsec:nh}

\begin{figure*}
\includegraphics[width=\linewidth]{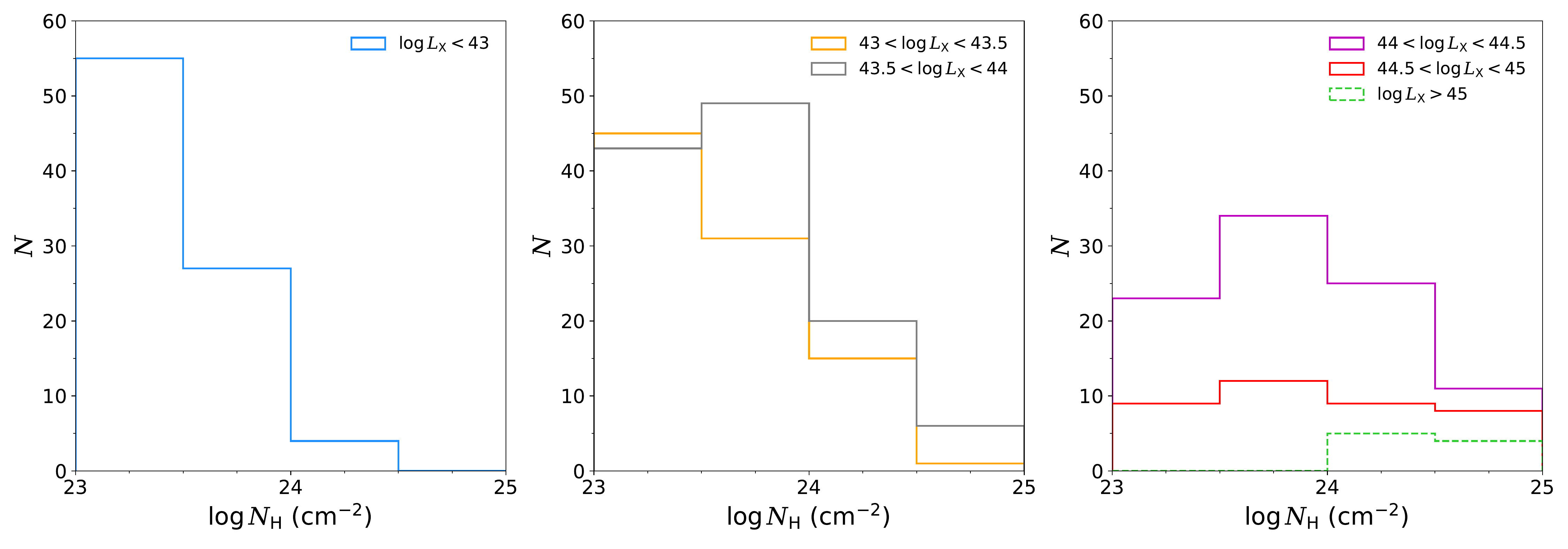}
\caption{The observed \nh~distributions in six \lx~bins. The small number of high \nh~sources in low \lx~bins is due to that such sources are difficult to detect. This bias will be corrected in Section \ref{subsec:malm}.}
\label{fig:observed_nH}
\end{figure*} 
 
 \begin{figure*}
 \centering
\includegraphics[width=\linewidth]{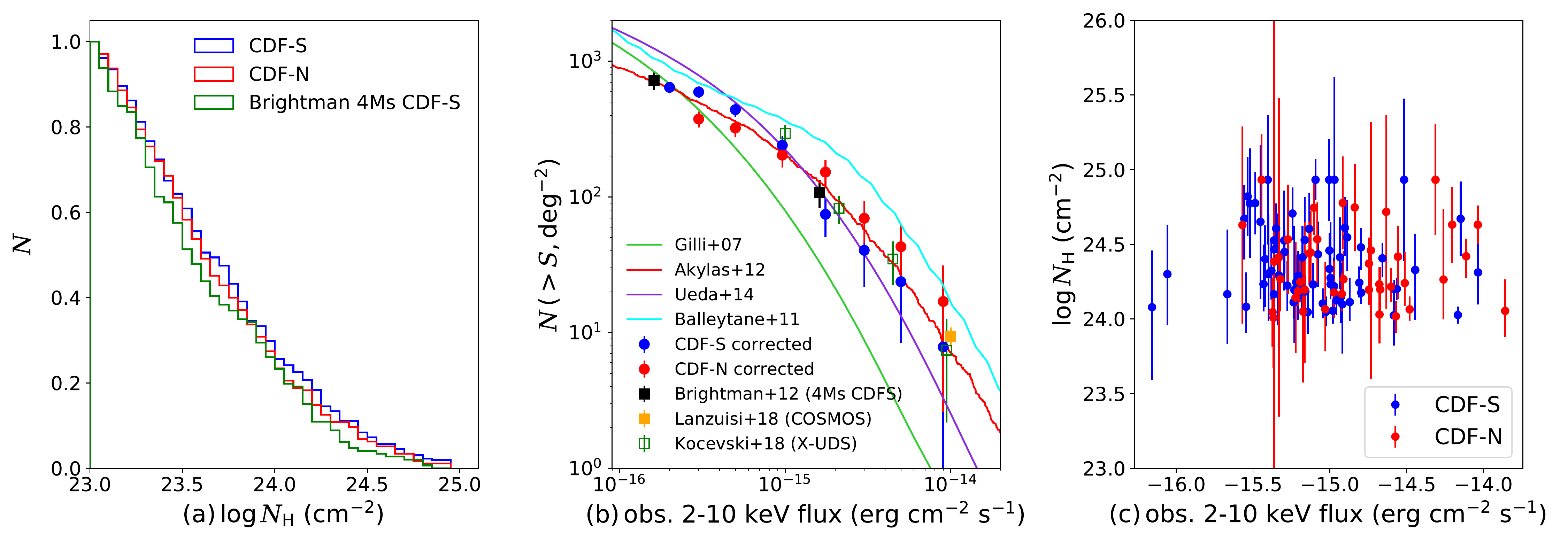}
\caption{Left: Normalized cumulative \nh~distribution for our highly obscured sample. The 4Ms CDF-S result adopted from \cite{Brightman2014} is also plotted for comparison. Middle: The observed log$\,N-$ log$\,S$ for CT candidates compared with previous number counts in the 4~Ms~CDF-S \citep{Brightman2012}, X-UDS \citep{Kocevski2018} and COSMOS \citep{Lanzuisi2018} as well as several CXB model predictions \citep{Gilli2007, Ballantyne2011, Akylas2012, Ueda2014}. Right: Column density of CT AGNs as a function of observed 2--10~keV flux in the CDF-S (blue) and CDF-N (red), respectively.}
\label{fig:logN_logS}
\end{figure*}

The observed distributions of the best-fit \nh~in six luminosity bins of the HOS are shown in Figure \ref{fig:observed_nH}. The small amount of high \nh~sources observed in the smaller \lx~bins will be clearly illustrated in Section \ref{subsec:sky} (see Figure \ref{fig:boundary}) that the minimum detectable luminosity significantly increases with \nh. 
We show the normalized cumulative \nh~distributions in Figure \ref{fig:logN_logS}a. Sources with best-fit \nh~\>~$\rm 1.0 \times 10^{24}\ cm^{-2}$ account for $\sim$25\% (108/436) of the HOS. The distribution obtained in \cite{Brightman2014}, which utilized the 4~Ms CDF-S data, is shown in the same plot for comparison. The K-S tests imply that the \nh~distributions from the CDF-S and the CDF-N are consistent with $p$-value = 0.96, but are different from that in \cite{Brightman2014} with $p$-value $\ll 0.001$, due to that we identify more sources between \lognh~= 23.5 -- 24.0 $\rm cm^{-2}$ while the distribution in \cite{Brightman2014} peaks at \lognh~= 23.0 -- 23.5 $\rm cm^{-2}$. 

Although we have significantly improved the photon statistics for each source using the deepest data, we should caution that most of CT AGNs discovered still have counts \<~200 and their \nh~might be poorly constrained. Therefore we adopt a more conservative criterion that a source will be regarded as  CT candidate only if it has the best-fit $\nh > \ct$ and the 1$\sigma$ lower limit on \nh~is constrained to be greater than $\rm 5 \times 10^{23}\ cm^{-2}$. This criterion  selects 102 sources, with 66 from CDF-S and 36 from CDF-N, accounting for $\sim$23\% of the HOS, to be CT candidates.  Note that there are some well-studied CT candidates in the literature that are not classified as CT AGNs in our sample due to their best-fit \nh~\<~\ct, such as XID 328, XID 551  \citep[XIDs 153, 202 in][respectively]{Comastri2011}, XID 539 \citep[XID 403 in ][]{Gilli2011} and XID 375 \citep[$BzK$8608 in ][]{Feruglio2011} in the CDF-S. However, all these sources have best-fit \nh~\>~$8 \times$ \cm, which shows good consistency with previous studies, thus we confirm their heavily obscured nature using deeper \chandra~data. 

We note that at $f_{\rm 2-10\, keV} > 10^{-15}\ \ergcms$, the number of our CT AGNs in the CDF-N (22 sources) is higher than  \cite{Georgantopoulos2009} which presented 10 CT candidates in the same field. This may because of \cite{Georgantopoulos2009} mainly focused on searching reflection-dominated CT AGNs and possibly missed some transmission-dominated sources. We also fit the 22 CT candidates using the BNtorus model \citep{Brightman2011} and find that 20 sources have best-fit $\nh > \ct$ and $f_{\rm 2-10\ keV}$ \>~$10^{-15}$ \ergcms, which confirms the MYTorus result. 

\subsection{Number Counts for Compton-thick AGNs}
\label{subsec:logn_logs}
We calculate the observed number counts (log$\,N-$ log$\,S$) for CT candidates selected in Section \ref{subsec:nh}. The observed log$\,N-$ log$\,S$ is defined as the observed source amount divided by survey area. However, as we will discuss in Section  \ref{subsec:sky}, the sky coverage is not a constant value across the whole flux and \nh~ranges, hence we assign a weighting factor $1 / \omega_{\rm area}$ (see Section  \ref{subsec:sky} for details) for each source while calculating their cumulative number counts. The corrected results are shown in Figure \ref{fig:logN_logS}b, and several number counts measurements presented in previous works in the 4 Ms CDF-S \citep{Brightman2012}, X-UDS \citep{Kocevski2018} and COSMOS \citep{Lanzuisi2018} fields are also displayed for comparison. Note that \cite{Brightman2012} reported the result in the 0.5--8 keV band and we convert the 0.5--8 keV flux into 2--10 keV by assuming a $\gm = -0.4$ power-law which is the median effective photon index for our CT AGNs.

The corrected number counts for the CDF-S and CDF-N are generally consistent within the error bars at higher fluxes, despite that we obtain six CT AGNs in the CDF-N but only three in the CDF-S at $f_{\rm 2-10\ keV}$ \>~$4 \times 10^{-15}\ \ergcms$, possibly caused by the low number statistics and the cosmic variance due to the small sky coverage of CDFs \citep[e.g.,][]{Harrison2012}. While at the faint end, both CDF-S and CDF-N number counts flatten due to the limited detection ability, and the CDF-N number counts are significantly smaller than the CDF-S, because the three times shallower exposure limits its detectability of faint CT sources (see Figure \ref{fig:logN_logS}c). Our results are also in  agreement with the 4 Ms CDF-S result at the faint end, the COSMOS result at the bright end, and the X-UDS result at moderate fluxes.

We also compare the observed log$\,N-$ log$\,S$ with several CXB model predictions in $z = 0-5$ \citep{Gilli2007, Ballantyne2011, Akylas2012, Ueda2014}. 
For the \cite{Gilli2007}\footnote{\url{http://www.bo.astro.it/~gilli/count.html/}. We assume a high-$z$ declined LF \citep{Vito2018}.}, \cite{Akylas2012}\footnote{\url{http://indra.astro.noa.gr/xrb.html}. We assume a 40\% CT fraction and the default 4.5\% reflection fraction.} and  \cite{Ueda2014}\footnote{\url{http://www.kusastro.kyoto-u.ac.jp/\~yueda/xrb2014.html}} CXB models, the luminosity range used to derive the predicted number counts is $\loglx = 42 - 45.5\ \ergs$, similar to our sample; while for the \cite{Ballantyne2011} model, the predicted result is presented in $\loglx = 41.5 - 48\ \ergs$. 
As can be seen from Figure \ref{fig:logN_logS}b, our observed log$\,N-$ log$\,S$ prefers the moderate CT number counts as predicted by \cite{Akylas2012} and \cite{Ueda2014}, while other models more or less overestimate or underestimate the number counts.

In summary, the observed parameter distributions and relationships presented in this section provide a basic description of the highly obscured AGN population and are crucial for distinguishing various CXB models. However, these observed distributions, in particular, the observed \nh~distribution, are influenced by several biases. We will discuss more details about these biases and reconstruct the intrinsic \nh~distribution representative for the highly obscured AGN population in Section  \ref{sec:intrin_nh}.

\section{The Origin of Heavy X-ray Obscuration}
\label{sec:ref}

\subsection{X-ray Highly-Obscured Broad-Line AGNs}

\begin{table*}
\caption{Information of X-ray highly obscured BLAGNs in the CDF-S}
\footnotesize
\centering
\begin{tabular}{c c c c c c c c c c c c}
\hline
\hline
XID & model & \fr & Oclass & $z$ & ztype & zqual & counts & \lognh (\nhu) & $z$-range & $z_{\rm x}$ & ${\rm log}\,{N_{\rm H, x}}$ (\nhu)\\
(1) & (2) & (3) & (4) & (5) & (6) & (7) & (8) & (9) & (10) & (11) & (12)\\
\hline
 882 & A & $\ll 1$ &    BLAGN &   3.19 & zspec &     Secure &   80 & 23.1 & -- & -- & --\\
 944 & C &    1 &    BLAGN &   0.96 & zphot &   Insecure &   69 & 23.9 & all & 2.00 & 24.1\\
  16 & A & $\ll 1$ &    BLAGN &  3.10 & zphot &   Insecure &  409 & 23.7 & $z > 1.4$ & 1.66 & 23.2\\
 399 & C &    1 &    BLAGN &   1.73 & zspec &     Secure & 1077 & 23.2 & -- & -- & --\\
 968 & C &    1 &    BLAGN & 2.03 &   zspec &     Secure &  500 & 23.8 & -- & -- & --\\
 977 & C &    1 &    BLAGN &  4.64 & zspec &   Insecure &  225 & 23.9 & $z > 1.2$ & 3.25 & 23.9\\
\hline
\end{tabular}

\vspace{5.0 pt}
{\sc \bf{Notes.}}
Column 4: optical classification result from \cite{Silverman2010}. 31 X-ray highly obscured AGNs in our CDF-S sample have optical classification results and 6 of them are identified as BLAGNs. Column 10: the redshift range for sources with insecure redshifts to be determined as being X-ray highly obscured. Column 11: the best-fit X-ray redshift by setting redshift as a free parameter during spectral fitting. Column 12: best-fit \lognh~when adopting the X-ray best-fit redshift. Other columns have the same meaning as Table \ref{table:fitting_result}.

\label{table:BLAGN}
\end{table*}

We collect optical classification results for our CDF-S sample by cross-matching their optical/NIR/IR/radio counterpart positions  presented in the X-ray source catalogs (CP\_RA and CP\_DEC) with the \cite{Silverman2010} E-CDF-S optical spectroscopic catalog using a 0.5$''$ matching radius. Among the 55 matched sources, 31 sources have been classified. To our surprise, 19\% (6/31) of them are labeled as broad-line AGN (BLAGN). The detailed information of these sources are summarized in Table \ref{table:BLAGN}. 

Most of the six sources have sufficient counts and reliable redshift measurements to constrain \nh. Three sources have insecure spectroscopic redshifts although they are labeled as BLAGN, and the 7 Ms main catalog chooses to adopt the photometric redshift as ZFINAL for two of them. This could be due to that they have low S/N optical spectra or only one emission line is available for  sources within particular redshift ranges, which makes it difficult to conclusively determine the line nature, thus giving an insecure redshift.
For three sources with insecure redshifts, we set redshift as a free parameter in the spectral fitting and obtain the best-fit X-ray redshift and corresponding \nh. All of them are still best-fitted by $\nh > \cm$. The redshift range in which the source will remain X-ray highly obscured is also listed. In general, the \nh~estimates for the six sources are robust. This result suggests that the heavy X-ray absorption in a fraction of sources is largely a l.o.s effect caused by some compact clumpy clouds  obscuring the central X-ray emitting region, but the global covering factor (CF) for the high-\nh~materials is limited thus the BLR is not blocked; or maybe the heavy X-ray obscuration is produced by the BLR itself.

\subsection{Soft Excess Fraction Dependences} \label{subsec:softexcess}
There are 80 sources in our sample that require a second power law to fit the soft excess component. The origin of this component is still a puzzle. Many different models, e.g., warm Comptonization or blurred ionized reflection from the disk, partially ionized absorption in a wind, or power-law continuum in other directions scattered into the line of sight from large-scale Compton-thin matters, are proposed to explain the diverse situations found in different sources \citep[e.g.,][]{Boissay2016}. However, for highly obscured AGNs, the scattered mechanism is preferred since either blurred reflection or warm Comptonization from the accretion disk will be significantly attenuated by the obscuring materials.

\begin{figure}
\includegraphics[width=\linewidth]{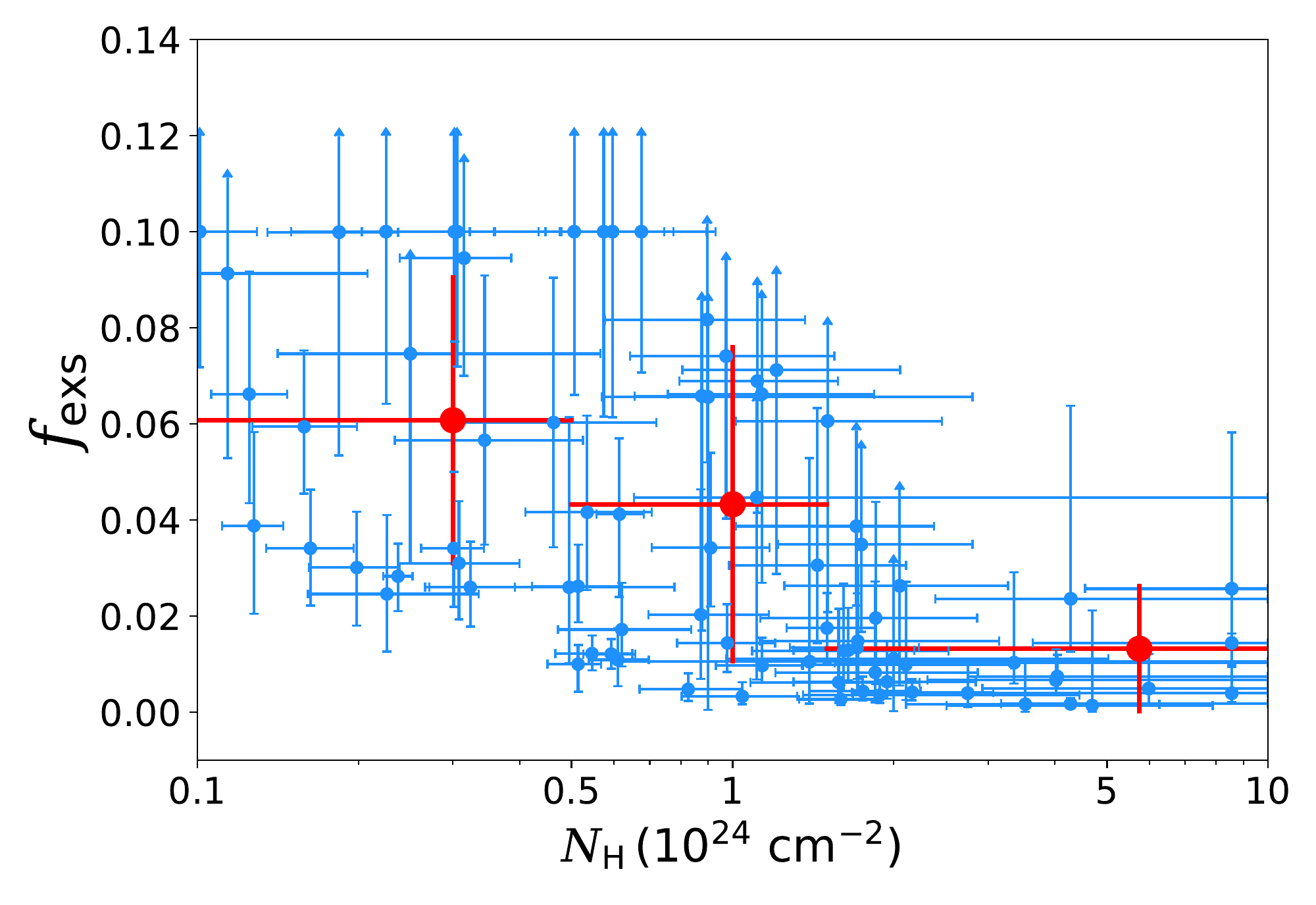}
\caption{The soft excess fraction \fs~as a function of \nh. The blue points show the individual sources and the red points show the binned results. The negative correlation between \fs~and \nh~indicates that a portion of high \nh~sources might have higher ``torus'' covering factors that makes the soft excess photons hard to escape, under the assumption that the excess originates from the scattered-back continuum in highly obscured AGNs.}
\label{fig:excess}
\end{figure}

We show the \fs, which represents the relative normalization of the soft component with respect to the intrinsic power law, as a function of \nh~in Figure \ref{fig:excess}. 
We find a significant anti-correlation between the two parameters: Spearman's $\rho = -0.66$, $p$-value = $\ll 0.001$. The mean \fs~is 3.7\% for the total sample. The scattered fraction decreases from 6.1\% to 4.3\% and 1.3\% for \nh~\<~5 $\times$ \cm, 5 $\times$ \cm~\<~\nh~\<~ 1.5 $\times$ \ct~and \nh~\>~1.5 $\times$ \ct, respectively. 

To verify that this anti-correlation is not caused by the parameter degeneracy that sources with a large soft excess fraction and a high \nh~might be misclassified as low \nh, low \fs~and low intrinsic luminosity sources owing to the low S/N, we assume that CT AGNs have exactly the same \fs~as CN AGNs (here we simply choose \fs~= 5\%). Then we generate fake spectra using model D for a sample of simulated sources, which have redshifts and \nh~uniformly distributed between $z = 0-4$ with an interval of $\Delta z = 0.5$ and $\rm 23\ cm^{-2} < \lognh < 25\ cm^{-2}$ with an interval of $\Delta \lognh = 0.25$. For each source we simulate 100 spectra with 0.5--7~keV counts $\sim$ 50, 100 and 150, respectively. We then fit the fake spectra and perform Spearman rank correlation test for the output \nh~and \fs. The Spearman's $\rho \approx 0.1$ (insensitive to the simulated spectral counts), indicating no apparent correlation, thus the observed anti-correlation is intrinsic. 

By assuming that the second power law originates from the scattered-back continuum, \cite{Brightman2012} found that the scattered fraction depends on the opening angle, i.e., sources with small opening angles have lower \fs. Moreover, \cite{Ueda2015} and \cite{Kawamuro2016} found that the [O III] and [O IV] to hard X-ray luminosity ratios are lower in low scattering fraction sources, inferring that the low \fs~AGNs are buried in small opening angle torus. Therefore, the anti-correlation between \fs~and \nh~indicates that high \nh~sources might preferentially reside in high CF toruses. This is also consistent with the results from \cite{Brightman2012} and \cite{Lanzuisi2015}, who found a similar anti-correlation between \fs~and \nh~and explained it as highly obscured AGNs being also heavily geometrically obscured (but see the discussion in the next section). 

\begin{figure*}
\includegraphics[width=0.59\linewidth]{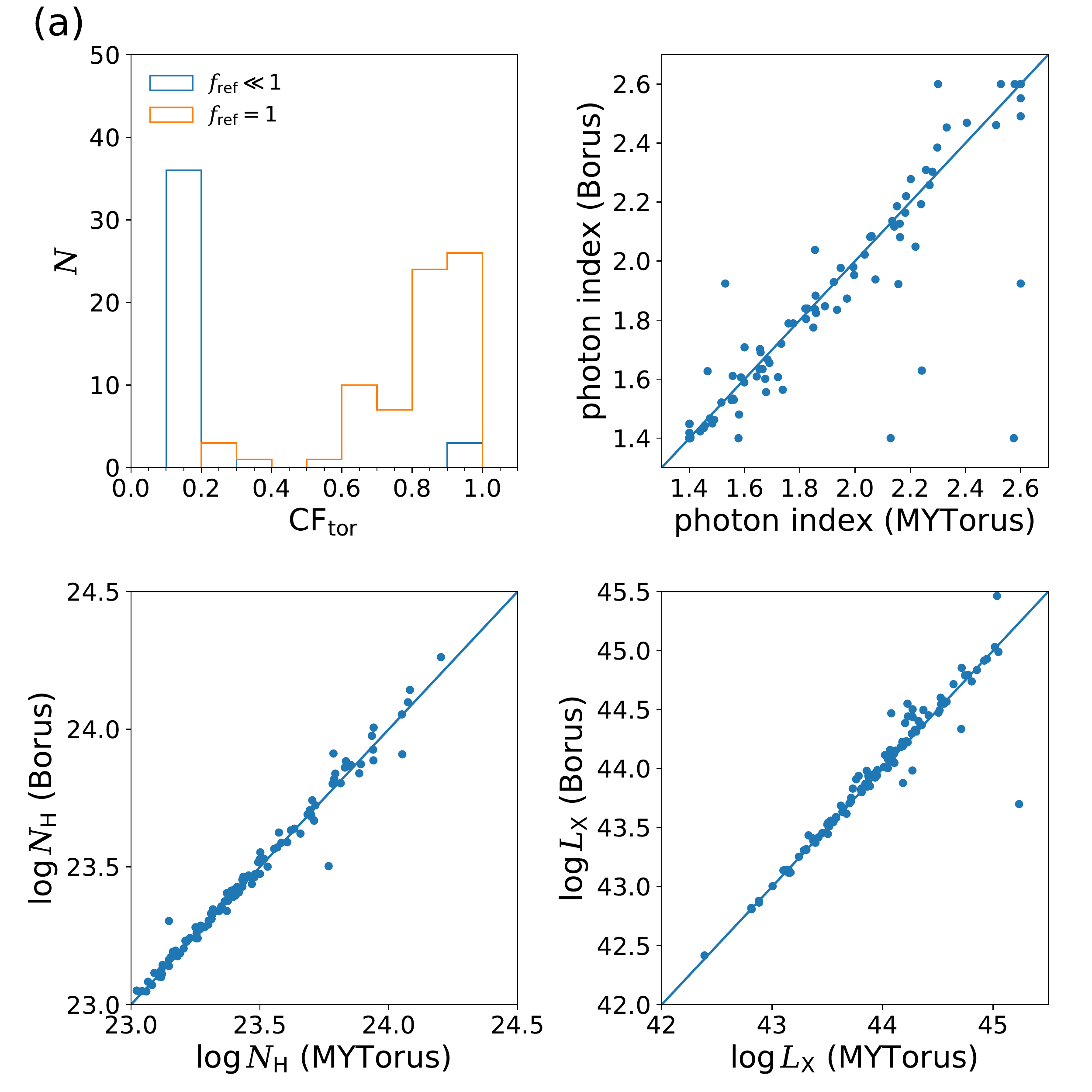}
\includegraphics[width=0.412\linewidth]{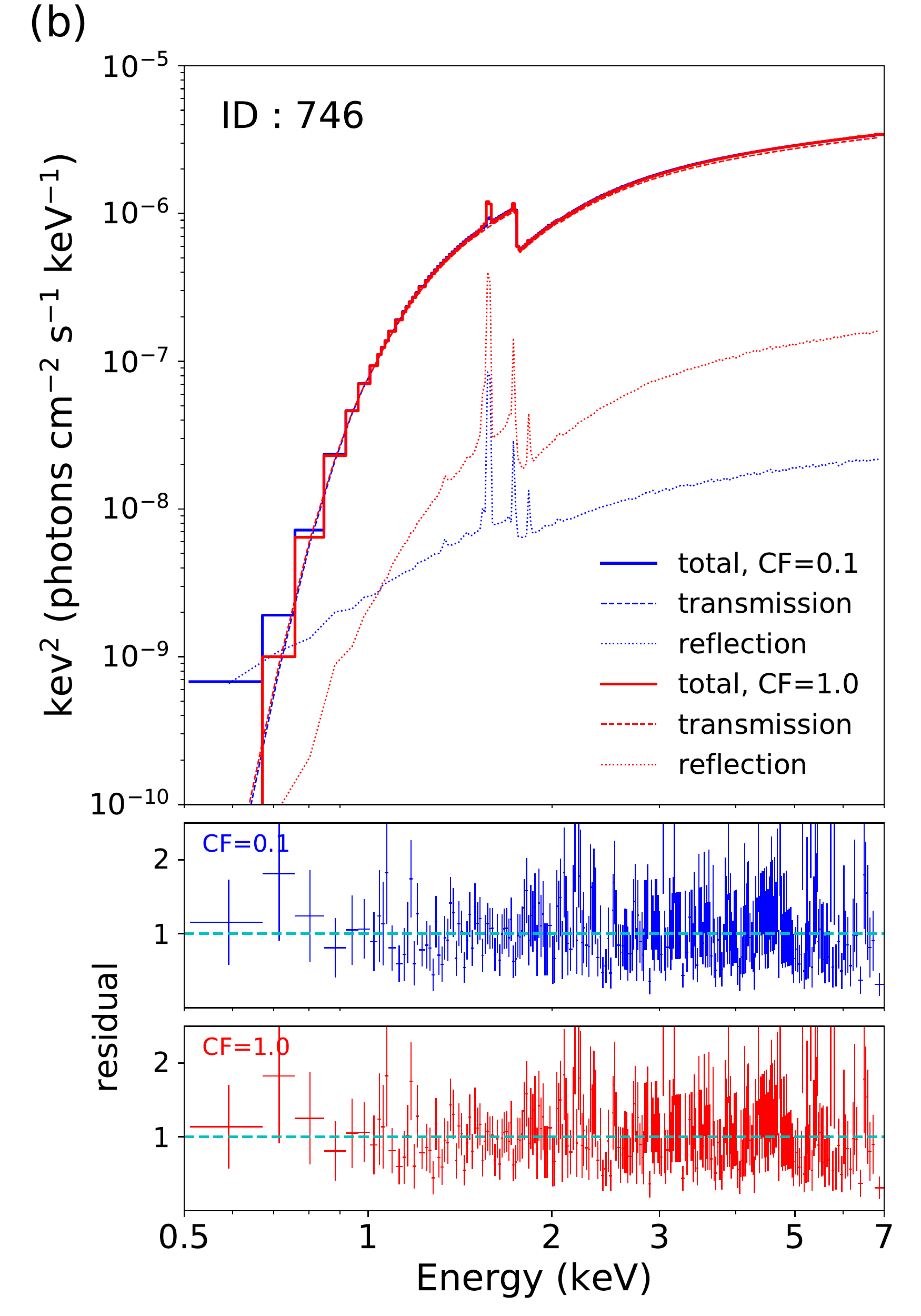}
\caption{(a) The covering factor derived through the Borus model \citep{Balokovic2018} and the comparisons between photon index, \nh, and \lx~obtained from the Borus and MYtorus models. The photon indexes in this plot are the original results without fixing \gm~at 1.8 for some extreme cases. (b) The spectral fitting result for CDF-S XID 746 using the Borus model. Both the fully buried model ($\cf = 1.0$; Cstat = 942.3) and the weak torus model ($\cf = 0.1$; Cstat = 941.8) can well reproduce the data. The contribution from the reflection component to the total spectrum in both models are largely negligible.}
\label{fig:CF}
\end{figure*}

\subsection {Reprocessed Components and Covering Factor}
\label{subsec:ref}

Unlike local CT AGNs in which the prominent neutral Fe lines and reflection hump (hereafter the reprocessed components) are prevalently detected in the high-quality hard X-ray spectra and can be served as unambiguous signatures of being Compton-thick \citep[e.g.,][]{Tanimoto2018, Zhao2019, Marchesi2019}, there are 41\% of our highly obscured sources and 38\% of the CT candidates having negligible \fr (see Table \ref{table:model}), respectively. It should be noted that the low detection rate of reprocessed components is not in conflict with their highly obscured nature. The poor S/N for low-count sources and the narrow \emph{Chandra} spectral coverage make it challenging to detect narrow iron lines and to distinguish the transmission and reflection components, which may cause a significant overestimation of the fractions. 

But we note that it is possible that highly obscured AGNs may have weak reprocessed components if the global CFs of high-\nh~materials are low or the central AGN is geometrically fully buried by CT materials such that even the reflected photons cannot escape \citep[e.g.,][]{Brightman2014}.
To distinguish different scenarios, we use the Borus model \citep{Balokovic2018}\footnote{In the Borus model, $\rm CF_{\rm tor} = 1.0$ corresponds to a fully covered torus while $\rm CF_{\rm tor} = 0.1$ represents a typical disk-like covering (see \url{http://www.astro.caltech.edu/~mislavb/download/}).}, which allows CF to vary freely, to fit the spectra for all sources with counts \>~200. This time we treat photon index as a free parameter in order to make a direct comparison between the two models. The distribution of the derived $\rm CF_{\rm tor}$, which is defined as the cosine value of the opening angle $\theta_{\rm tor}$ measured from the symmetry axis towards the equatorial plane, is shown in Figure \ref{fig:CF}a. We also show the comparisons of the main spectral fitting parameters with those obtained from the MYTorus model and find good consistency. 

For sources with negligible reprocessed components in the MYTorus model ($f_{\rm ref} \ll 1$), the Borus results are consistent with the weak torus scenario ($\cf \approx 0.1$), which might suggest that their heavy obscuration is simply a l.o.s effect (i.e., some high-column-density clouds on various scales along our sightline may obscure the compact X-ray emitter), without the necessity to invoke a strongly-buried nuclear environment. In contrast, sources with \fr~= 1.0 are mostly best fitted by a highly-covered model ($0.8 < \cf < 1.0$) that have torus covering angles ($90^\circ - \theta_{\rm tor}$) between $65^\circ - 90^\circ$. We emphasize that though the l.o.s and the global torus \nh~are linked in the spectral fitting,  those fully-buried sources do not need to be covered by very high \nh~materials in all directions, since the best-fit \nh~will more likely converge to the values determined by the l.o.s component because the photoelectric absorption is a much stronger spectral feature.
Moreover, we find that the average \cf~for $\fs < 0.05$ sources is 0.60; while for $\fs > 0.05$ sources, the $\overline{\cf}$ is 0.32. This might provide evidence for that the lower soft-excess fraction in some sources is caused by a  geometrically more buried structure (see Section \ref{subsec:softexcess}).

However, we caution here that the current S/N and spectral coverage do not allow us to constrain CF from X-ray spectral analysis. If we fix \cf~at 0.1 for sources that are best fitted by a fully-buried model, we may still obtain acceptable fits with Cstat values only slightly increasing (an example is shown in Figure \ref{fig:CF}b). Conclusively disentangling several scenarios is beyond the scope of this work, and multi-wavelength data are needed to further shed light on the nature of the obscuring materials (Li et al., in preparation).

\section{Intrinsic \nh~distribution for highly obscured AGNs}\label{sec:intrin_nh}
In Section  \ref{subsec:nh}, we present the observed \nh~distribution which is affected by several sample incompleteness. In order to derive the intrinsic \nh~distribution, we should take into account the errors on best-fit \nh~and correct for several survey biases \citep{Liu2017}. To perform such corrections, we restrict our analysis to the 394 sources with off-axis angle \<~10.0$'$, for the sake of avoiding large background contamination, extremely small sky coverage (see Section  \ref{subsec:sky}) and limited detectable fraction (see Section  \ref{subsec:malm}). 

\subsection {Errors on Best-fit \nh}
To consider the uncertainty of the spectral fitting, we perform a resampling procedure to the best-fit \nh. Given the asymmetric errors, we assume that the errors on \nh~obey the ``half-gaussian'' distribution. We first generate 1000 \nh~for each source with the $\sigma$ of the half-gaussian distribution equals to the lower 1$\sigma$ error, and then generate another 1000 \nh~with $\sigma$ equals to the upper 1$\sigma$ error. The mean value $\mu$ is set to the best-fit \nh. The resampled \nh~distribution (calculated by averaging the 2000 resampled distributions and also corrected for the sky coverage effect; see Section  \ref{subsec:sky}) is shown in the shaded region of Figure \ref{fig:mytorus_nh}. It has an extended tail down to $\nh < 10^{23}\ \nhu$ which contains only 4.0\% of the resampled data, suggesting that even if considering the spectral fitting errors, most of our sources are still consistent with being highly obscured.  

 \begin{figure*}
\centering
\begin{minipage}[t]{0.4\linewidth}  
\centering  
\includegraphics[width=\linewidth]{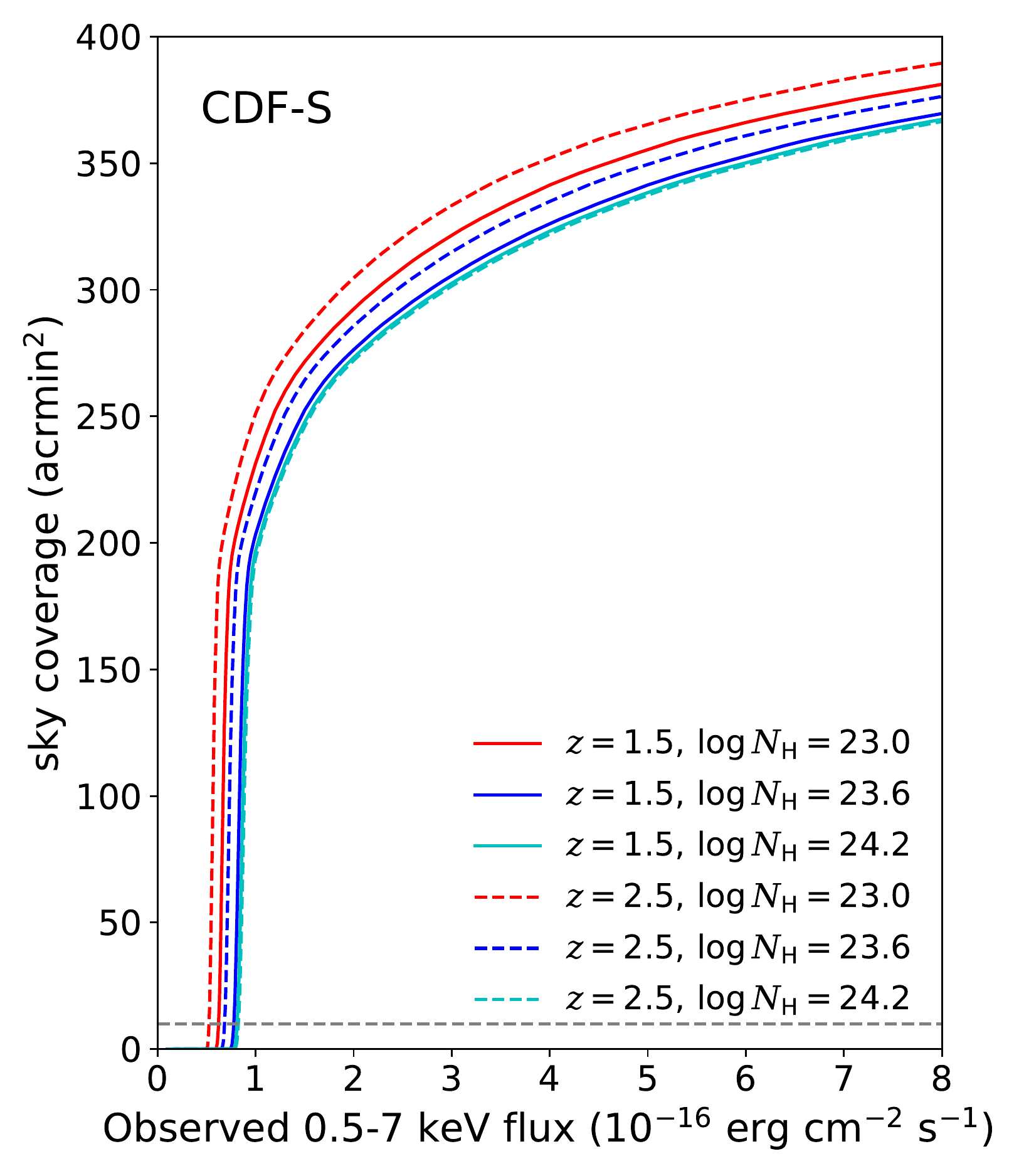}  
\end{minipage}  
\begin{minipage}[t]{0.4\linewidth}  
\centering  
\includegraphics[width=\linewidth]{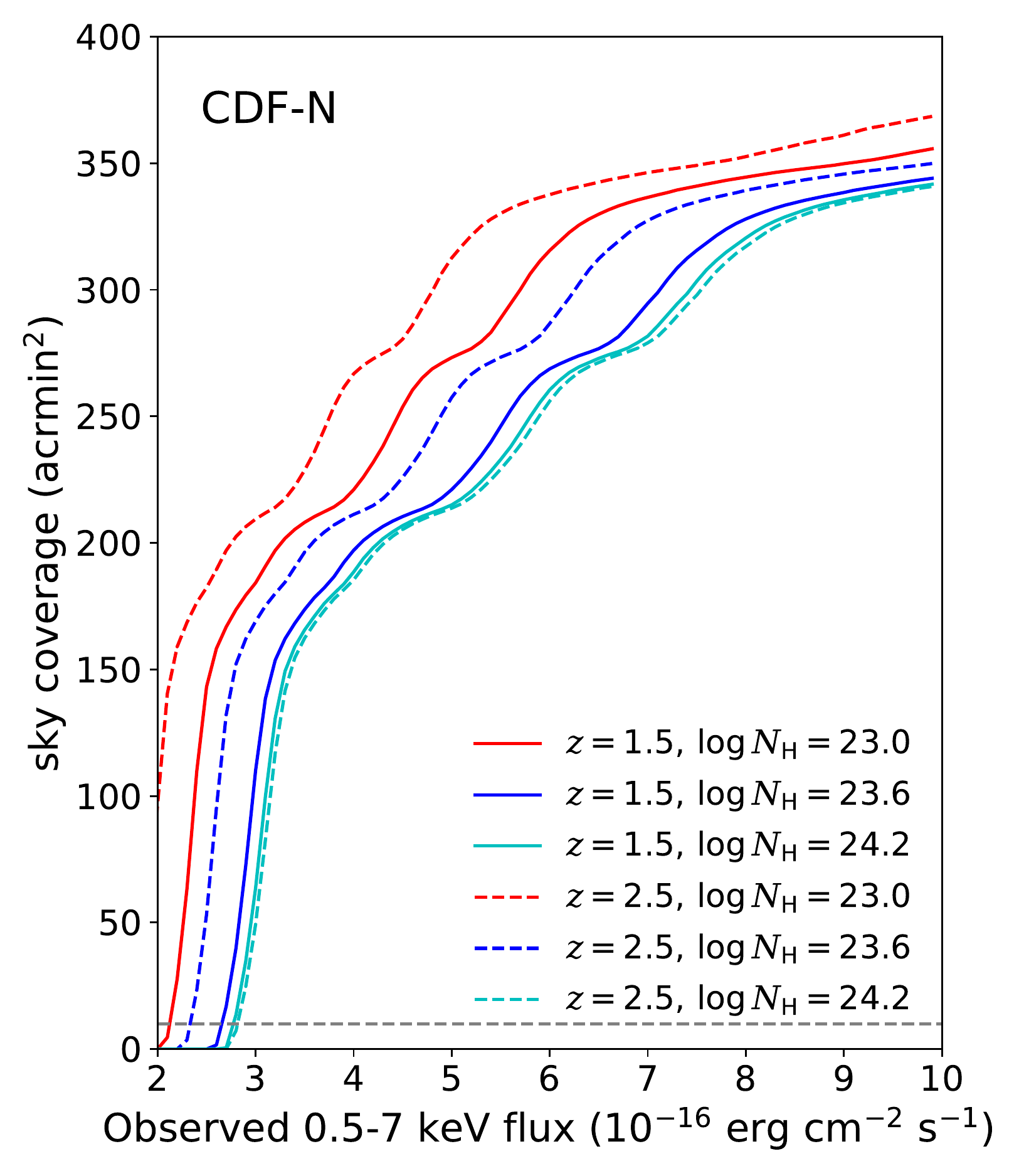}  
\end{minipage}  
\caption{Sky coverage as a function of the observed 0.5--7~keV flux corresponding to the count cut 20 in the CDF-S and CDF-N, shown for three different \nh~values at redshifts, respectively.}
\label{fig:sky}
\end{figure*} 

\subsection {Sky Coverage Effect}\label{subsec:sky}

Due to \emph{Chandra}'s instrument features, the point-source PSF size increases and the effective exposure time dramatically decreases toward large off-axis angles. 
The sensitivities of detecting faint sources reduce prominently at the outskirt of the field, leading to small sky coverage while the observed flux is low. 
Therefore, a correction must be made. 

We calculate the energy flux to count rate conversion factor (ECF) by assuming the soft-excess model (model D) with \gm, \fr~and \fs~fixed at 1.8, 1.0 and 1\%, respectively. Combining ECF and the exposure map, we build a sensitivity map that represents the flux limits corresponding to the 20-count cut. Using this map, we calculate the sky coverage as a function of observed 0.5--7~keV flux, \nh~and redshift as shown in Figure \ref{fig:sky}. Then we measure the sky coverage for each source based on their observed flux, \nh~and redshift obtained from spectral fitting. We define $\omega_{\rm area}$ as the ratio between the source sky coverage and the maximum sky coverage of the two \emph{Chandra} surveys (484.2 arcmin$^2$ and 447.5 arcmin$^2$ in the CDF-S and the CDF-N, respectively). To correct for the sky coverage effect, we simply weigh each source by $1 / \omega_{\rm area}$ while resampling the observed \nh~distribution. To avoid extremely large weights, we apply another cut for the 22 sources with sky coverage less than  100 $\rm arcmin^2$, setting their weighting factors to the median factor 1.3, since these sources lie close to the count cut and we cannot rule out the possibility that their extremely small sky coverage may be a result from inappropriate spectral modeling.

\begin{figure*}
\centering
\begin{minipage}[t]{0.4\linewidth}  
\centering  
\includegraphics[width=\linewidth]{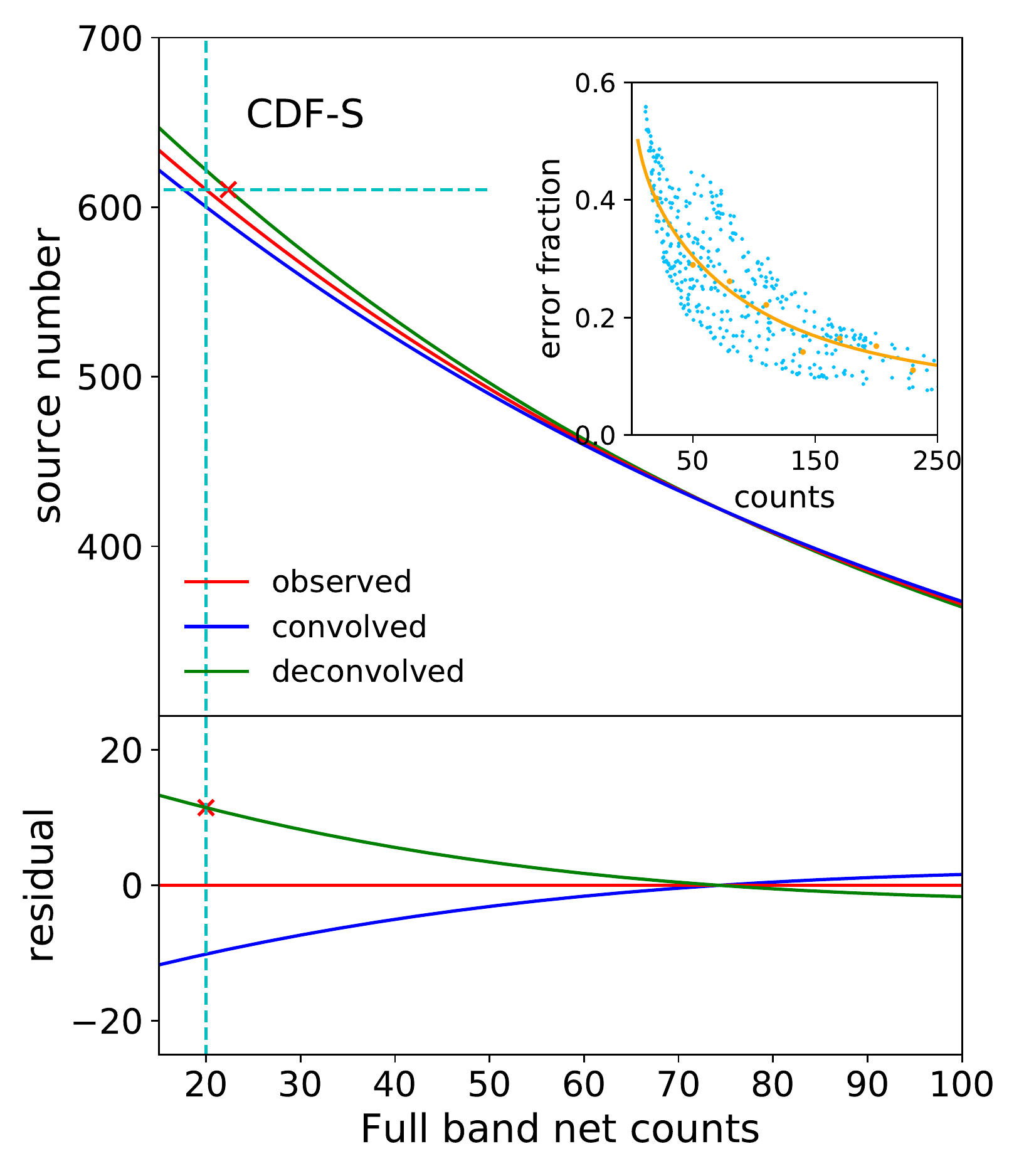}  
\end{minipage}  
\begin{minipage}[t]{0.4\linewidth}  
\centering  
\includegraphics[width=\linewidth]{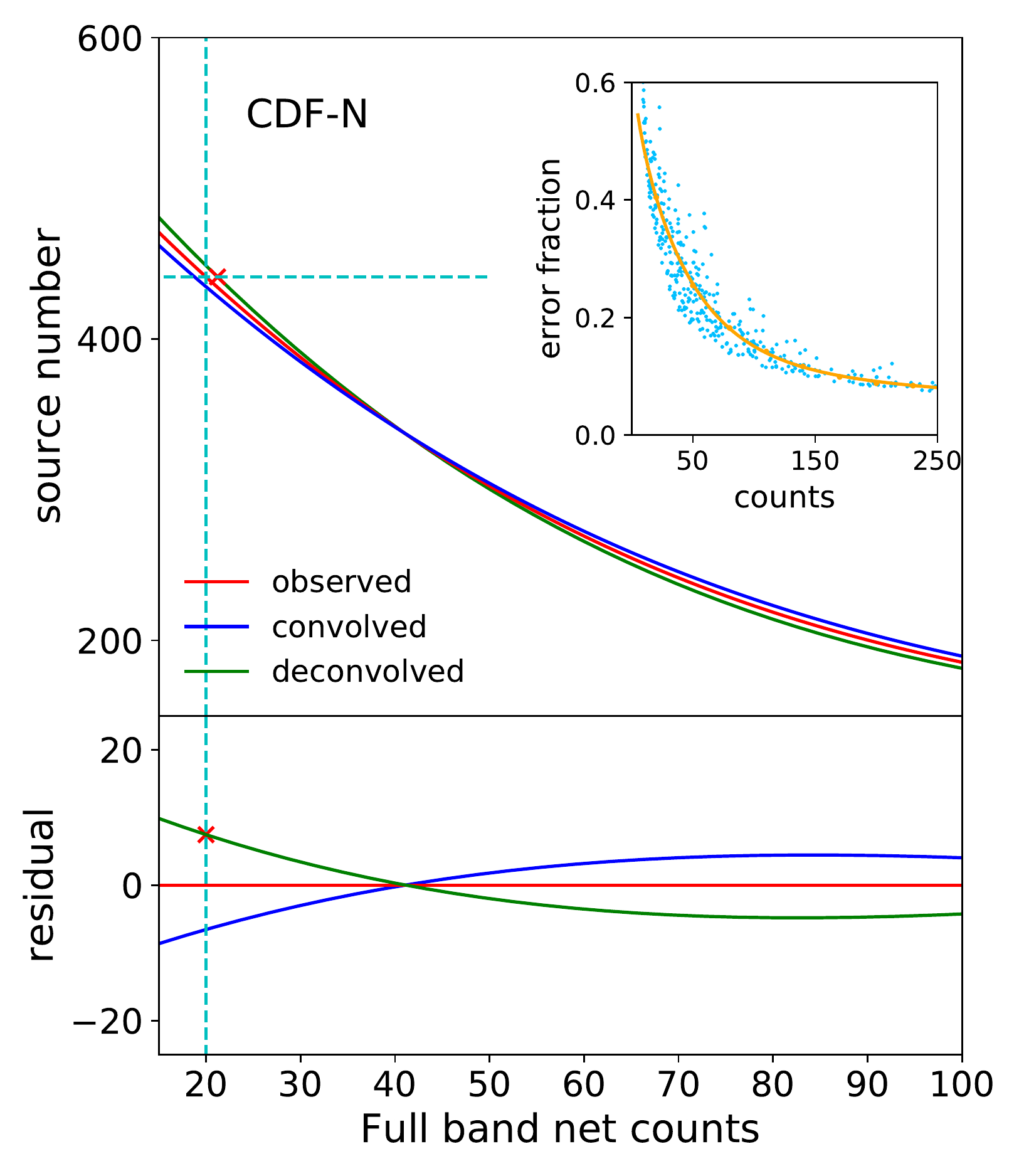}  
\end{minipage}  
\caption{Illustration of the Eddington bias in the CDF-S and CDF-N, respectively. In each panel/field, the red curve shows the smoothed cumulative count distribution for all AGNs with counts \>~20 and off-axis angle \<~10$'$. 
The blue curve is obtained by Poisson-sampling the observed count distribution by considering the measurement errors. For sources without count error measurements in the catalogs, we fit the mean error fraction $f_{\rm err}$ ($f_{\rm err} = \rm count_{error} / count$) in a given count range (shown as orange points) as a function of full-band net counts using sources with errors; the fitted curve is shown in the inset. Then we simply assign an error to sources without errors, which is given by their total net counts times the corresponding $f_{\rm err}$. We apply a pseudo-deconvolved method to obtain the deconvolved intrinsic count distribution by shifting the blue curve leftward or rightward (based on the location on the right or left side of the node of the two curves) with the value equal to the displacement between the blue and red curves, as shown as the green curve. 
From the residual between the observed and deconvolved curves, we can see that we missed 11.6 and 6.5 sources (shown as red crosses in the bottom panels), which correspond to the effective count cuts of 22.4 and 22.1 (shown as red crosses in the top panels) in the CDF-S and CDF-N, respectively. 
}
\label{fig:eddington}
\end{figure*} 

\subsection {Eddington Bias}

Considering the measurement error of net counts, sources with intrinsically low counts may exceed our count cut (20 photons), and sources with high counts may be missed in our sample. Since the number of faint and bright sources are not equal, this error leads to the Eddington bias. To correct for this bias, we convolve the cumulative count distribution for AGNs in the two \emph{Chandra} survey catalogs with the count errors using Poisson sampling in order to obtain the intrinsic count distribution. We follow the `pseudo-deconvolved' method proposed in L17 (see Section  5.1.3 of L17 for details), by shifting the observed curve leftward with the value equal to the displacement between the observed and convolved curves. We then obtain the deconvolved curve that represents the intrinsic cumulative count distribution. 

The difference between the deconvolved and the observed curves can be treated as the number of sources missed or mis-included (depending on the adopted count cut) in our sample. As shown in Figure \ref{fig:eddington}, at a 20-photon count cut, we miss 11.6 and 6.5 sources in the CDF-S and CDF-N, respectively. This is due to the fact that though sources with counts \<~20 may outnumber the bright sources, such faint sources are not detected in the source catalogs. By controlling the sample size (i.e., the cumulative source number in the $y$-axes) to be the same, we obtain the effective net counts of 22.4 and 21.1 for the CDF-S and CDF-N, respectively. Thus the 20-photon count cut will be replaced by the new effective count cut while performing other corrections.

\begin{figure*}
\centering
\begin{minipage}[t]{0.4\linewidth}  
\includegraphics[width=\linewidth]{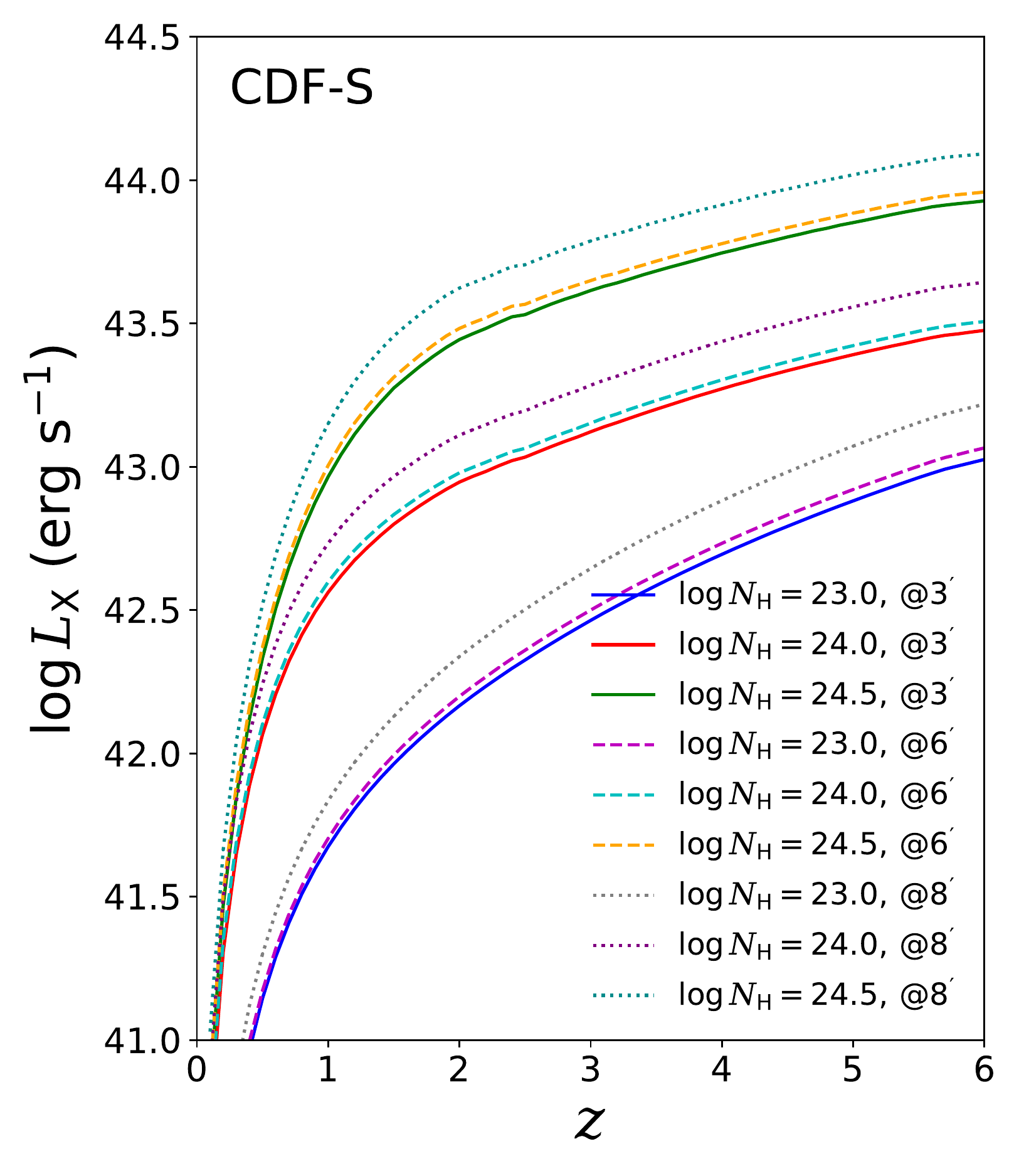}  
\end{minipage}  
\begin{minipage}[t]{0.4\linewidth}  
\includegraphics[width=\linewidth]{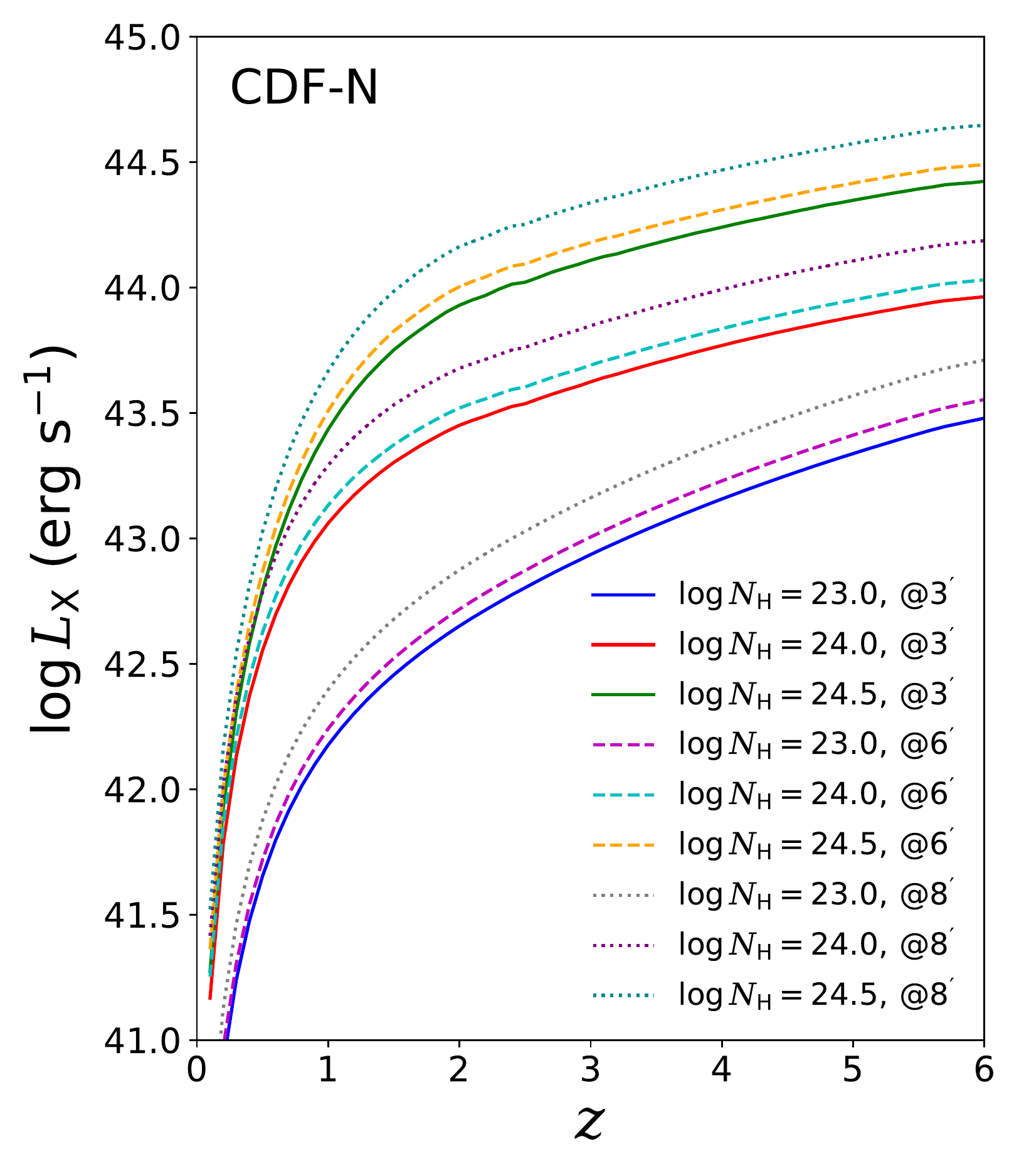}  
\end{minipage}  

\caption{Detection boundary curves as a function of redshift for different \nh~and off-axis angles, corresponding to the effective count cut 22.4 in the CDF-S and 21.1 in the CDF-N, respectively.}
\label{fig:boundary}
\end{figure*}

\begin{figure*}
\begin{minipage}[t]{0.33\linewidth}  
\includegraphics[width=\linewidth]{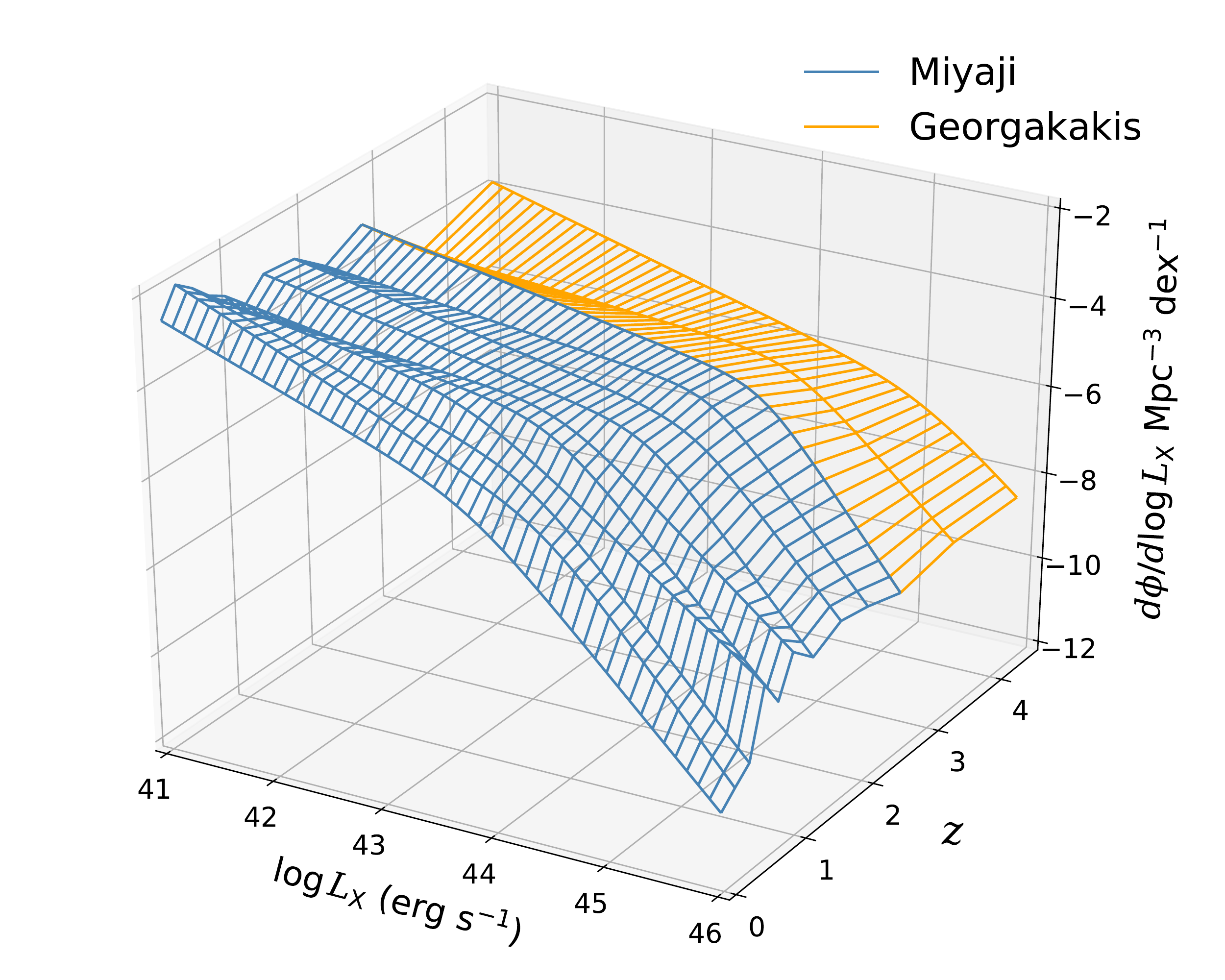}  
\end{minipage}  
\begin{minipage}[t]{0.33\linewidth}  
\includegraphics[width=\linewidth]{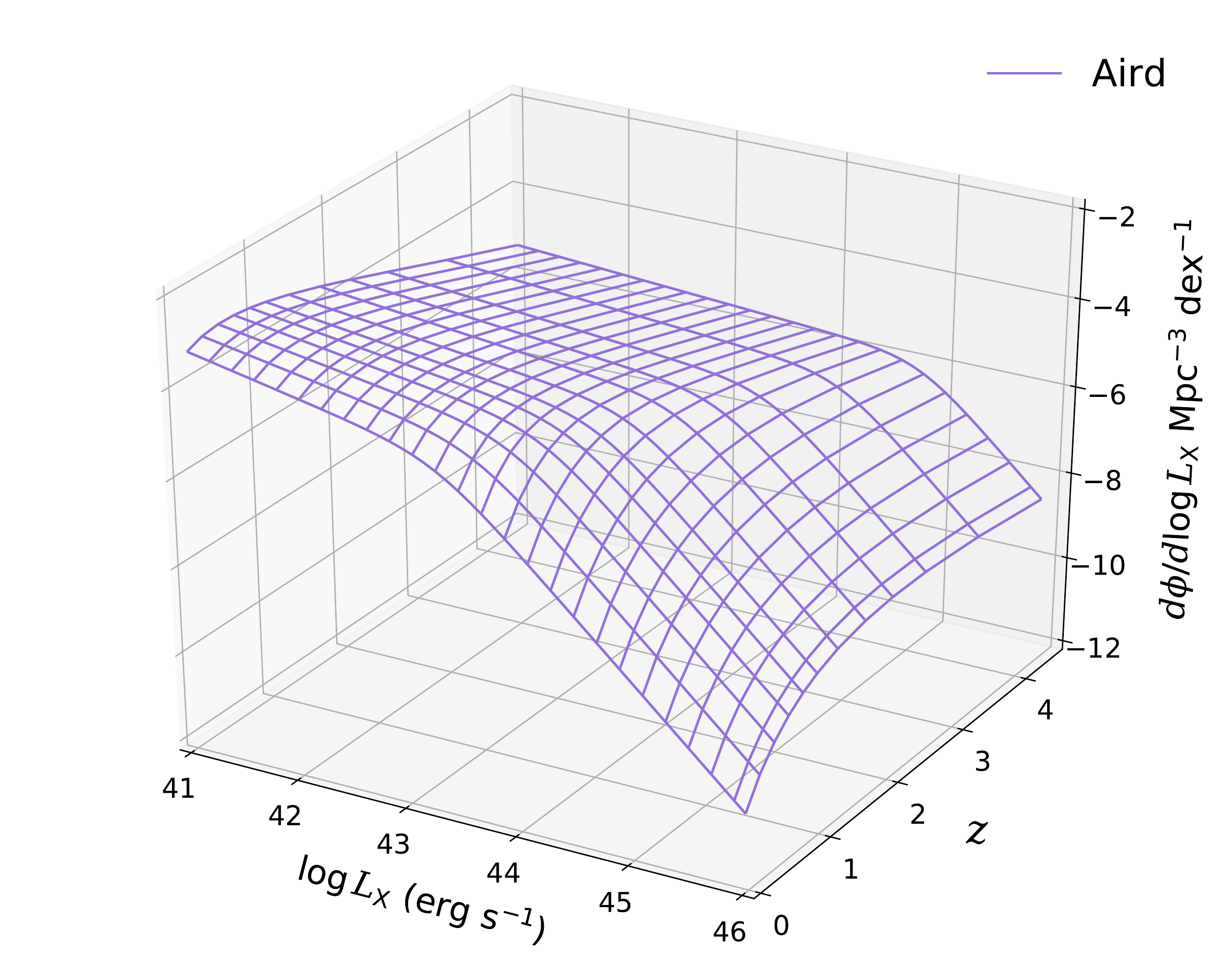}  
\end{minipage}  
\begin{minipage}[t]{0.33\linewidth}  
\includegraphics[width=\linewidth]{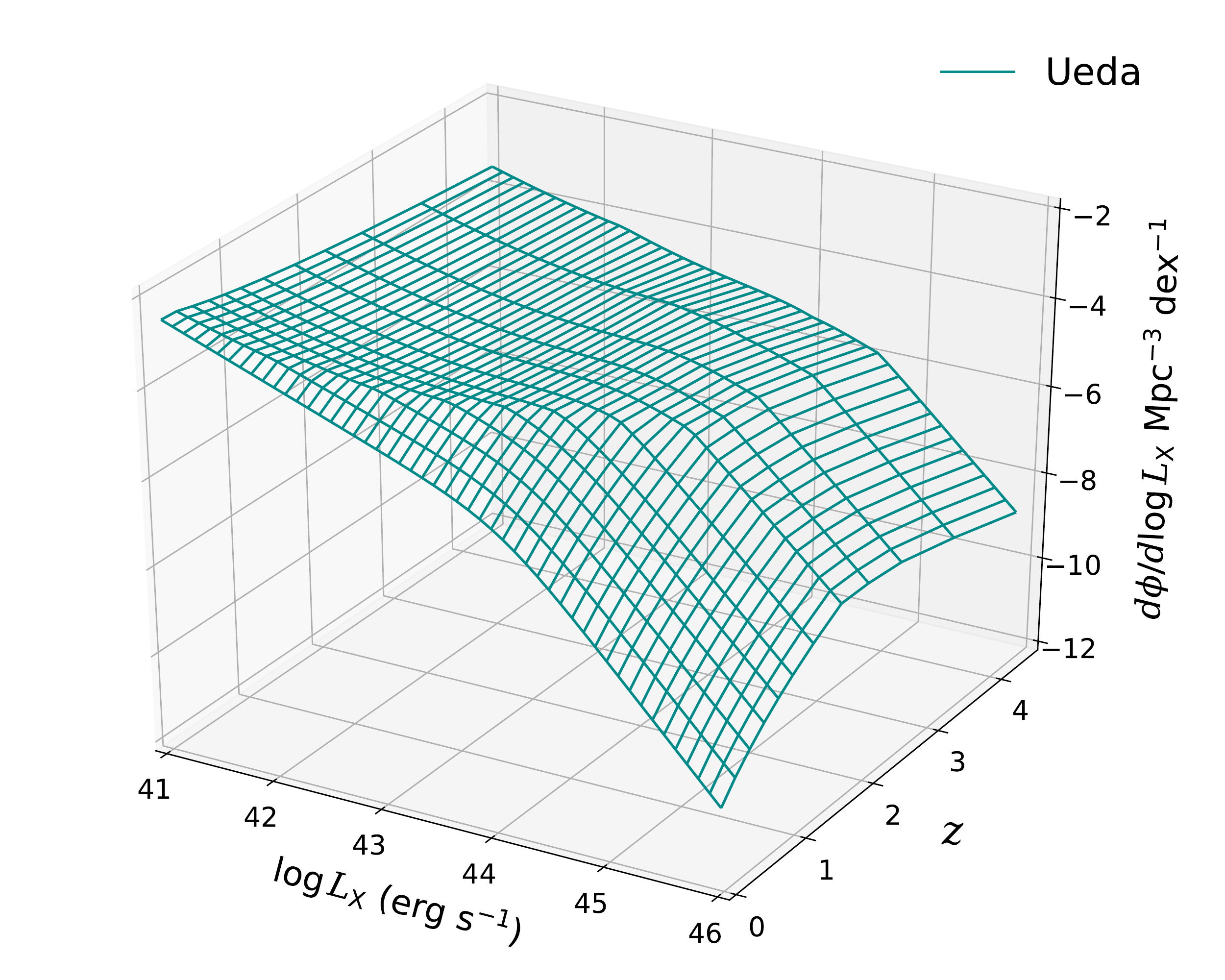}  
\end{minipage} 
\caption{Left: The combined X-ray LFs taken from \cite{Miyaji2015} at $z$ \<~3 and \cite{Georgakakis2015} at $z$ \>~3. Middle: The obscured (22 $\rm cm^{-2}$ \<~\lognh~\<~24 $\rm cm^{-2}$) AGN LFs from \cite{Aird2015}. Right: The Compton-thin AGN (\lognh~\<~24 $\rm cm^{-2}$) LFs from \cite{Ueda2014}.}
\label{fig:LF}
\end{figure*} 

\subsection {Malmquist Bias}
\label{subsec:malm}

So far we have only obtained the corrected observed \nh~distribution for our sample. In order to derive the intrinsic \nh~distribution for the highly obscured AGN population, we should also consider the undetectable part of the \emph{Chandra} survey, which is know as the Malmquist bias that sources with lower luminosities will be missed by a flux-limited survey.
More specifically, given the count rate limit of the survey (the effective count cut in our situation), whether a source is detectable or not depends on its redshift, spectral shape, column density, intrinsic luminosity as well as its position in the field, since large off-axis angles lead to significant lower sensitivities.

To quantify this bias, we generate fake X-ray spectra to find out the minimum \lx~($L_{\rm cut}$, corresponding to our effective count cut) in different redshift and \nh~bins, which can be detected at a specific position in the field.
We divide the two CDFs into 10 annuli, with each corresponding to an interval of 1$'$. 
While running XSPEC simulations, we use the averaged exposure time, rmf file and arf file calculated from all sources in each annulus. Model D is used in the simulation with the parameters set as the same as in Section \ref{subsec:sky}. 

We show some of the detectable boundary curves in Figure \ref{fig:boundary}. We define the detectable region as the area between the boundary curve and the maximum log \lx~= 46 \ergs. As we can clearly see, sources with high \nh~and large off-axis angles have significantly small detectable regions. 
We follow Equation 3 in T06, 
\begin{gather}
\nonumber F(N_{\rm H}) dN{\rm_H} = f(N_{\rm H}) dN_{\rm H} \int_0^{z_{\rm max}} \frac{dV}{dz}dz\\
\times \int_{L_{{\rm cut} (N_{\rm H}, z)}}^{L_{\rm max}} N(L_X, z)\, dL_X,
\label{eq:tozzi3}
\end{gather}
where $f$(\nh) represents the intrinsic \nh~distribution, $F$(\nh) represents the observed \nh~distribution, and $N$(\lx, $z$) is the X-ray LF. The detectable source fraction for each \nh~bin relative to the \lognh~= 23 $\rm cm^{-2}$ bin in a given redshift interval can be calculated as:
\begin{equation}
f=\frac{\int_{L_{{\rm cut} (N_{\rm H}, z)}}^{L_{\rm max}} N(L_{\rm X}, z)\, dL_{\rm X}}{ \int_{L_{\rm cut (23, z)}}^{L_{\rm max}} N(L_{\rm X}, z)\, dL_{\rm X}}.
\label{eq:factor}
\end{equation}
Thus for the bin we choose,
\begin{equation}  
f=
\left\{  
            \begin{array}{lr}  
             1, &{\rm if\ 23.0\ \rm cm^{-2} \leq log\, }N_{\rm H} < 23.25\ \rm cm^{-2},\\  
             <1, &{\rm if\ log\, }N_{\rm H} \geq  23.25\ \rm cm^{-2}.\\      
             \end{array}  
\right.  
\end{equation}

Note that T06 derived the above Equation \ref{eq:tozzi3} by assuming that $f(N_{\rm H}, L_{\rm X}, z)$, which represents the possibility of detecting a source with column density in the range of \nh~to  \nh~+ d\nh~for a given $L_X$ and $z$, varying slowly as \lx~and $z$ change. 
There is evidence that the obscured fraction of AGNs evolves with both luminosity and redshift. However, we do not consider such a complicated evolution and simply extract $f(N_{\rm H}, L_{\rm X}, z)$ from the integral.
Therefore, by multiplying $1/f$ to the observed \nh~distribution $F(N_{\rm H})\, dN_{\rm H}$ in each bin, sources with different \nh~are corrected to cover the same observable space with respect to the \lognh$\,$=$\,23 - 23.25\ \rm cm^{-2}$ bin and the Malmquist bias is corrected. To avoid the case of the corrected distribution being dominated by sources with extremely small detectable fractions, we apply a cut that while $f$ \<~0.25, the weighting factor $1/f$ will be simply set to 1/0.25. Adopting a lower (0.2) or higher (0.3) detectable fraction cut will lead to $\sim 2\%$ systematic difference of the final intrinsic CT-to-highly-obscured fraction. 

Apparently, the correction depends on both spectral models and the detailed shape of X-ray LFs. Therefore we compare the results using different X-ray LFs. 
We combine the $z$ \<~3 LF from \cite{Miyaji2015} and $z$ \>~3 LF from \cite{Georgakakis2015} as our first LF model, which is the same as in L17. As for the second model, we adopt \cite{Aird2015} LFs for obscured AGNs (22 $\rm cm^{-2}$ \<~\nh~\<~24 $\rm cm^{-2}$). We do not use their LFs for CT AGNs, since they obtained a systematically lower CT fraction than other works and our result in Section \ref{subsec:logn_logs} prefers higher CT number counts (but see the discussion in Section  6.4 of \cite{Aird2015} that this low CT fraction should not be ruled out) and the spectral model they used may not be appropriate for CT AGNs (see Section \ref{subsec:cmp_pre}). The third LF we used is from \cite{Ueda2014}. The three LF models are shown in Figure \ref{fig:LF}. We note that the adopted obscured AGN LFs may not be suitable for CT AGNs, but an extensive exploration of this issue is beyond the scope of this work.

\begin{figure*}
\begin{minipage}[t]{0.32\linewidth}  
\includegraphics[width=\linewidth]{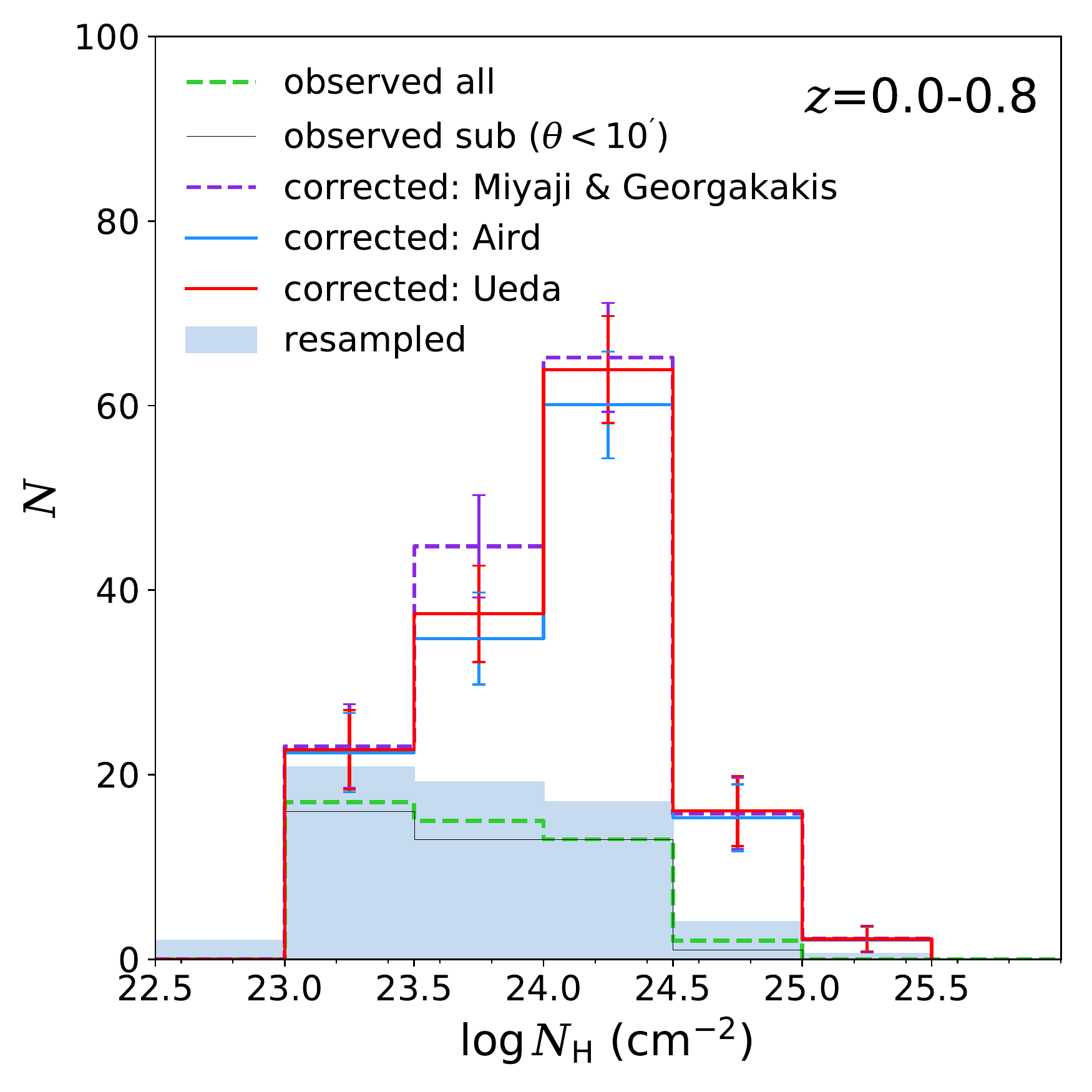}  
\end{minipage}  
\begin{minipage}[t]{0.32\linewidth}  
\includegraphics[width=\linewidth]{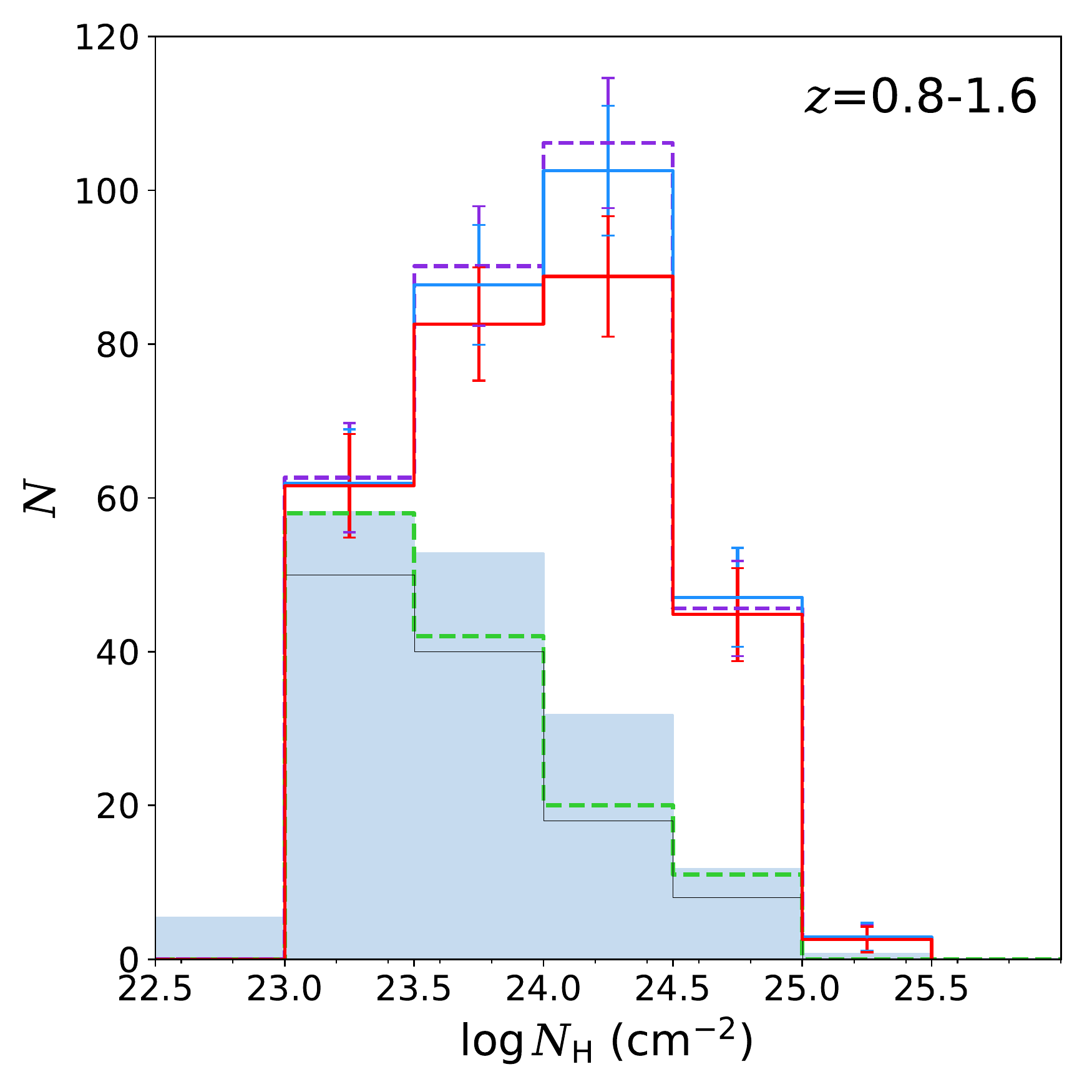}  
\end{minipage}  
\begin{minipage}[t]{0.32\linewidth}  
\includegraphics[width=\linewidth]{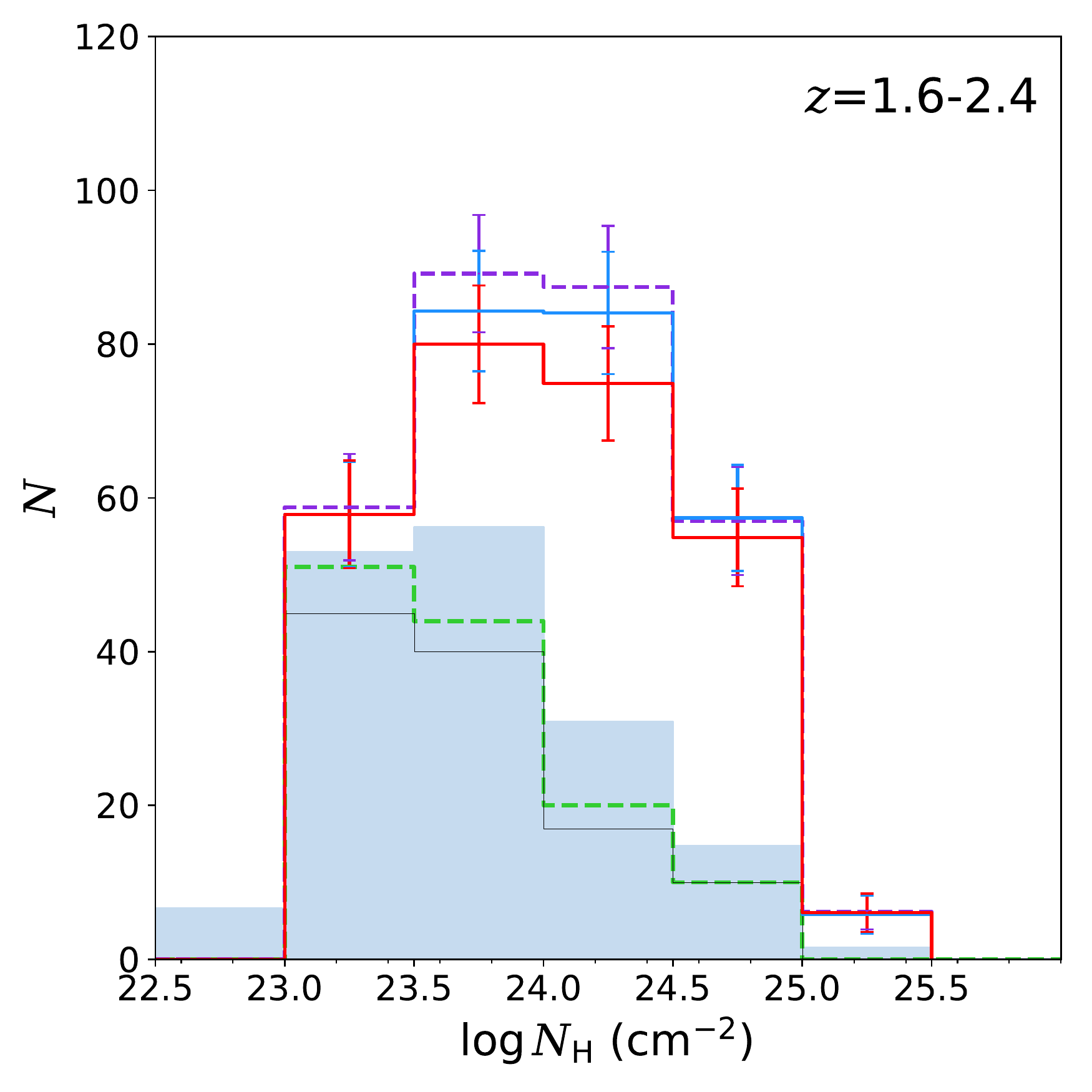}  
\end{minipage}  
\begin{minipage}[t]{0.32\linewidth}  
\includegraphics[width=\linewidth]{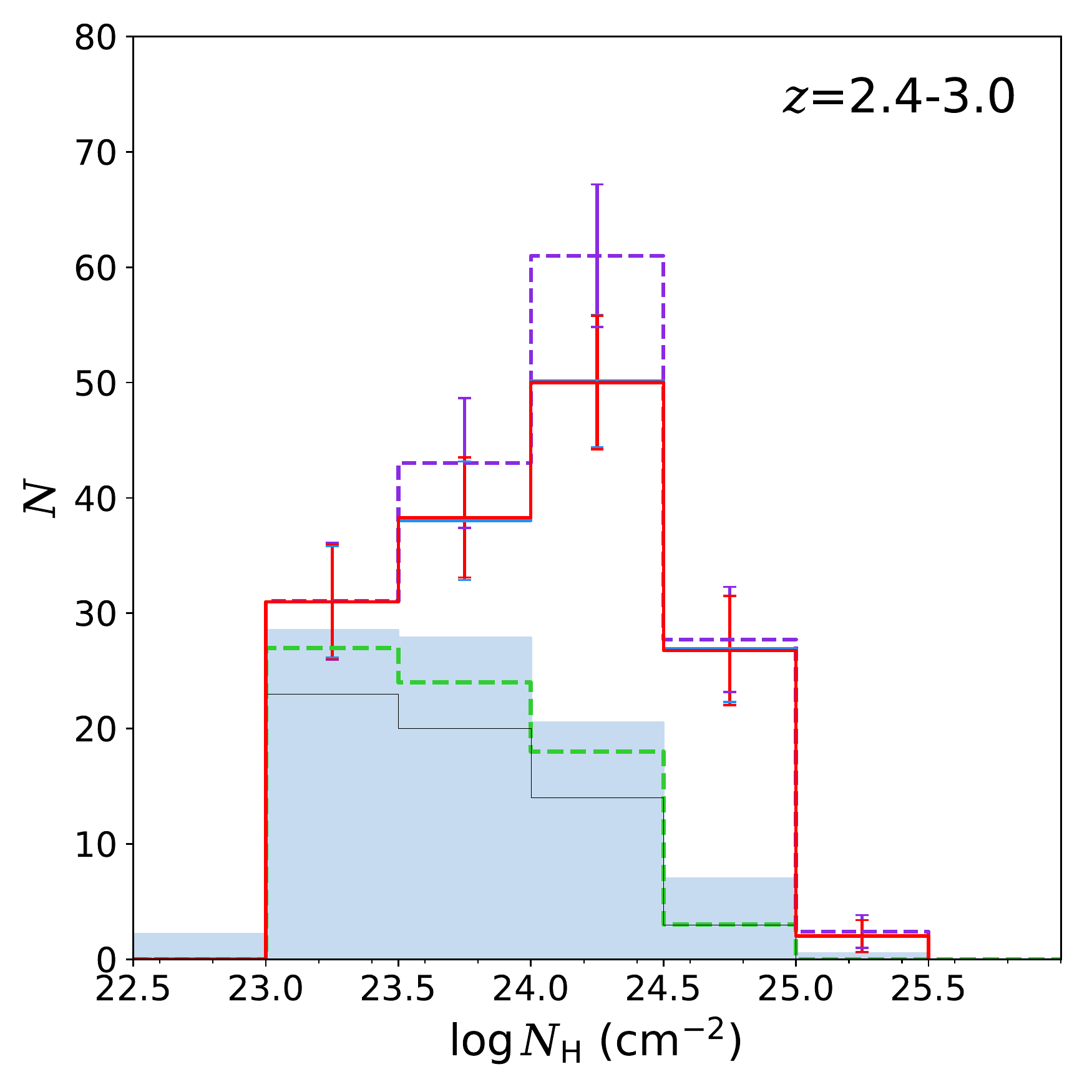}  
\end{minipage}  
\centering
\begin{minipage}[t]{0.32\linewidth}  
\includegraphics[width=\linewidth]{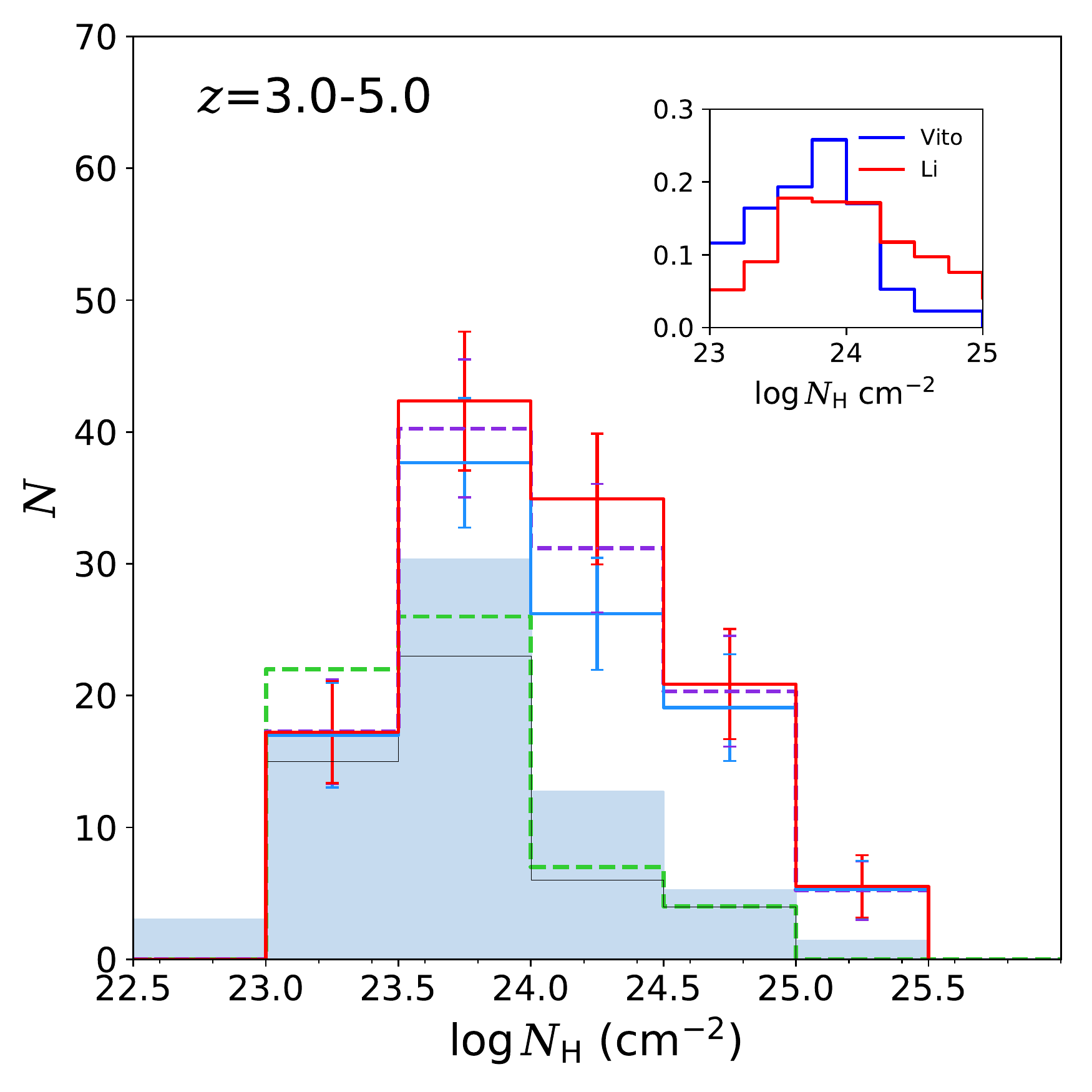}  
\end{minipage} 
\begin{minipage}[t]{0.32\linewidth}  
\includegraphics[width=\linewidth]{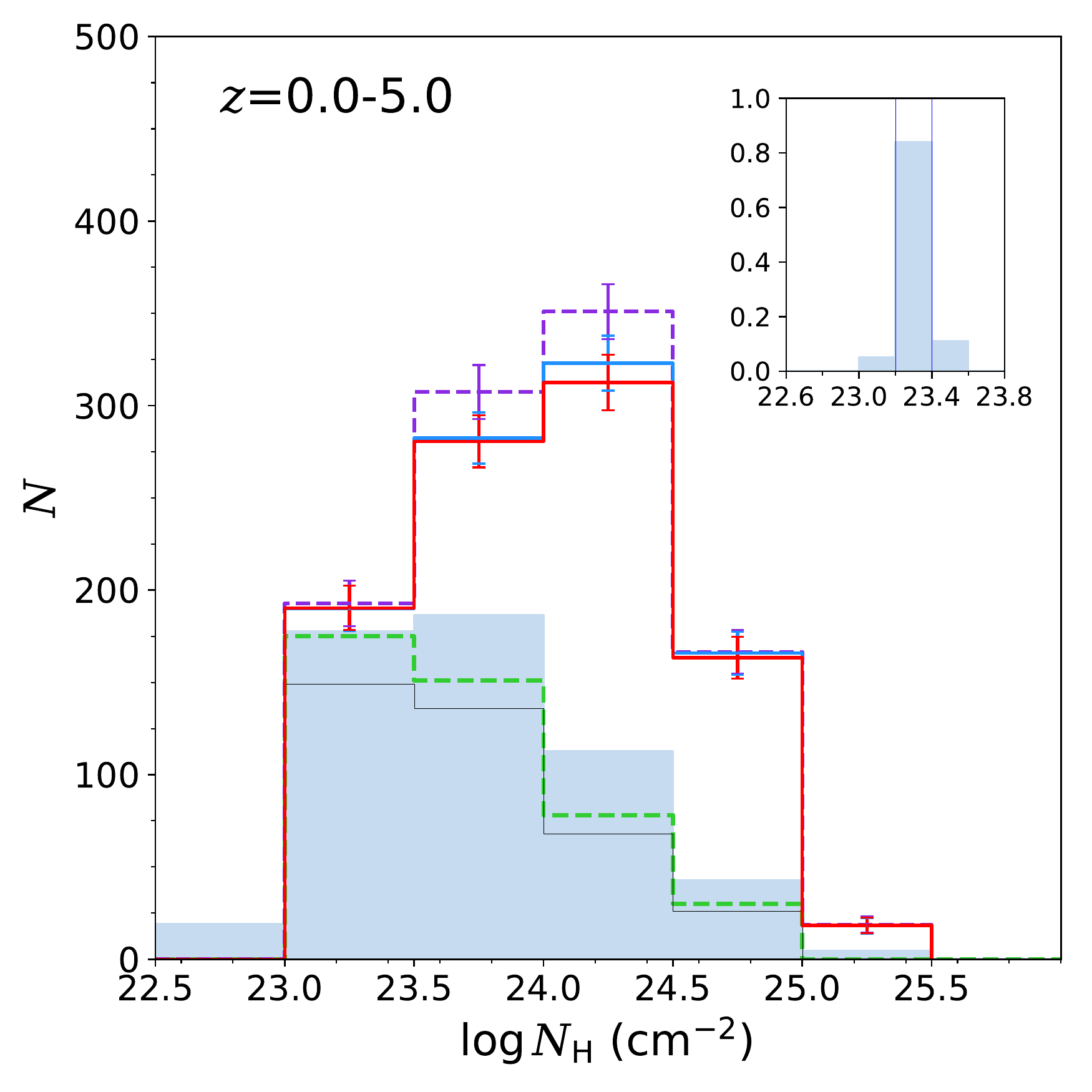}  
\end{minipage} 

\caption{\nh~distributions for our HOS in five redshift bins and the entire redshift range ($z = 0-5$). The error bars are estimated through bootstrapping. 
In each panel, the green dashed histogram shows the observed best-fit \nh~distribution; 
the black histogram shows the observed \nh~distribution for sources with off-axis angle \<~10$'$;
the blue shaded area shows the resampled \nh~distribution that takes into account the spectral fitting errors and sky coverage effect; and the red and blue solid histograms and purple dashed histogram show the intrinsic \nh~distributions after correcting for the undetectable parts of the two surveys using the \cite{Ueda2014}, \cite{Aird2015}, as well as \cite{Miyaji2015} \& \cite{Georgakakis2015} LFs, respectively. 
The comparison of the normalized \nh~distribution of highly obscured AGNs at $z$ \>~3 presented in \cite{Vito2018} and this work are shown in inset of the $z = 3 \sim 5$ panel.
The inset in the $z = 0-5$ panel illustrates the effect of the resampling procedure, where the observed \nh~distribution shown in blue is resampled into the shaded histogram by considering \nh~errors.}
\label{fig:mytorus_nh}
\end{figure*} 

\subsection {Intrinsic \nh~Distribution}
\label{subsec:intrinsic_nh}
Figure \ref{fig:mytorus_nh} shows the intrinsic \nh~distributions in five redshift bins after correcting for all the aforementioned biases. We define the intrinsic CT-to-highly-obscured source fraction $f_{\rm CH}$ as 
\begin{equation}
f_{\rm CH} = \frac{N ({\rm log}\, N {\rm_H} \geq 24)}{N ({\rm log}\, N \rm_H \geq 23)},\\
\end{equation}
and list its evolution with redshift in Table \ref{table:ct_fraction}. 

As we can see, different LFs give consistent results. It is obvious that the corrected \nh~distributions are quite different from the observed ones, especially in the CT regime. This highlights the importance of carefully correcting for the observational biases in order to uncover the underlying distributions.
Though the intrinsic \nh~distributions are different between high-redshift and low-redshift bins, the CT-to-highly-obscured fraction is in accordance with no evident redshift evolution given the uncertainties (i.e., $f_{\rm CH} \approx 0.52$ for all redshift bins; see Table \ref{table:ct_fraction}). 
Note that \cite{Buchner2015} also claimed that the CT fraction is constant across cosmic time, but the CT fraction in their work is defined as the number of CT AGNs to the total AGN population. To avoid possible biases induced by the fact that different redshift bins are sensitive to different luminosity ranges, we also select two subsamples with $43.0\ \ergs < \loglx < 44.5 \ \ergs$ at $z =  1-2$ and $z = 2-3$, respectively, which have similar \lx~distributions (see Figure \ref{fig:corner}), and perform the same corrections. The results are listed in Table \ref{table:ct_fraction} and are still consistent with no redshift evolution.

\cite{Vito2018} presented the \nh~distribution for the high-redshift ($z$ \>~3) AGNs also in the CDFs. Their analyses were based on modeling the X-ray spectra using the $wabs \times zwabs \times zpow$ model for the $z$ \>~3 sources by carefully taking into account the photometric redshift errors. The comparison of the two works is shown in the $z = 3.0 - 5.0$ bin in Figure \ref{fig:mytorus_nh}. Our result shows a higher CT fraction, which might be ascribed to the different redshift values adopted in the two works as well as the large weighting factors while we correcting for the undetectable part (see Section \ref{subsec:malm}). 

\begin{table}
\footnotesize
\centering
\caption{The evolution of the observed ($f_{\rm obs}$) and the corrected intrinsic ($f_{\rm int}$) CT-to-highly-obscured fraction with redshift. The values listed below are calculated by averaging the results from the three LF models.}
\begin{tabular}{c c c c }
\hline
\hline
$z$ & $f_{\rm obs, CH}$ & $f_{\rm int, CH}$ & $\overline{\loglx}$ (\ergs)\\
(1) & (2) & (3) & (4)\\
\hline
\multicolumn{4}{c}{Total sample}\\
\hline
0.0 $-$ 0.8 & 0.32 & $0.57 \pm 0.07$ & 42.9\\
0.8 $-$ 1.6 & 0.24 & $0.50 \pm 0.05$ & 43.5\\
1.6 $-$ 2.4 & 0.24 & $0.50 \pm 0.05$ & 43.7\\
2.4 $-$ 3.0 & 0.29 & $0.54 \pm 0.06$ & 43.9\\
3.0 $-$ 5.0 & 0.18 & $0.50 \pm 0.09$ & 44.1\\
0.0 $-$ 5.0 & 0.25 & $0.52 \pm 0.06$ & 43.6\\
\hline
\multicolumn{4}{c}{Subsample}\\
\hline
1.0 $-$ 2.0 & 0.25 & $0.50 \pm 0.05$ & 43.7\\
2.0 $-$ 3.0 & 0.26 & $0.51 \pm 0.05$ & 43.8\\
\hline
\end{tabular}

\vspace{10.0 pt}
{{\sc \bf{Notes.}} Top table: the results for the total sample. Bottom table: the results for two subsamples with $43.0\ \ergs < \loglx < 44.5\ \ergs$ at $z = 1.0 - 2.0$ and at $z = 2.0 - 3.0$, which have similar \lx~distributions (see Figure \ref{fig:corner}).}
\label{table:ct_fraction}
\end{table}

However, we note that all these aforementioned corrections are based on sources we do observe, including the correction of the undetectable part in a given \nh~bin for which we have to make assumptions regarding the unknown AGN population. The series of corrections can be simply understood as multiplying an evolved weighting factor to the observed
distribution, which have several limitations. First of all, as mentioned before, different redshift bins sample different luminosity ranges. Several authors have found that the obscured fraction changes with luminosity \citep{Buchner2015, Lusso2013}, thus the \nh~distribution might be luminosity-dependent. However, in Equation \ref{eq:tozzi3}, we ignore this effect to simplify the calculation.
Second, for the extremely buried sources (\nh~\>~$\rm 10^{25}\ cm^{-2}$), or intrinsically very faint sources, the corrections cannot be made since there are hardly any such sources observed due to the limit of the survey and the restrictions imposed by our sample-selection count cut. Therefore, the part of the \nh~distribution of sources that are under the detection limit is actually still missed in our final result. Furthermore, our identification method of CT AGNs is based on the absorption turnover imprinted in the spectra. For a source with \nh~\>~$\rm 10^{25}\ cm^{-2}$, the turnover occurs at rest-frame $\sim$ 20 keV. This implies that for low-redshift sources, the characteristic feature of CT AGNs may not be observable in the \chandra~spectra of limited energy range, which is coupled with the uncertainties induced by the photometric redshifts, leading to great challenges in correctly measuring the obscuration level. 
It is plausible that \nh~\>~$\rm 10^{25}\ cm^{-2}$ sources may contribute significantly to heavily obscured AGN population \citep{Risaliti1999}, but such sources are missed in our sample and the corrections at this part is beyond our attainment. 
Therefore, the \nh~distribution at the highest \nh~bin displayed in Figure \ref{fig:mytorus_nh} is highly uncertain and incomplete so that the CT-to-highly-obscured fractions in Table \ref{table:ct_fraction} should be better treated as lower limits.

\section{Spectral variability analyses}\label{sec:var}
X-ray variability is a ubiquitous feature among AGNs that can provide useful information about AGN properties and relevant underlying physical processes \citep[e.g.,][]{Vaughan2003, McHardy2006, Gonzalez2012, Soldi2014, Zheng2017}. 
In this section, we aim to study the main driving mechanism that causes the variability of highly obscured AGNs by investigating the variability of some main spectral parameters, such as the luminosity, column density and reflection strength.

To perform detailed long-term spectral variability analyses of highly obscured AGNs and expand the sample size, we select our sources from the HOS based on only one additional criterion, i.e., the net 0.5--7~keV counts in the total $\rm 7\ Ms$ exposure in the CDF-S or in the total $\rm 2\ Ms$ exposure in the CDF-N should be larger than 700 or 900, respectively.  
We bin data from neighboring observations as one epoch to enhance the S/N. For sources in the CDF-S that have counts $>$ 1500, we bin the data into 4 epochs; for sources having 700 $<$ counts $<$ 1500, we bin the data into 3 epochs. The data for all the sources in the CDF-N are binned into 3 epochs. Detailed binning information can be found in Table \ref{table:variab_bin}. These binning strategies and the adopted count criteria can roughly ensure that the average counts in each epoch are at least $\sim 200$. 
Thirty-one sources, which satisfy our selection criteria, are selected for subsequent variability analyses. The redshift distribution of this variability sample is shown in Figure \ref{fig:corner} and the count distribution is shown in Figure \ref{fig:cnts}. The mean counts for the variability sample is 1615. The binning process is carried out using the \texttt{combine\_spectra} tool in CIAO to generate the source spectra, background spectra, rmf and arf files in each epoch.

\begin{table}
\scriptsize
\caption{Observational epochs, variability sample and binning information}
\begin{center}
\begin{tabular}{c c c | c c c}
\hline
\hline
CDF-S & Start time & End  time & CDF-N & Start time & End time\\
\hline
I & 1999 Oct & 2000 Dec & I & 1999 Nov & 2000 Nov\\ 
II & 2007 Sep &  2007 Nov & II & 2001 Feb & 2001 Nov\\
III & 2010 Mar & 2010 Jul & III & 2002 Feb & 2002 Feb\\
IV & 2014 Jun & 2015 Jan\\
\hline
\end{tabular}
\vspace{10.0 pt}

\begin{tabular}{c c c c c c c c c c}
\hline  
XID & $z$ & $N$ & $M$ & counts & $N_{\rm obs}$ & I & II & III & IV\\
(1) & (2) & (3) & (4) & (5) & (6) & (7) & (8) & (9) & (10)\\
\hline
\multicolumn{10}{c}{CDF-S 4 epochs}\\
\hline
49 & 2.394 &  4 &   A &  2345 & 98 & 0.93 & 0.96 & 1.96 & 2.70\\
 81 & 3.309 &  4 &   A &  3579 & 102 & 0.93 & 0.96 & 1.96 & 2.88\\
 98 & 1.412 &  4 &   B &  3689 & 102 & 0.93 & 0.96 & 1.96 & 2.88\\
186 & 2.810 &  4 &   A &  2532 & 102 & 0.93 & 0.96 & 1.96 & 2.88\\
214 & 3.740 &  4 &   A &  2863 & 102 & 0.93 & 0.96 & 1.96 & 2.88\\
236 & 2.562 &  4 &   A &  1578 & 102 & 0.93 & 0.96 & 1.96 & 2.88\\
308 & 1.097 &  4 &   B &  2484 & 102 & 0.93 & 0.96 & 1.96 & 2.88\\
458 & 2.291 &  4 &   C &  2093 & 102 & 0.93 & 0.96 & 1.96 & 2.88\\
746 & 3.064 &  4 &   A &  2144 & 102 & 0.93 & 0.96 & 1.96 & 2.88\\
760 & 3.350 &  4 &   B &  1974 & 102 & 0.93 & 0.96 & 1.96 & 2.88\\
876 & 3.470 &  4 &   A &  4191 & 102 & 0.93 & 0.96 & 1.96 & 2.88\\
933 & 1.654 &  4 &   A &  1792 & 102 & 0.93 & 0.96 & 1.96 & 2.88\\
\hline
\multicolumn{10}{c}{CDF-S 3 epochs}\\
\hline
 63 & 0.737 &  3 &   B &   848 & 29 & 0.45 & 0.56 & 1.04 \\
 91 & 2.256 & 3 & A & 756 & 29 & 0.45 & 0.56 & 1.04 \\
 752 & 0.733 &  3 &   A &   866 & 55 & 0.80 & 1.27 & 1.69\\
785 & 1.600 &  3 &   C &  1345 & 22 & 0.31 & 0.36 & 0.60\\
 73 & 2.509 &  3 &   A &   879 & 102 & 1.89 & 1.96 & 2.88\\
 240 & 1.185 &  3 &   C &   785 & 102 & 1.89 & 1.96 & 2.88\\ 
 249 & 0.735 &  3 &   A &   830 & 102  & 1.89 & 1.96 & 2.88\\ 
 328 & 1.536 &  3 &   C &  1432 & 102  & 1.89 & 1.96 & 2.88\\ 
 367 & 0.604 &  3 &   C &   948 & 102  & 1.89 & 1.96 & 2.88\\ 
 399 & 1.730 &  3 &   C &  1077 & 102  & 1.89 & 1.96 & 2.88\\ 
551 & 3.700 &  3 &   C &   756 & 102  & 1.89 & 1.96 & 2.88\\ 
621 & 1.213 &  3 &   A &   784 & 102   & 1.89 & 1.96 & 2.88\\ 
658 & 1.845 &  3 &   A &   839 & 102   & 1.89 & 1.96 & 2.88\\ 
733 & 2.404 &  3 &   D &  973 & 102 &  1.89 & 1.96 & 2.88\\
818 & 2.593 &  3 &   C &  711 & 102   & 1.89 & 1.96 & 2.88\\ 
840 & 1.220 &  3 &   C &  1321 & 102   & 1.89 & 1.96 & 2.88\\ 
846 & 2.483 &  3 &   A &   974 & 102   & 1.89 & 1.96 & 2.88\\ 
\hline
\multicolumn{10}{c}{CDF-N 3 epochs}\\
\hline
66 & 0.959 &  3 &   A &   934 & 20 & 0.49 & 0.87 & 0.58\\
143 & 1.727 &  3 &   A &  1516 & 20 & 0.49 & 0.87 & 0.58\\ 
\hline
\end{tabular}
\end{center}

\vspace{10.0 pt}
{\sc \bf{Notes.}}
Top: Definition of the observational epochs. Bottom: Variability sample and the binning information. Column (3): number of binning epochs. Column (4): best-fit model. Column (5): total 0.5--7~keV net counts. Column (6): number of individual observations of the source during the total CDF-S $\rm 7\ Ms$ or CDF-N $\rm 2\ Ms$ exposure. Columns (7 -- 10): exposure time in units of Ms in each epoch. 
\label{table:variab_bin}
\end{table}

\subsection {Method}
We use the same spectral fitting models as described in Section  \ref{subsec:model} to perform spectral variability analyses, except for removing the constants \fr~and \fs~in the model and untying the normalizations of the intrinsic power law and other spectral components. We simultaneously fit the spectra in each epoch using the best-fit model for each source obtained in Section  \ref{sec:spec_result}, but this time we allow $norm_{\rm ref}$ to vary freely to search for possible variation in the Compton-scattered continuum. Considering the relatively low counts in each epoch, the fitting strategies are adopted as follows:

\begin{enumerate}[1.]
\item The uncertainties on $\Gamma$ can cause large degeneracies if we set it free in all epochs. 
Given that some sources may have large variability in $\Gamma$, we first let \gm~in each epoch vary freely and obtain $C_{\rm free}$. Then for the first two adjacent epochs (i.e., epoch1 and epoch2), we link \gm~and measure the $\Delta C$ with respect to $C_{\rm free}$. If $\Delta C$ \>~3.84 (for $\Delta \rm d.o.f = 1$), \gm~is considered to be different between the two epochs at \>~95\% confidence level, and we will set \gm~free; otherwise, we will link \gm. Then for the second epoch pair (i.e., epoch2 and epoch3), we link their \gm~and compare the Cstat value with the last step. After traversing all the epoch pairs, if no variability is detected, \gm~is linked together.

\item The reflection and soft excess components are often considered to be produced in large-scale clouds, e.g., torus,  for highly obscured AGNs. Since the timespan in the rest-frame is (1+$z$) times less than in the observed frame and the typical redshift of our sample is relatively high, it is reasonable to assume that in such a short timescale, the large-scale components may not vary dramatically. To better constrain the normalization of the intrinsic power law, we tie the normalizations of the reflection component and the soft excess component in each epoch, respectively, unless setting them free leading to $\Delta C$ \>~3.84. Finally, only one source shows a significant change in the reflection component, and the soft excess fluxes remain constant for all sources.

\item For those sources with a weak and constant reflection component confirmed in the last step, we delete the MYTS and MYTL components from the model and only use MYTZ to fit the spectra, in order to reduce the number of free parameters. This procedure may break the self-consistency of the MYTorus model to some extent but will not influence the variability analysis, since we only remove a small constant component in each epoch, which has very little influence on the calculated $\chi^2$.

\item The \nh~and normalization of the intrinsic power law always vary freely.
\end{enumerate}

The simultaneous fitting yields the best-fit model parameters $\Gamma$, $norm_{\rm ref}$ and $norm_{\rm scat}$. Then we fit the spectra in each epoch with $\Gamma _i$, $norm_{\rm ref,i}$ and $norm_{\rm scat,i}$ fixed at the best-fit value obtained in the simultaneous fitting to obtain the \nh, observed flux, intrinsic flux and the corresponding errors in each epoch. The spectral fitting parameters of each epoch are listed in Table \ref{table:multi-epoch}. No significant photon index variability is detected for all sources.

\begin{table*}
\caption{Multi-epoch spectral fitting results for the variability sample}
\centering
\footnotesize
\begin{tabular}{r c c c c c c c r r r}
\\
\hline\hline
XID & $\Gamma$ & $N_{\rm H1}^a$ & $N_{\rm H2}$ & $N_{\rm H3}$ & $L \rm_{X1}^b$ & $L \rm_{X2}$ & $L \rm_{X3}$ & flux$_1^c$ & flux$_2$ & flux$_3$\\
\hline
\multicolumn{11}{c}{CDF-S}\\
\hline
 63 &  1.54 & 0.06$_{-0.01}^{+0.01}$ & 0.14$_{-0.02}^{+0.02}$ & 0.19$_{-0.03}^{+0.04}$ & 0.47 & 0.66 & 0.41 & 16.0$_{ -1.8}^{ +1.9}$ & 16.6$_{ -1.5}^{ +1.5}$ &  8.8$_{ -0.9}^{ +0.9}$\\
 73 &  1.81 & 0.64$_{-0.07}^{+0.08}$ & 0.56$_{-0.11}^{+0.11}$ & 0.67$_{-0.07}^{+0.07}$ & 3.37 & 1.87 & 3.78 &  2.7$_{ -0.3}^{ +0.3}$ &  1.7$_{ -0.3}^{ +0.3}$ &  2.9$_{ -0.3}^{ +0.3}$\\
 91 &  2.08 & 0.16$_{-0.03}^{+0.03}$ & 0.12$_{-0.03}^{+0.03}$ & 0.14$_{-0.03}^{+0.03}$ & 1.29 & 1.03 & 1.25 &  2.3$_{ -0.3}^{ +0.3}$ &  2.1$_{ -0.3}^{ +0.3}$ &  2.4$_{ -0.3}^{ +0.3}$\\
240 &  1.40 & 1.80$_{-0.15}^{+0.17}$ & 1.41$_{-0.15}^{+0.17}$ & 1.34$_{-0.11}^{+0.12}$ & 1.76 & 1.21 & 1.04 &  3.7$_{ -0.4}^{ +0.5}$ &  3.4$_{ -0.5}^{ +0.5}$ &  3.3$_{ -0.3}^{ +0.3}$\\
249 &  1.51 & 0.50$_{-0.12}^{+0.14}$ & 0.28$_{-0.06}^{+0.08}$ & 0.14$_{-0.01}^{+0.02}$ & 0.25 & 0.16 & 0.18 &  2.1$_{ -0.3}^{ +0.3}$ &  2.4$_{ -0.3}^{ +0.3}$ &  4.2$_{ -0.3}^{ +0.3}$\\
328 &  2.12 & 1.91$_{-0.12}^{-1.02}$ & 1.31$_{-0.10}^{+0.11}$ & 1.57$_{-0.06}^{+0.06}$ & 4.37 & 1.69 & 4.94 &  4.2$_{ -0.4}^{ +0.4}$ &  3.4$_{ -0.4}^{ +0.5}$ &  6.0$_{ -0.4}^{ +0.4}$\\
367 &  2.07 & 0.42$_{-0.02}^{+0.03}$ & 0.38$_{-0.02}^{+0.02}$ & 0.47$_{-0.04}^{+0.06}$ & 0.14 & 0.13 & 0.07 &  3.7$_{ -0.3}^{ +0.2}$ &  3.8$_{ -0.3}^{ +0.3}$ &  2.2$_{ -0.2}^{ +0.2}$\\
399 &  1.40 & 0.24$_{-0.02}^{+0.02}$ & 0.23$_{-0.03}^{+0.03}$ & 0.32$_{-0.02}^{+0.02}$ & 0.38 & 0.13 & 0.32 &  2.9$_{ -0.3}^{ +0.3}$ &  1.5$_{ -0.3}^{ +0.3}$ &  2.6$_{ -0.2}^{ +0.2}$\\
551 &  1.84 & 1.56$_{-0.10}^{+0.11}$ & 1.62$_{-0.14}^{+0.15}$ & 1.54$_{-0.09}^{+0.09}$ & 4.02 & 2.31 & 3.89 &  2.0$_{ -0.2}^{ +0.2}$ &  1.5$_{ -0.2}^{ +0.2}$ &  1.9$_{ -0.2}^{ +0.2}$\\
621 &  2.11 & 0.18$_{-0.02}^{+0.02}$ & 0.15$_{-0.02}^{+0.02}$ & 0.17$_{-0.02}^{+0.02}$ & 0.40 & 0.32 & 0.30 &  2.3$_{ -0.2}^{ +0.2}$ &  2.0$_{ -0.3}^{ +0.3}$ &  1.8$_{ -0.2}^{ +0.2}$\\
658 &  1.67 & 0.15$_{-0.02}^{+0.02}$ & 0.18$_{-0.03}^{+0.03}$ & 0.15$_{-0.02}^{+0.02}$ & 0.54 & 0.57 & 0.52 &  2.0$_{ -0.2}^{ +0.2}$ &  2.0$_{ -0.2}^{ +0.2}$ &  2.0$_{ -0.2}^{ +0.2}$\\
733 &  1.84 & 0.21$_{-0.03}^{+0.04}$ & 0.17$_{-0.03}^{+0.03}$ & 0.12$_{-0.02}^{+0.02}$ & 0.94 & 0.96 & 0.92 &  1.8$_{ -0.2}^{ +0.2}$ &  2.0$_{ -0.2}^{ +0.2}$ &  2.2$_{ -0.2}^{ +0.2}$\\
752 &  2.24 & 0.15$_{-0.02}^{+0.02}$ & 0.15$_{-0.02}^{+0.02}$ & 0.11$_{-0.01}^{+0.01}$ & 0.28 & 0.33 & 0.25 &  4.6$_{ -0.7}^{ +0.7}$ &  5.4$_{ -0.7}^{ +0.7}$ &  5.0$_{ -0.5}^{ +0.5}$\\
785 &  1.97 & 0.54$_{-0.05}^{+0.05}$ & 0.56$_{-0.04}^{+0.05}$ & 0.53$_{-0.02}^{+0.03}$ & 4.73 & 5.97 & 9.26 & 16.2$_{ -2.7}^{ +2.8}$ & 18.8$_{ -2.1}^{ +2.1}$ & 29.7$_{ -1.8}^{ +1.8}$\\
818 &  1.97 & 0.14$_{-0.02}^{+0.02}$ & 0.11$_{-0.03}^{+0.03}$ & 0.19$_{-0.02}^{+0.03}$ & 0.61 & 0.46 & 0.93 &  1.1$_{ -0.1}^{ +0.1}$ &  0.9$_{ -0.1}^{ +0.1}$ &  1.5$_{ -0.1}^{ +0.1}$\\
840 &  1.60 & 0.58$_{-0.04}^{+0.05}$ & 0.59$_{-0.04}^{+0.04}$ & 0.40$_{-0.02}^{+0.02}$ & 0.53 & 0.49 & 0.59 &  3.9$_{ -0.3}^{ +0.3}$ &  3.8$_{ -0.3}^{ +0.3}$ &  5.1$_{ -0.3}^{ +0.3}$\\
846 &  1.56 & 0.14$_{-0.02}^{+0.03}$ & 0.11$_{-0.02}^{+0.03}$ & 0.15$_{-0.03}^{+0.03}$ & 0.84 & 0.86 & 0.90 &  2.1$_{ -0.2}^{ +0.2}$ &  2.2$_{ -0.2}^{ +0.2}$ &  2.2$_{ -0.2}^{ +0.2}$\\
\hline
\multicolumn{11}{c}{CDF-N}\\
\hline
 66 &  1.92 & 0.34$_{-0.06}^{+0.09}$ & 0.24$_{-0.02}^{+0.02}$ & 0.29$_{-0.03}^{+0.03}$ & 1.19 & 1.49 & 1.10 &  7.6$_{ -1.0}^{ +1.0}$ & 12.7$_{ -0.7}^{ +0.7}$ &  8.1$_{ -0.7}^{ +0.7}$\\
143 &  1.46 & 0.15$_{-0.02}^{+0.02}$ & 0.15$_{-0.01}^{+0.01}$ & 0.11$_{-0.01}^{+0.01}$ & 2.07 & 2.79 & 2.42 & 10.2$_{ -0.9}^{ +0.9}$ & 13.6$_{ -0.7}^{ +0.6}$ & 13.1$_{ -0.7}^{ +0.7}$\\
\hline
\end{tabular}

\begin{tabular}{r c c c c c c c c c r r r r}
\\
\hline\hline
XID & $\Gamma$ & $N_{\rm H1}$ & $N_{\rm H2}$ & $N_{\rm H3}$ & $N_{\rm H4}$ & $L \rm_{X1}$ & $L \rm_{X2}$ & $L \rm_{X3}$ & $L \rm_{X4}$ & flux$_1$ & flux$_2$ & flux$_3$ & flux$_4$\\
\hline
\multicolumn{14}{c}{CDF-S}\\
\hline
49 &  1.66 & 0.19$_{-0.02}^{+0.02}$ & 0.15$_{-0.03}^{+0.04}$ & 0.30$_{-0.03}^{+0.04}$ & 0.16$_{-0.02}^{+0.02}$ & 3.90 & 1.85 & 2.69 & 3.44 &  8.3$_{ -0.6}^{ +0.6}$ &  4.4$_{ -0.6}^{ +0.6}$ &  4.7$_{ -0.4}^{ +0.4}$ &  7.8$_{ -0.4}^{ +0.4}$\\
 81 &  1.83 & 0.17$_{-0.02}^{+0.02}$ & 0.14$_{-0.02}^{+0.02}$ & 0.17$_{-0.02}^{+0.02}$ & 0.19$_{-0.02}^{+0.02}$ & 7.32 & 9.21 & 5.15 & 6.37 &  7.8$_{ -0.4}^{ +0.5}$ & 10.4$_{ -0.6}^{ +0.6}$ &  5.4$_{ -0.3}^{ +0.4}$ &  6.5$_{ -0.3}^{ +0.3}$\\
 98 &  0.98 & 0.18$_{-0.02}^{+0.03}$ & 0.17$_{-0.02}^{+0.02}$ & 0.18$_{-0.02}^{+0.02}$ & 0.17$_{-0.01}^{+0.01}$ & 1.06 & 1.18 & 1.29 & 1.11 & 10.3$_{ -0.8}^{ +0.8}$ & 11.4$_{ -1.0}^{ +1.0}$ & 12.4$_{ -0.8}^{ +0.8}$ & 11.1$_{ -0.6}^{ +0.6}$\\
186 &  1.79 & 0.22$_{-0.02}^{+0.02}$ & 0.27$_{-0.03}^{+0.03}$ & 0.22$_{-0.02}^{+0.02}$ & 0.23$_{-0.02}^{+0.02}$ & 3.92 & 4.68 & 3.20 & 4.23 &  5.2$_{ -0.3}^{ +0.4}$ &  5.7$_{ -0.5}^{ +0.4}$ &  4.3$_{ -0.3}^{ +0.3}$ &  5.6$_{ -0.3}^{ +0.2}$\\
214 &  1.87 & 0.26$_{-0.03}^{+0.03}$ & 0.35$_{-0.04}^{+0.05}$ & 0.30$_{-0.03}^{+0.03}$ & 0.30$_{-0.03}^{+0.03}$ & 9.25 & 9.36 & 8.61 & 8.81 &  6.3$_{ -0.4}^{ +0.4}$ &  5.6$_{ -0.4}^{ +0.5}$ &  5.5$_{ -0.4}^{ +0.4}$ &  5.7$_{ -0.3}^{ +0.2}$\\
236 &  1.62 & 0.15$_{-0.02}^{+0.03}$ & 0.20$_{-0.03}^{+0.03}$ & 0.22$_{-0.03}^{+0.04}$ & 0.21$_{-0.03}^{+0.03}$ & 1.68 & 2.16 & 1.61 & 1.65 &  3.6$_{ -0.3}^{ +0.3}$ &  4.2$_{ -0.5}^{ +0.5}$ &  3.0$_{ -0.3}^{ +0.3}$ &  3.1$_{ -0.2}^{ +0.2}$\\
308 &  1.69 & 0.21$_{-0.01}^{+0.02}$ & 0.24$_{-0.02}^{+0.03}$ & 0.21$_{-0.01}^{+0.02}$ & 0.22$_{-0.02}^{+0.02}$ & 1.45 & 0.90 & 1.11 & 0.76 & 11.9$_{ -0.7}^{ +0.5}$ &  6.8$_{ -0.6}^{ +0.6}$ &  9.2$_{ -0.5}^{ +0.5}$ &  6.2$_{ -0.3}^{ +0.3}$\\
458 &  1.83 & 0.19$_{-0.02}^{+0.02}$ & 0.25$_{-0.02}^{+0.02}$ & 0.31$_{-0.02}^{+0.02}$ & 0.27$_{-0.02}^{+0.02}$ & 0.96 & 1.75 & 2.11 & 1.57 &  3.8$_{ -0.4}^{ +0.4}$ &  5.7$_{ -0.5}^{ +0.5}$ &  4.7$_{ -0.3}^{ +0.3}$ &  3.7$_{ -0.2}^{ +0.2}$\\
746 &  1.68 & 0.41$_{-0.04}^{+0.04}$ & 0.46$_{-0.06}^{+0.06}$ & 0.45$_{-0.04}^{+0.04}$ & 0.55$_{-0.03}^{+0.04}$ & 5.17 & 4.75 & 5.81 & 6.93 &  4.9$_{ -0.3}^{ +0.2}$ &  4.2$_{ -0.4}^{ +0.4}$ &  5.2$_{ -0.4}^{ +0.3}$ &  5.4$_{ -0.3}^{ +0.3}$\\
876 &  1.87 & 0.15$_{-0.02}^{+0.02}$ & 0.16$_{-0.02}^{+0.02}$ & 0.15$_{-0.02}^{+0.02}$ & 0.16$_{-0.02}^{+0.02}$ & 7.18 & 7.98 & 8.99 & 8.84 &  6.9$_{ -0.4}^{ +0.4}$ &  7.6$_{ -0.5}^{ +0.5}$ &  8.7$_{ -0.4}^{ +0.4}$ &  8.3$_{ -0.3}^{ +0.3}$\\
933 &  1.71 & 0.31$_{-0.03}^{+0.04}$ & 0.23$_{-0.03}^{+0.03}$ & 0.25$_{-0.02}^{+0.02}$ & 0.23$_{-0.02}^{+0.02}$ & 1.81 & 1.46 & 1.77 & 1.29 &  5.6$_{ -0.6}^{ +0.6}$ &  5.3$_{ -0.5}^{ +0.5}$ &  6.2$_{ -0.4}^{ +0.4}$ &  4.7$_{ -0.3}^{ +0.3}$\\
760 &  1.49 & 0.81$_{-0.10}^{+0.12}$ & 0.67$_{-0.10}^{+0.10}$ & 0.68$_{-0.06}^{+0.06}$ & 0.59$_{-0.05}^{+0.05}$ & 9.18 & 6.51 & 8.99 & 7.33 &  6.5$_{ -0.8}^{ +0.9}$ &  5.4$_{ -0.8}^{ +0.9}$ &  7.0$_{ -0.6}^{ +0.6}$ &  6.5$_{ -0.4}^{ +0.4}$\\
\hline
\end{tabular}

\vspace{10.0 pt}
{\bf Notes.}
a: in units of $\rm 10^{24}\ cm^{-2}$. b: $2-10$ keV intrinsic luminosity, in units of $\rm 10^{44}\ erg\ s^{-1}$. c: observed 0.5--7~keV flux, in units of $\rm 10^{-15}\ erg\ cm^{-2}\ s^{-1}$.
\label{table:multi-epoch}
\end{table*}

\subsection {$L_{\rm X}$, \nh~and Observed Flux Variability}
We identify \lx, \nh~and observed flux variable sources based on the classical $\chi^2$ test. 
For illustration, we explain how we determine \lx-variable sources. First, we use the $cflux$ model in XSPEC to calculate the observed and absorption-corrected $\rm 0.5-7$ keV fluxes and corresponding errors. Then we derive the intrinsic rest-frame $\rm 0.5-7$ keV luminosity using equation $L_{\rm 0.5-7\,keV} = 4 \pi d_L^2 f_{\rm 0.5-7\, keV,int}\, (1+z)^{\Gamma_{\rm int} - 2}$, where the ``int'' subscript represents the intrinsic value. Since the photon index and redshift are all fixed during spectral fitting of each single epoch, the error on $L_{\rm 0.5-7\,keV}$ is totally attributed to the error of the intrinsic flux $f_{\rm 0.5-7\,keV, int}$. Thus the $\chi^2$ of the rest-frame $L_{\rm 0.5-7\,keV}$ actually equals the $\chi^2$ of the observed-frame $f_{\rm 0.5-7\,keV, int}$. If a source does not vary, i.e., the intrinsic luminosity (or \nh, flux) is constant, assuming that the errors obey the Gaussian distribution, the $\chi^2$ value calculated from
\begin{equation}
\chi^2 = \sum_{i=1}^{N} \frac{(x_i - \overline{x})^2}{\sigma_{x, i}^2}
\label{eq:chi}
\end{equation}
should obey the $\chi^2$ distribution with ($N-1$) d.o.f, where $x$ represents the tested parameter (\lx, \nh~or flux), $\sigma_{x, i}$ means the 1$\sigma$ error in the $i$th epoch, and $N$ is the number of epochs. The $\chi^2$ test results are listed in Table \ref{table:exc}. When the source with 3 (4) binning epochs satisfies $\chi^2$ \>~7.8 (9.8), we regard it as being variable at \>~98\% confidence level.\footnote{The choice of this confidence level is because Y16 showed that it roughly corresponds to $\Delta \rm AIC = 4$, so that we can directly compare the $\chi^2$ results with those obtained using the AIC method.}  
We also check the \lx- and \nh-variable identification results using the Akaike information criterion (AIC) method as described in Y16 that does not need to assume the Gaussian errors (see Section  3.2.3 of Y16 for details). The \aic~values for each source are also listed in Table \ref{table:exc}. Sources with $\aic > 4.0$ are assigned as being variable. It can be seen that 30/31 and 30/31 of sources have consistent results in identifying \lx~and \nh~variability while using the $\chi^2$ test and the AIC method, respectively.\footnote{We decide to use the $\chi^2$ test results in the following analysis since for the two discrepant sources:  CDFS XID 933 and CDFN XID 66, the AIC results show neither \lx~nor \nh~variability, which is inconsistent with their flux-variable nature.}

Based on the $\chi^2$ test results, sources that show observed flux variability account for $55\pm13$\% (17/31) of the entire sample. The resulting \lx~and \nh~variable source fractions are $29\pm10\%$ (9/31) and $19\pm8\%$ (6/31), respectively.
We do not find any correlation between $L_{\rm X}$ and \nh~variability patterns, suggesting that the main reason that causes the variation of \nh~is likely not the change of ionization parameter induced by the variable \lx, but is likely the occultation of the clumpy clouds moving in/out of our l.o.s.

\begin{figure}
\includegraphics[width=\linewidth]{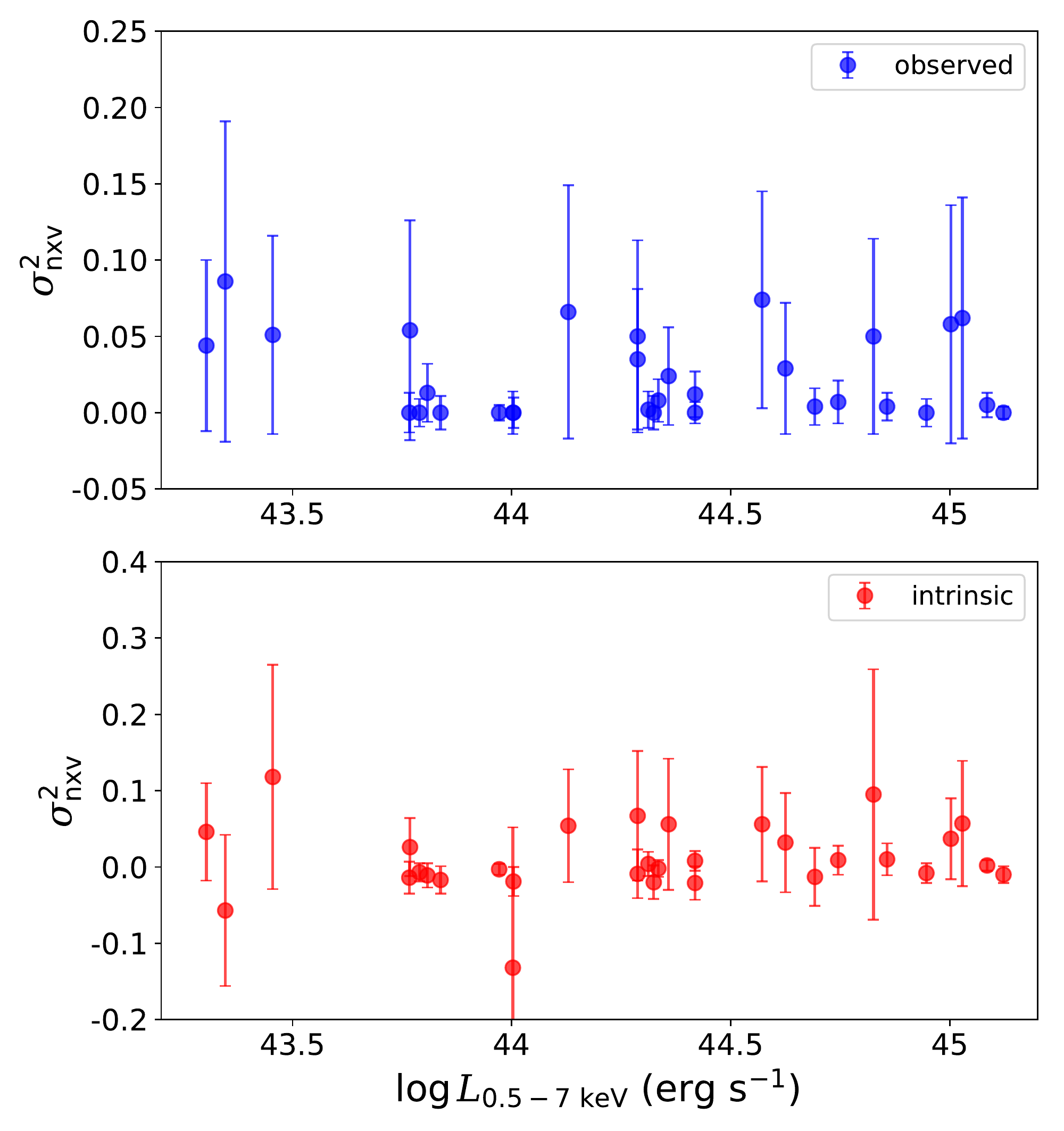}
\caption{Normalized excess variance (\nxv) of the observed (top) and intrinsic (bottom) 0.5--7~keV flux versus the intrinsic 0.5--7~keV luminosity for the variability sample. The original values without fixing negative \nxv~at 0 are displayed. The error bars are calculated using Equation \ref{eq:excerr}.}
\label{fig:exc}
\end{figure}

\subsection {Variability Amplitude Estimation}

\label{subsec:exc}
We use the normalized excess variance \nxv~(nxv) to estimate the variability amplitude. \nxv~ is defined as
\begin{equation}
\nxv~= \frac{1}{N \langle x\rangle^2} \sum_{i=1}^N[ (x_i - \langle x\rangle)^2 - (\delta x_i)^2 ],
\end{equation}
where $N$ is the number of binning epochs (3 or 4), $x_i$ and $\delta x_i$ are the best-fit parameters (observed flux, \lx, \nh) and their 1$\sigma$ errors, $\langle x\rangle$ is the unweighted mean of the calculated parameter. The error on \nxv~is estimated as $s_{\rm D} / (\langle x\rangle^2 \sqrt N)$, where
\begin{equation}
s_{\rm D}^2= \frac{1}{N - 1} \sum_{i=1}^N[ (x_i - \langle x\rangle)^2 - (\delta x_i)^2 - \nxv \langle x\rangle^2] ^2.
\label{eq:excerr}
\end{equation}
We calculate \nxv~and corresponding errors on the intrinsic rest-frame 0.5--7~keV luminosity (\nxv$_{,L}$), observed 0.5--7~keV flux (\nxv$_{,F}$) and \nh~(\nxv$_{,N}$) for each source. The results are listed in Table \ref{table:exc}. To check whether our calculation of $\sigma^2_{\rm nxv}$ is affected by the limited counts (e.g., \citealt{Zheng2017}; hereafter Z17), we perform Spearman rank correlation tests and find no significant correlation between \nxv$_{,F}$,  \nxv$_{,N}$, \nxv$_{,L}$ and counts with Spearman's $\rho = 0.07$, $p$-value = 0.70; $\rho = -0.11$, $p$-value = 0.56 and $\rho = 0.32$, $p$-value = 0.08, respectively. 

\begin{table}
\caption{The Chi-squared value $\chi^2$,  the \aic~value $\Delta$, and the normalized excess variance $\sigma^2$ of the observed 0.5--7~keV flux, intrinsic rest-frame $L \rm_{0.5-7\,keV}$ and \nh.}
\scriptsize
\begin{center}
\begin{tabular}{r r r r r r r r r r}
\hline \hline
XID & $\chi^2_{flux}$ & $\chi^2_\lx$ & $\Delta_\lx$ & $\chi^2_\nh$ & $\Delta_\nh$ & $\sigma^2_{flux}$ & $\sigma^2_\lx$ & $\sigma^2_\nh$\\
\hline
 \multicolumn{8}{c}{CDF-S 4 Epochs}\\
\hline
  49 &     53.3 &     21.9 &     11.9 &     17.3 &     11.3 &    0.074 &    0.056 &    0.073\\
  81 &     69.1 &     23.7 &     13.9 &      3.1 &     $-$3.2 &    0.058 &    0.037 &    0\\
  98 &      3.6 &      2.9 &     $-$3.1 &      0.4 &     $-$5.6 &    0 &    0 &    0\\
 186 &     13.4 &      9.6 &      2.6 &      2.5 &     $-$3.9 &    0.007 &    0.009 &    0\\
 214 &      2.9 &      0.4 &     $-$6.0 &      3.0 &     $-$3.3 &    0 &    0 &    0\\
 236 &      7.0 &      2.7 &     $-$3.6 &      4.4 &     $-$2.1 &    0.008 &    0 &    0\\
 308 &     86.7 &     44.9 &     32.4 &      1.4 &     $-$4.7 &    0.066 &    0.054 &    0\\
 458 &     21.7 &     28.4 &     13.5 &      7.5 &      2.8 &    0.024 &    0.056 &    0.013\\
 746 &      6.8 &      7.7 &      2.0 &      7.8 &      1.9 &    0.004 &    0.010 &    0.003\\
 876 &     12.0 &      5.5 &     $-$0.9 &      0.4 &     $-$5.6 &    0.005 &    0.002 &    0\\
 933 &     10.3 &     10.3 &      2.2 &      4.9 &     $-$1.4 &    0.002 &    0.004 &    0.005\\
 760 &      2.6 &      3.0 &     $-$3.1 &      5.6 &     $-$1.7 &    0 &    0 &    0\\
  \hline
 \multicolumn{8}{c}{CDF-S 3 Epochs}\\
 \hline
  63 &     34.2 &      7.2 &      2.8 &     65.2 &    29.45 &    0.054 &    0.026 &    0.147\\
  73 &      7.7 &      5.7 &      0.7 &      0.9 &    $-$3.27 &    0.029 &    0.032 &    0\\
 240 &      0.6 &      1.6 &     $-$0.9 &      1.6 &    $-$0.52 &    0 &    0 &    0\\
 249 &     31.0 &      2.2 &     $-$2.9 &    130.8 &    14.38 &    0.086 &    0 &    0.150\\
 328 &     20.8 &     18.3 &      8.5 &      2.4 &     1.14 &    0.050 &    0.095 &    0\\
 367 &     33.8 &      9.6 &      1.4 &      1.4 &    $-$0.38 &    0.044 &    0.046 &    0\\
 399 &     13.5 &     14.6 &      8.3 &      2.6 &    $-$1.31 &    0.051 &    0.118 &    0\\
 551 &      3.6 &      2.4 &     $-$2.2 &      0.1 &    $-$3.95 &    0.004 &    0 &    0\\
 658 &      0.1 &      0.2 &     $-$3.8 &      0.8 &    $-$3.14 &    0 &    0 &    0\\
 752 &      0.7 &      1.9 &     $-$2.3 &      6.8 &     1.16 &    0 &    0 &    0.002\\
 785 &     26.0 &     10.6 &      4.3 &      0.2 &    $-$3.87 &    0.062 &    0.057 &    0\\
 840 &     12.0 &      1.7 &      1.1 &     16.1 &     9.59 &    0.013 &    0 &    0.012\\
 846 &      0.3 &      0.1 &     $-$3.9 &      1.1 &    $-$2.96 &    0 &    0 &    0\\
 621 &      2.4 &      2.0 &     $-$2.0 &      1.3 &    $-$2.83 &    0 &    0 &    0\\
  91 &      0.7 &      1.0 &     $-$3.1 &      1.1 &    $-$2.92 &    0 &    0 &    0\\
   818 &      13.9 &      12.8 &     8.1 &      3.2 &    $-$0.7 &    0.035 &    0.067 &    0.004\\
   733 &  2.0 & 0.04 & -3.9 & 4.9 & 0.2 & 0 & 0 & 0.012\\
\hline
 \multicolumn{8}{c}{CDF-N 3 Epochs}\\
 \hline
  66 &     25.8 &      4.2 &     $-$0.4 &      8.6 &     0.29 &    0.050 &    0 &    0\\
 143 &     11.8 &      6.8 &      2.4 &      8.8 &     4.28 &    0.012 &    0.008 &    0.011\\

 \hline
\end{tabular}
\end{center}
{\bf Notes.}
Negative values of $\sigma^2$ are set to 0. Sources with $\chi^2 > 7.8$ for 3 epochs, $\chi^2 > 9.8$ for 4 epochs or $\aic > 4.0$ will be regarded as being variable at $> 98\%$ confidence level (Y16).
\label{table:exc}
\end{table}

We also calculate the fractional root-mean-square (frms) variability amplitude, which is defined as $\sqrt{\sigma^2_{\rm nxv}}$ for sources having $\sigma^2_{\rm nxv}$ \>~0 and can be treated as the percentage of the variability amplitude. It is obvious that our sample lacks sources that display large variability amplitudes. 
The mean frms values of the total sample for the observed flux, intrinsic luminosity and \nh~are 12\%, 10\% and 6\%, respectively. While for sources having positive nxv, the mean (maximum) frms values are 17\% (29\%), 19\% (34\%), and 16\% (39\%), respectively.

In addition, we do not find any correlation between \nxv$_{,F}$,  \nxv$_{,L}$ and \lx~(Figure \ref{fig:exc}) whereas other studies suggested a negative trend, i.e., luminous AGNs show relatively small variations \citep[Y16; Z17;][]{Paolillo2017}. We perform a Spearman correlation test for the 25 common sources using the \nxv$_{,F}$ derived in Z17 which is obtained from the light curve analysis, the result is still consistent with no correlation between \lx~and \nxv$_{,F}$ (Spearman's $\rho = -0.31$, $p$-value = 0.13). As pointed out by \cite{Allevato2013}, the calculation of \nxv~is biased at low counts and uneven cadence and should be restricted to large samples. 
Therefore, the limited source number and the broad redshift span may prevent us from detecting such a correlation. But beyond that, since Y16 and Z17 mainly focused on the AGNs with largest counts available in the CDF-S rather than highly obscured AGNs, the variability behavior and its underlying physical drivers of our sources and those in Y16 and Z17 may be intrinsically different.

Recently, \cite{Gonzales2018} found a new X-ray variability plane for AGNs that links the characteristic break frequency of the PSD ($f_{\rm break}$) with bolometric luminosity and \nh. This makes \nh~play a non-negligible role in understanding AGN variability.  As we will show in the next section, although Z17 argued that the \nh~variation does not contribute significantly to the total variability in their sample, the \nh~variability does play a substantial role for highly obscured AGNs. Nevertheless, the small \nxv~values indicate that the state of highly obscured AGNs does not vary dramatically while concerning the average source properties on a 17-year timescale in the observed-frame. 

\subsection {Detailed Variability Analysis} \label{subsec:srcvariab}

\begin{figure*}  
\begin{minipage}[t]{0.31\linewidth}  
\centering  
\includegraphics[width=2.3in]{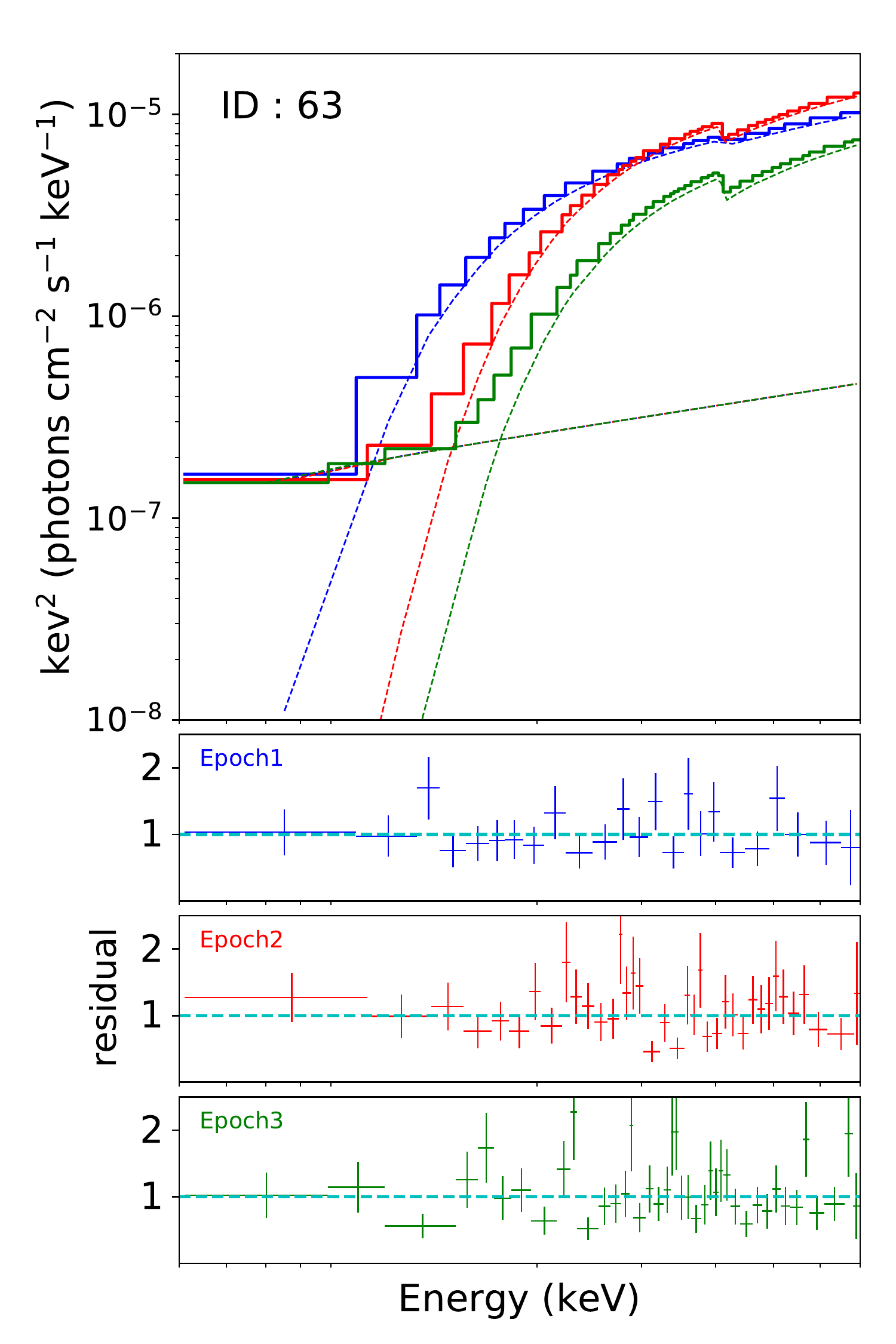}  
\label{fig:63}  
\end{minipage}
\begin{minipage}[t]{0.31\linewidth}  
\centering  
\includegraphics[width=2.3in]{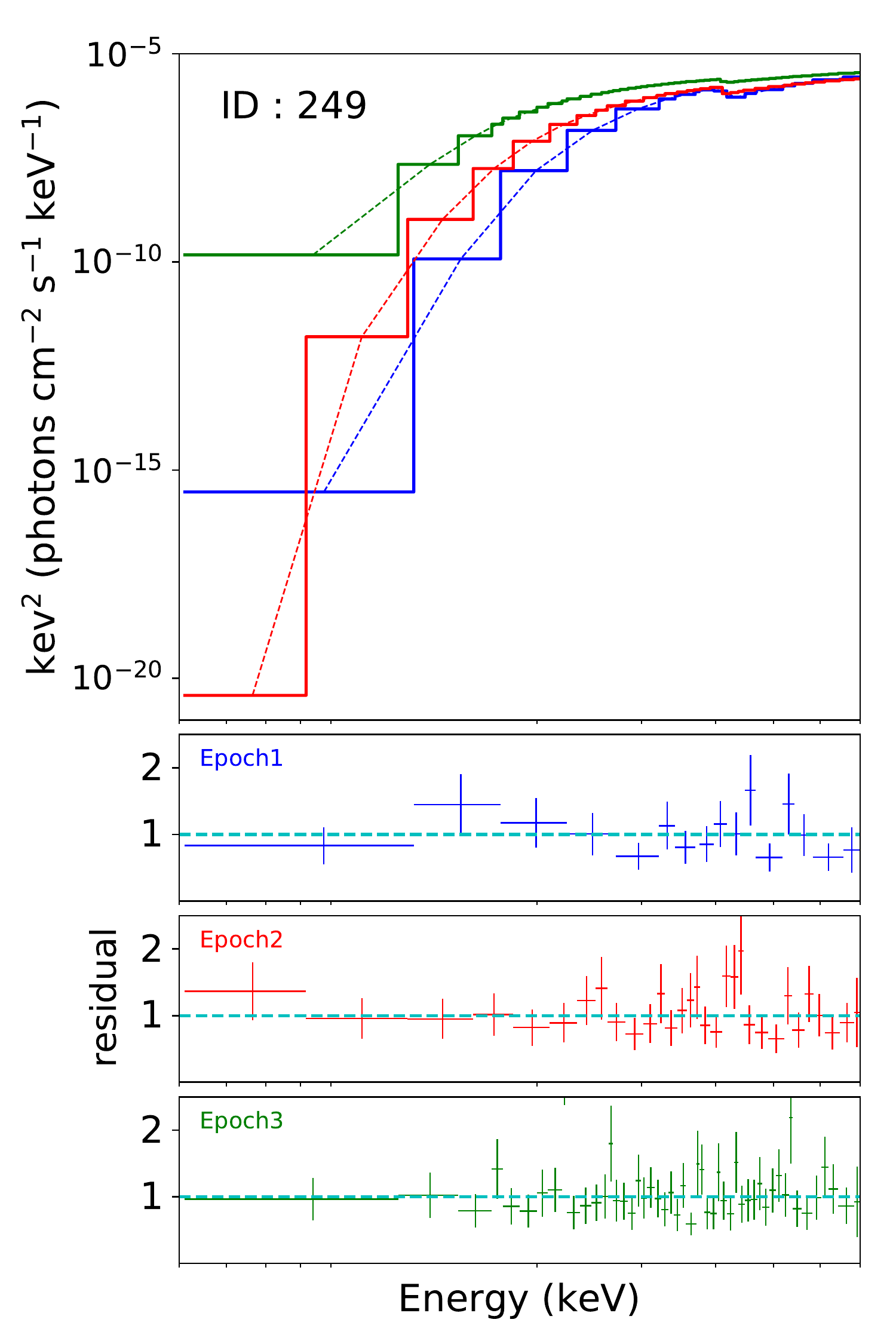}  
\label{fig:143}  
\end{minipage}  
\begin{minipage}[t]{0.31\linewidth}  
\centering  
\includegraphics[width=2.3in]{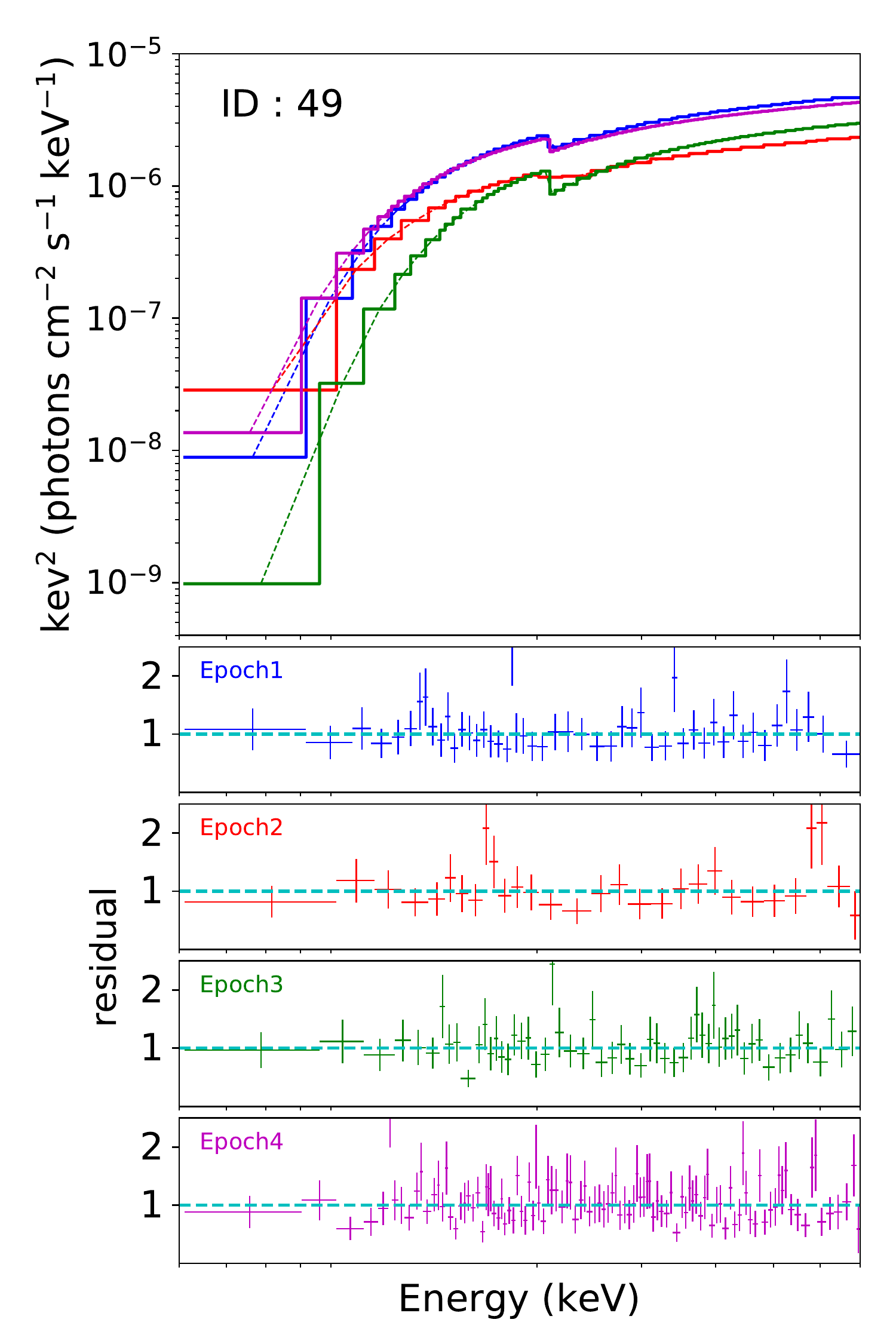}  
\label{fig:49}  
\end{minipage}  
 
\begin{minipage}[t]{0.31\linewidth}  
\centering  
\includegraphics[width=2.4in]{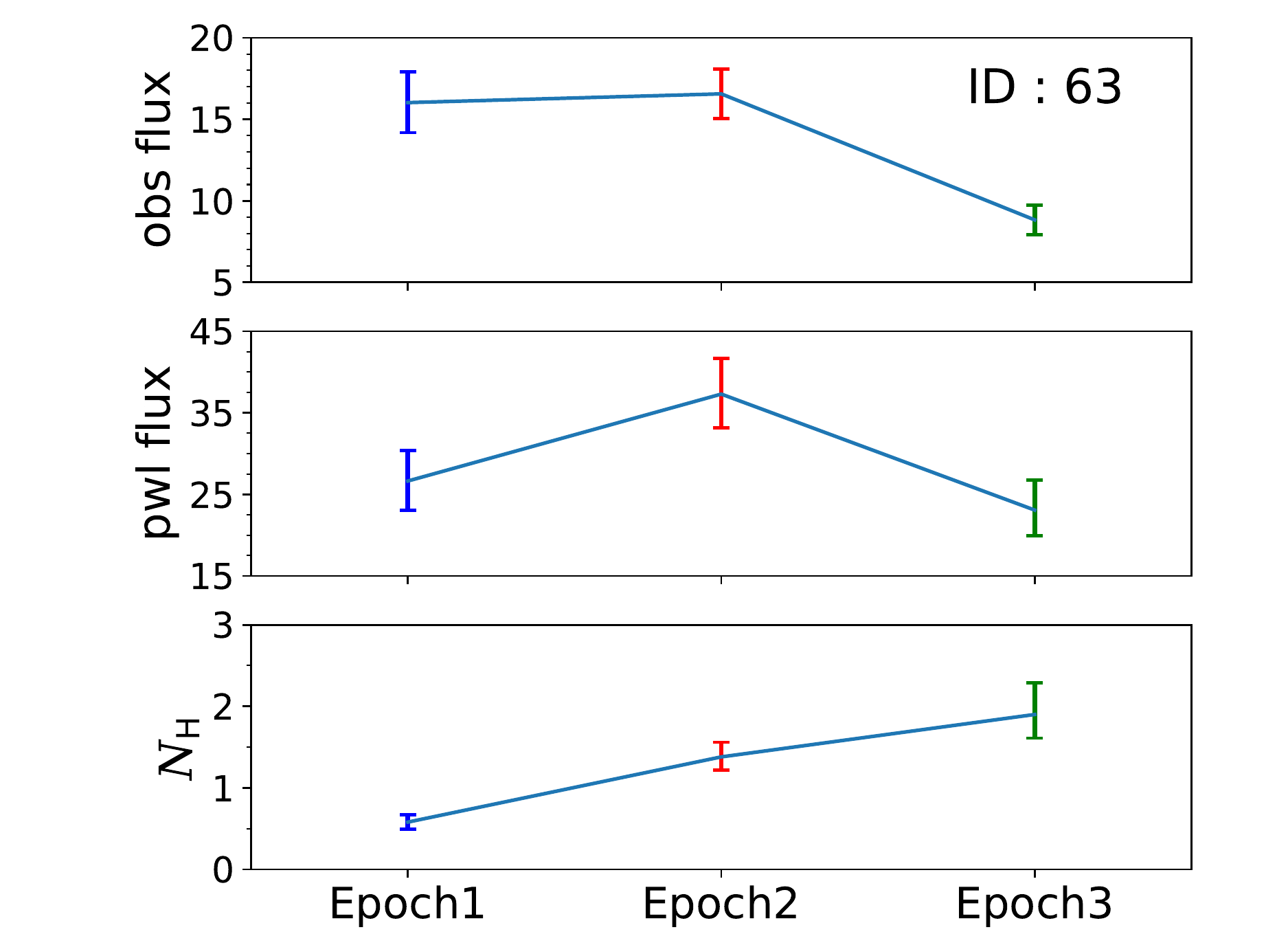}  
\label{fig:63fn}  
\end{minipage}
\begin{minipage}[t]{0.31\linewidth}  
\centering  
\includegraphics[width=2.4in]{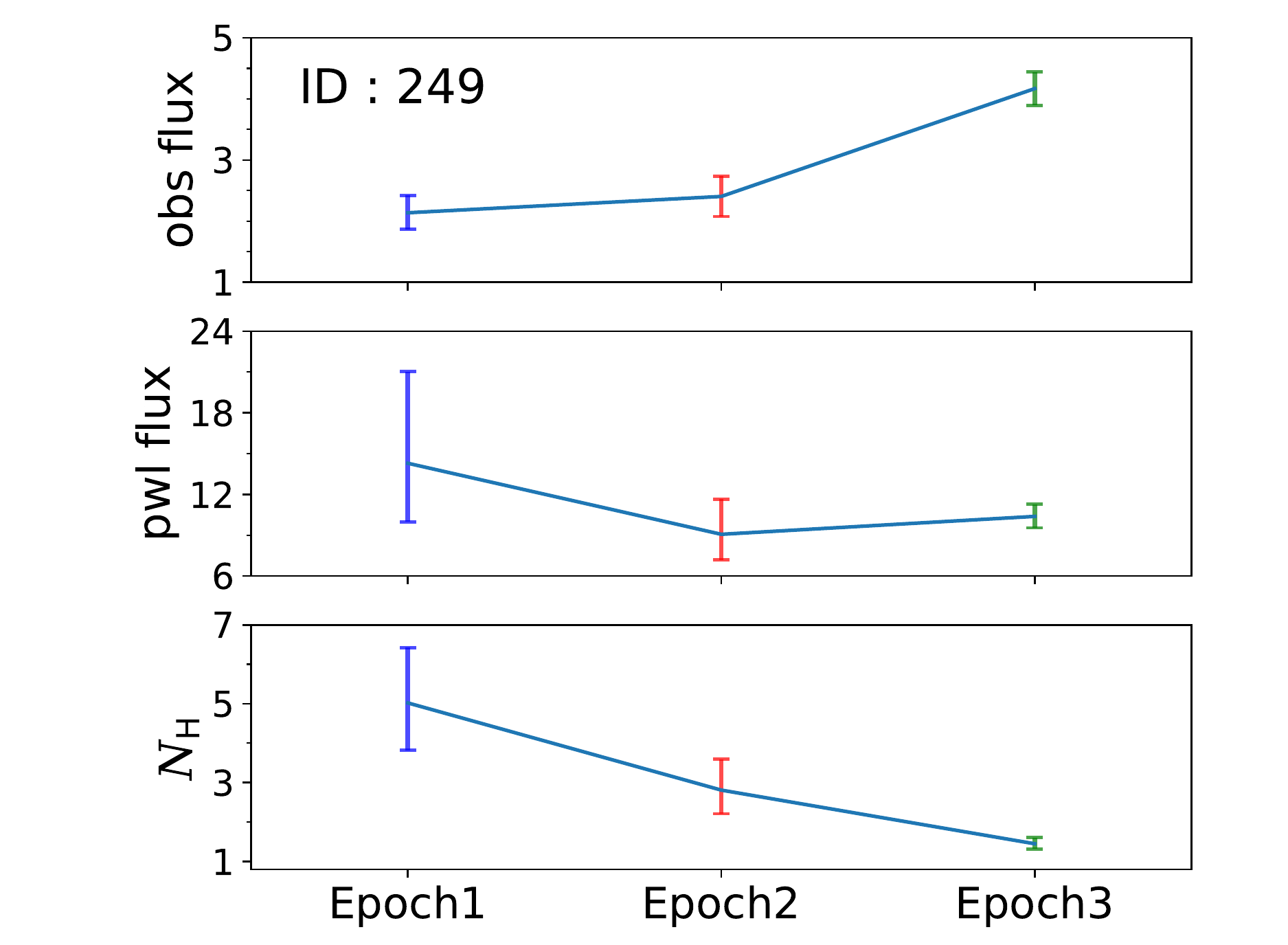}  
\label{fig:143fn}  
\end{minipage}  
\begin{minipage}[t]{0.31\linewidth}  
\centering  
\includegraphics[width=2.4in]{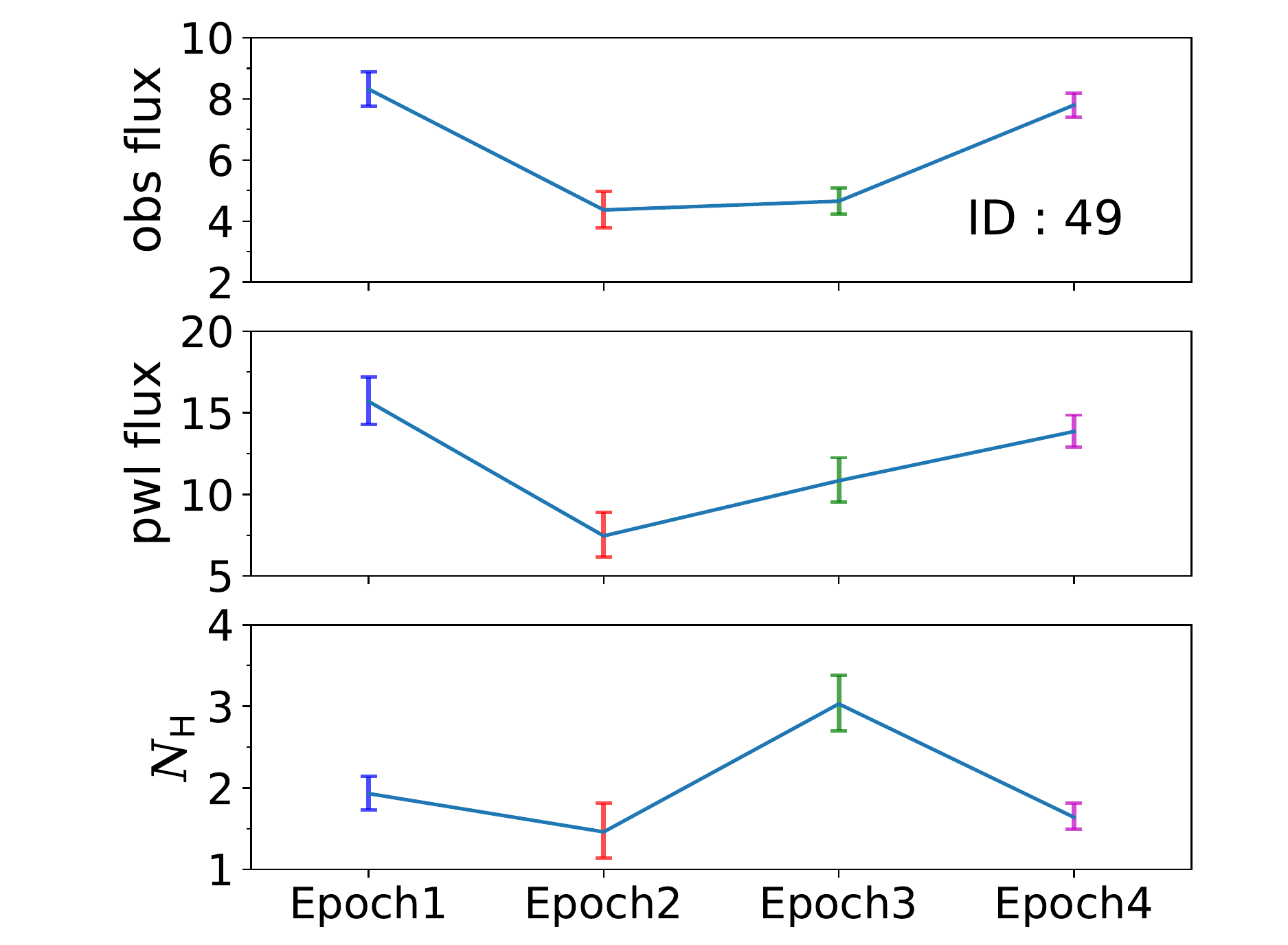}  
\label{fig:49fn}  
\end{minipage}  

\begin{minipage}[h]{0.31\linewidth}  
\centering  
\includegraphics[width=2.4in]{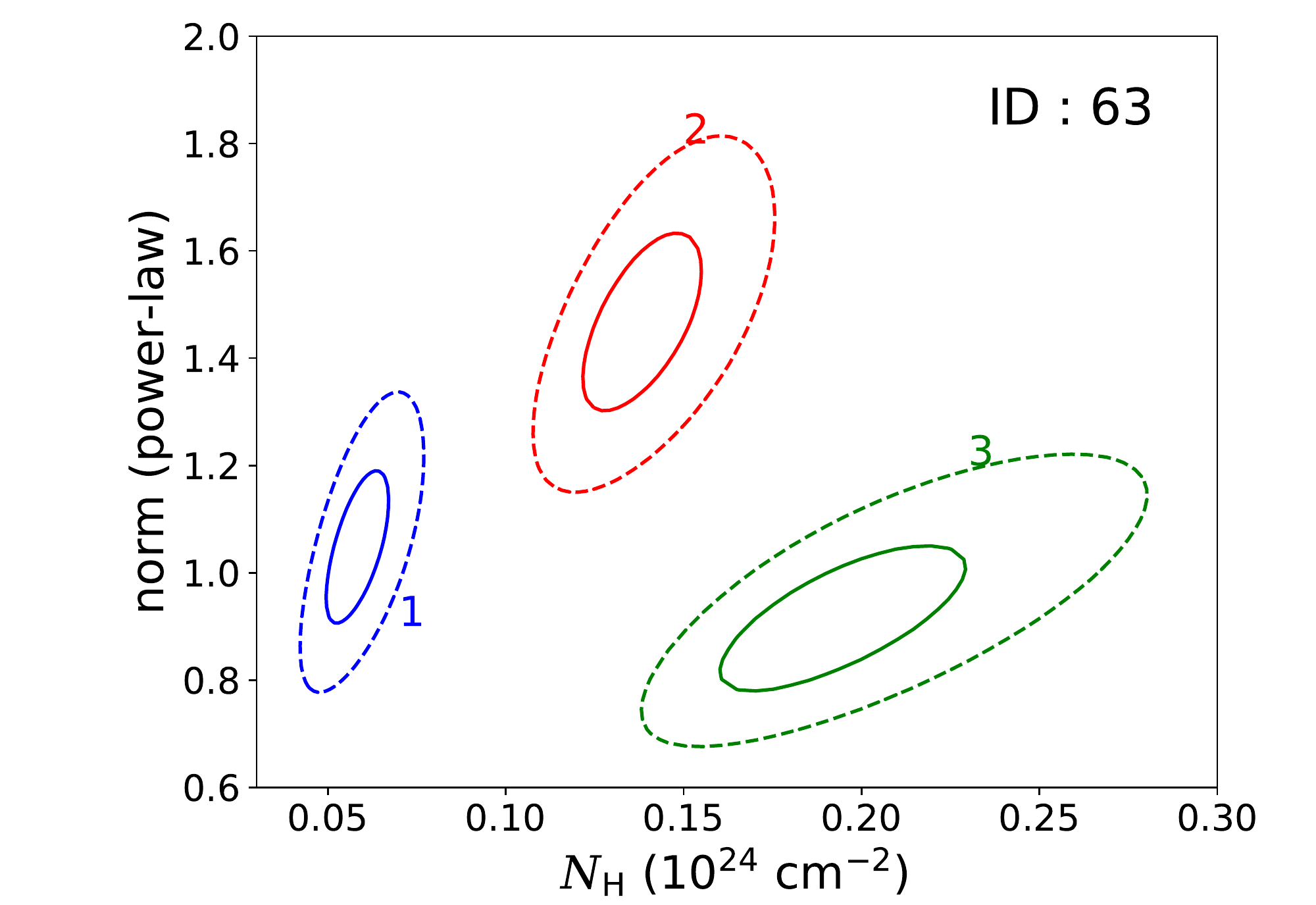}  
\label{fig:63fn}  
\end{minipage}
\begin{minipage}[h]{0.31\linewidth}  
\centering  
\includegraphics[width=2.4in]{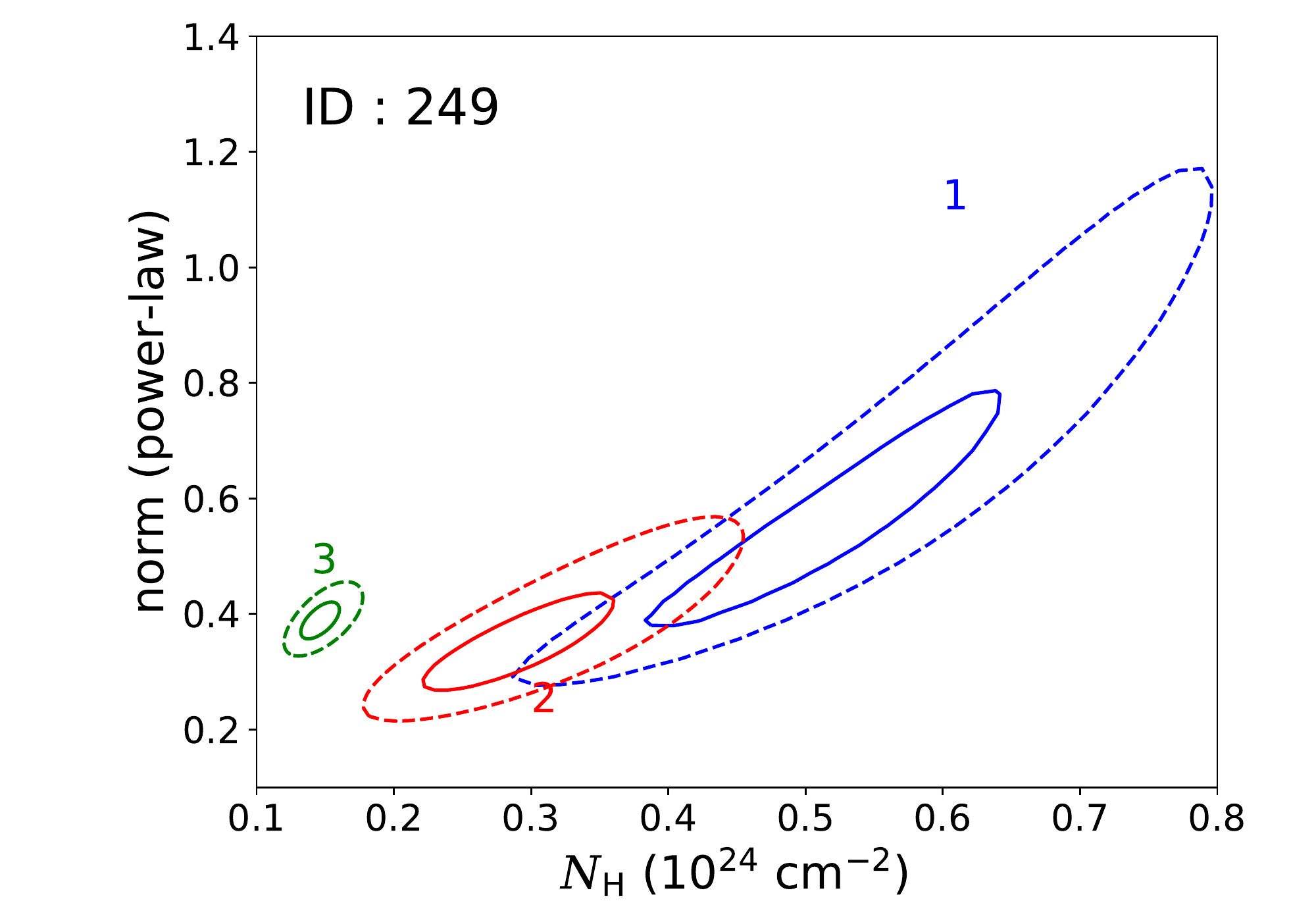}  
\label{fig:143fn}  
\end{minipage}  
\begin{minipage}[h]{0.31\linewidth}  
\centering  
\includegraphics[width=2.4in]{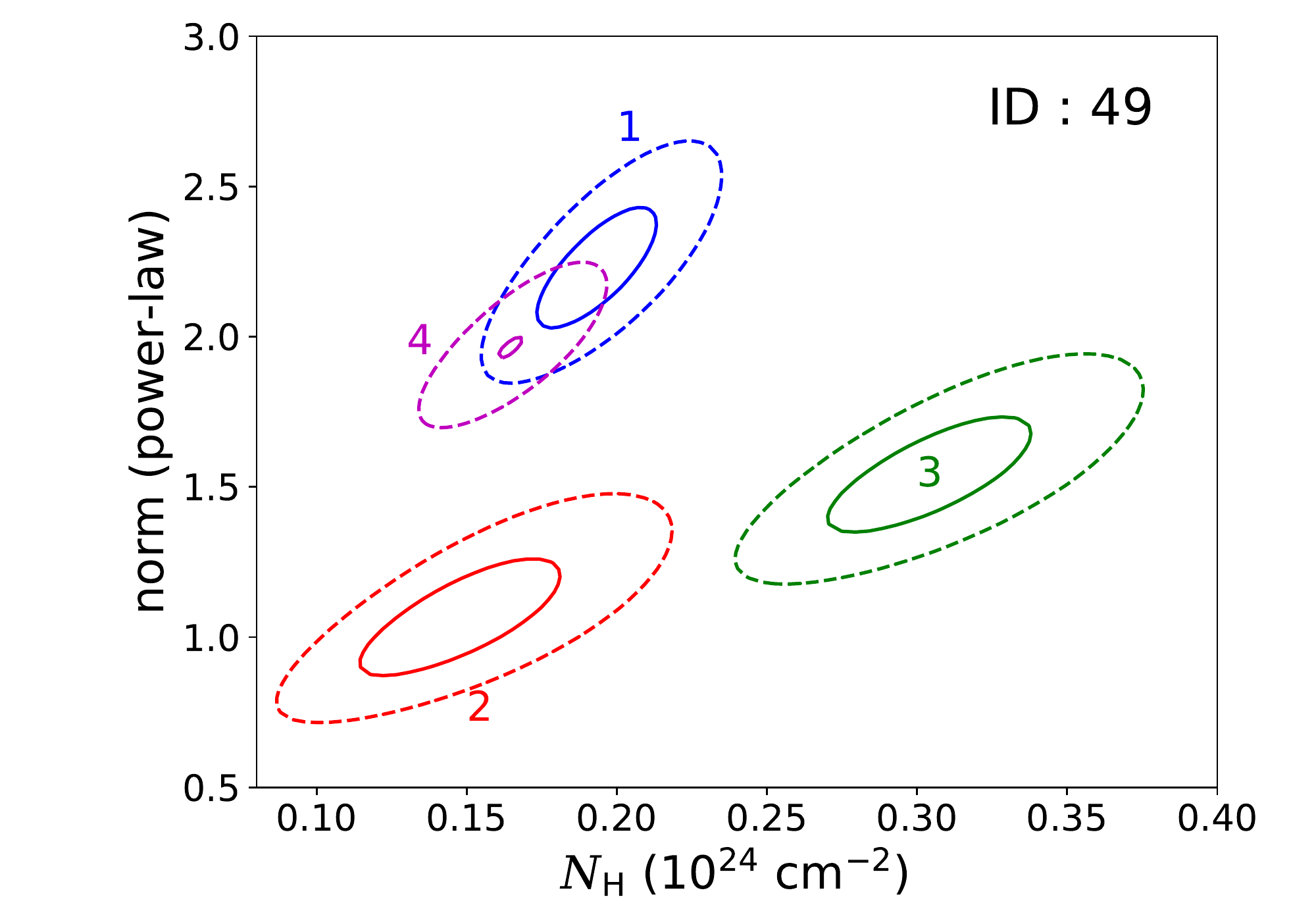}  
\label{fig:49fn}  
\end{minipage}  

\caption{\emph{first row}: The spectra for three highly variable sources XID 63, XID 249 and XID 49 in the CDF-S. The top panels show the unfolded spectra in each epoch and the bottom panels show the data-to-model ratios. \emph{second row}: The observed 0.5--7~keV flux, intrinsic 0.5--7~keV flux (in units of $\rm 10^{-16}\ erg\ cm^{-2}\ s^{-1}$) and \nh~(in units of $\rm 10^{23}\ cm^{-2}$) variability curves for the three sources, respectively. \emph{third row}: Confidence contours of the normalizations of the intrinsic power law and \nh~for the three sources. The solid and dashed curves indicate 1$\sigma$ and 2$\sigma$ confidence contours, respectively. The numbers annotated represent different epochs.}
\label{fig:variab_spectra}
\end{figure*}

\label{subsec:variabsrc}
There are some sources in our sample that show significant variability of the observed flux, intrinsic luminosity or \nh. In this section, we perform detailed analyses to study their variability behavior aiming at shedding light on the leading mechanism that drives the large variabilities, and try to understand the typical location and size of the obscuring clouds (all the following sources are in the CDF-S) . 

\subsection*{XID 63}
This source has 848 counts available and its data are binned to three epochs. It is a moderately luminous source ($\bar{L} \rm_X = \rm 5.1 \times 10^{43}\ erg\ s^{-1}$) at $z  = 0.737$. The best-fit model of this source is the absorbed power-law ($\Gamma = 1.54^{+0.21}_{-0.22}$) plus an additional soft excess component. \gm~and $norm_{\rm soft}$ do not show variations and are linked during all the epochs.  The unfolded spectra, light curves, as well as the 1$\sigma$ and 2$\sigma$ confidence contours of \nh~and normalization are displayed in the left column of Figure \ref{fig:variab_spectra}. It has $\chi_{N_{\rm H}}^2 = 65.2$, $\chi_{f, \rm obs}^2 = 34.2$ and $\chi_{L}^2 = 7.2$, indicating significant variabilities.
Although the photon index does not vary, the observed spectral shape changes prominently due to the large variation in column density. The absorption is weak in epoch1 with \nh~$\rm = 5.8 \times 10^{22}\ cm^{-2}$. Then it transforms into a highly obscured state with \nh~$= \rm 1.4 \times 10^{23}\ cm^{-2}$ in epoch2, and its \nh~continues to rise to $\rm 1.9 \times 10^{23}\ cm^{-2}$ in epoch3. The intrinsic flux increases about a factor of 1.4 from epoch1 to epoch2, and decreases about a factor of 1.6 from epoch2 to epoch3. Due to the large variability amplitude in \nh, the observed flux remains constant in the first two epochs, but significantly declines in epoch3 by a factor of 1.9. The \nh~variability may be caused by the clumpy cloud moving into the l.o.s. Since the X-ray obscuration type transition happens between epoch1 and epoch2, which corresponds to a rest-frame timescale $t < 6.0$ years ($\approx$ 10.4 years in the observed frame). 
By assuming this transition as an eclipse event, we can roughly estimate the distance and angular size of the cloud using the same method in Y16, thus to constrain the cloud size. We use the empirical relation between the inner torus radius $r$ and 14 - 195 keV luminosity from \cite{Koshida2014} to estimate the distance. The relation can be described as ${\rm log} \,r = -1.04 + 0.5\, {\rm log}\, (L_{\rm 14 - 195\,keV} /10^{44})$, where $r$ and $L_{\rm 14-195 keV}$ are in units of pc and \ergs, respectively. The $L \rm_{14 - 195\,keV} = 2.5 \times 10^{44}\ erg\ s^{-1}$ at epoch1 is obtained by extrapolating the power-law to the higher energy band. Thus the estimated $r$ from the inner torus to the central black hole is $\approx 0.14$ pc. By assuming Keplerian motion, the orbital period of the cloud $t_{\rm orbit} = 2\pi \frac{r^{\frac{3}{2}}}{(M_{\rm BH} G)^{\frac{1}{2}}} \approx 519\, (\frac{M_{\rm BH}}{10^8 M_\odot})^{-\frac{1}{2}}$ year. The angular size (viewed from the BH) of the cloud is estimated as $\frac{t_{\rm tran}}{t_{\rm orbit}} \times 360^\circ < 4.1^\circ (\frac{M_{\rm BH}}{10^8 M_\odot})^{\frac{1}{2}}$. By multiplying $r$, we estimate the cloud size to be $< 0.01\, (\frac{M_{\rm BH}}{10^8 M_\odot})^{\frac{1}{2}}\rm \,pc$.

\subsection*{XID 249}
This source can be well fitted by the absorbed power-law model with \gm~= $1.51^{+0.26}_{-0.28}$ and has the most prominent \nh~variability in our sample ($\chi^2_{N \rm_H}$ = 130.8) as shown in Figure \ref{fig:variab_spectra}. The \nh~declines from $\rm 5.0 \times 10^{23}\ cm^{-2}$ to $\rm 2.8 \times 10^{23}\ cm^{-2}$, and finally $\rm 1.5 \times 10^{23}\ cm^{-2}$ during the three epochs. The intrinsic flux of this source remains roughly constant ($\chi_{flux, \rm in}^2 = 2.2$), but the strong variation in \nh~causes a significant increase in the observed flux ($\chi_{flux, \rm obs}^2 = 31.0$). The variability analyses of each single epoch spectrum provide very different results compared with the $\rm 7\ Ms$ stacked spectrum (\gm~=~1.20 when using MYTZ alone which does not limit the range of \gm), also indicating strong \nh~variation. 
Using the same method as for XID 63, we estimate the distance and the cloud size to be $< 0.09 \rm \ pc$ and $< 0.006\, (\frac{M_{\rm BH}}{10^8 M_\odot})^{\frac{1}{2}}\rm \,pc$ for the occultation event from epoch2 to epoch3.

\subsection*{XID 328}
This source has the highest column density in our variability sample. The epoch-mean \lx~and \nh~are $\rm 3.7 \times 10^{44}\ erg\ s^{-1}$ and $\rm 8.5 \times 10^{23}\ cm^{-2}$, respectively, which are consistent with the stacked spectral analyses. The spectra can be well described by a power law (\gm~= $2.12^{+0.24}_{-0.24}$) plus Compton reflection continuum and strong Fe K lines. \nh~and the reflection flux remain roughly constant in the timespan of 5.8 years in the rest-frame, and the observed flux variations result from the intrinsic X-ray power variability. According to the stacked spectral analyses, the reflection flux accounts for 27.5\% of the observed 0.5--7~keV flux, but only accounts for 3.8\% of the intrinsic 0.5--7~keV flux. This source has also been reported in the literature as CT candidates (\citealt{Comastri2011}; Y16). Our results confirm its highly obscured nature but with higher \nh, \lx~and \gm.

\subsection*{XID 49}
This source was observed 98 times during the total $\rm 7\ Ms$ campaign and has 2345 counts available. The spectra can be explained by a simple absorbed power law ($\Gamma = 1.66^{+0.11}_{-0.12}$) with different column densities and normalizations (see Figure \ref{fig:variab_spectra}). The intrinsic flux decreases from $\rm 1.57\times 10^{-14}\ erg\ cm^{-2}\ s^{-1}$ to $\rm 7.45\times 10^{-15}\ erg\ cm^{-2}\ s^{-1}$ during the first two epochs and finally rises again to $\rm 1.39\times 10^{-14}\ erg\ cm^{-2}\ s^{-1}$ in the fourth epoch. \nh~has a wave-shape variability behavior and does not follow the intrinsic flux variability pattern, indicating that the varying absorption is not caused by the change of ionization state. 

\subsection*{XID 458}
This source has a redshift of 2.291 and can be described by a power law (\gm$=1.83^{+0.14}_{-0.13}$) with reflection hump and iron emission lines. Though the \nh~is not so high during all the epochs ($\bar{N} \rm _H \approx \rm 1.4 \times 10^{23}\ cm^{-2}$), the reflection component is required to fit the $\rm 7\ Ms$ stacked spectrum. The intrinsic flux increases from epoch1 to epoch3, and declines at epoch4. But the reflection flux remains roughly constant from epoch1 to epoch2, decreases at epoch3, and does not show any variability from epoch3 to epoch4. The different variability patterns indicate that there may be a time lag between the intrinsic continuum variability and the large-scale reflection flux variability. The decline of the reflection flux during epoch2 to epoch3 (from observed-frame 2007 September to 2010 July, corresponding to rest-frame 0.76 years) might result from significant flux decline before epoch1, and this decline has just propagated to the reflection medium (possibly the clumpy torus) and causes significant diminishment of the reflection flux. This means that the variable continuum signal needs at least 2.4 years (rest-frame timespan from 1999 October to 2007 September) to spread to the torus from the central emission region, which provides a rough lower-limit estimate, $\approx$ 0.7 pc, of the location of the reflecting cloud. 

\subsection*{Remaining sources}
By applying similar analyses to the remaining sources, we find that 17 sources in our sample that show observed flux variabilities can be classified into three types: 29\% (5/17) are caused by the change of \nh, 53\% (9/17) are caused by the AGN intrinsic luminosity variability, and 6\% (1/17) are driven by both effects. Note that there are two sources (CDF-S XIDs 186 and 876) identified to be flux-variable but showing neither \lx- nor \nh-variability according to the $\chi^2$ tests. 
However, their best-fit intrinsic luminosities both show obvious variations, but due to large errors, the corresponding $\chi^2_L$ values are smaller than the critical value.  The high \nh-variable fraction among flux-variable sources confirms our previous thoughts in Section \ref{subsec:exc} that the \nh~variability is a key ingredient for investigating the variability in highly obscured AGNs.

\section{Conclusions}
In this study, we present systematic X-ray spectral analyses of 436 highly obscured AGNs (\nh~\>~\cm) with \lx~\>~$10^{42}$ \ergs~identified in the \emph{Chandra} Deep Fields, which make up the largest dedicated highly obscured AGN sample to date, to explore their physical properties and evolution. We also carry out detailed long-term variability analyses for a subsample of 31 sources with largest counts available to investigate the main driver of their spectral variability and the typical variability amplitude. Below we summarize our main results.

\begin{enumerate}[1.]
\item We perform detailed X-ray spectral fitting for 1152 AGNs in the \emph{Chandra} Deep Fields with observed-frame 0.5--7~keV net counts \>~20 using physically appropriate MYTorus-based models, in order to identify heavily obscured ones. By limiting our analyses to sources with $\loglx > 42\ \ergs$ in order to avoid possible contamination from star-forming galaxies, 436 AGNs are confirmed to be highly obscured, with $\overline{z}=1.9$ and $\overline{\loglx} = \rm 43.6\ \ergs$.

\item We select 102 Compton-thick (CT) candidates with best-fit \nh~\>~$\rm 10^{24}\ cm^{-2}$ and 1$\sigma$ lower limit \>~$\rm 5 \times 10^{23}\ cm^{-2}$, accounting for $\sim$ 23\% of the highly obscured sample. The observed log$\,N$-log$\,S$ for CT AGNs prefers the moderate CT number counts as predicted by \cite{Akylas2012} and \cite{Ueda2014} cosmic X-ray background models, while other models \citep[e.g.,][]{Gilli2007, Ballantyne2011} more or less overestimate or underestimate the number counts.

\item We present a new hardness-ratio measure of the obscuration level as a function of redshift (the HR curve), which can be used to select heavily obscured AGN candidates without resorting to detailed spectral fitting. The completeness and accuracy by applying the HR curve on the CDFs AGN population to identify highly obscured ones are 88\% and 80\%, respectively.

\item We find a strong negative correlation between the soft excess fraction \fs~and \nh~with a Spearman rank correlation coefficient $\rho=-0.66$. By assuming that the soft excess originates from the scattered-back continuum and treating the small \fr~as an indicator of high covering factor of the obscuring materials, this result indicates that a portion of the most heavily absorbed AGNs reside in an extremely geometrically buried environment.

\item Among the 31 CDF-S highly obscured sources which have optical classification results, 19\% (6/31) of them are labeled as BLAGN, indicating that at least for part of the AGN population, the high-level X-ray obscuration is largely a line-of-sight effect, i.e., some high-column-density clouds on various scales (but not necessarily a dust-enshrouded torus) may obscure the compact X-ray emitter without blocking the entire BLR; alternatively, the heavy X-ray obscuration maybe produced by the BLR itself.

\item After considering the errors on the best-fit \nh~and correcting for the sky coverage effect,  the Eddington bias and the \nh-dependent Malmquist bias, we derive the intrinsic \nh~distribution representative of the highly obscured AGN population as well as its evolution across cosmic time. The intrinsic CT-to-highly-obscured fraction is roughly 52\% and is consistent with no evident redshift evolution.

\item We select 31 sources with 0.5--7~keV net counts \>~700 and \>~900 in the CDF-S and CDF-N, respectively,  to perform long-term ($\approx 17$ years in the observed frame) spectral variability analyses. We find that the flux-variable, \lx-variable and \nh-variable source fractions are $55\pm13\%$ (17/31), $29\pm10\%$ (9/31) and $19\pm8\%$ (6/31), respectively. The typical flux, \lx~and \nh~variability amplitudes for those variable sources are 17\%, 19\%, and 16\%, respectively.

\item We calculate the normalized excess variance ($\sigma_{\rm nxv}^2$) of \nh, observed 0.5--7~keV flux and \lx~to investigate the intrinsic variability amplitude. No correlation between $\sigma_{\rm nxv}^2$ and \lx~is detected, possibly due to the observed flux variability for a significant fraction of highly obscured AGNs being caused by the change of \nh~rather than the variation of \lx~alone, as well as the limited sample size and the broad redshift span. 

\item We discuss detailed variability behaviors for 5 sources that show significant \nh, \lx~or observed flux variability. The typical locations and sizes of the occultation and reflection clouds are estimated (see Section  \ref{subsec:srcvariab}). The main driver for the variability of the 17 flux-variable sources can be classified into three types: 29\% (5/17) are caused by the change of \nh, 53\% (9/17) are caused by the AGN luminosity variability and 6\% (1/17) are driven by both effects. Two sources are not classified due to large measurement errors. 
\end{enumerate}

Benefiting from the deepest X-ray surveys to date, our work provides meaningful constraints on the properties and evolution of the AGN obscuring materials over a broad redshift range, and quantifies the detailed variability behaviors of these hidden sources, which are crucial for us to better understand the role that highly obscured AGNs played in galaxy evolution. However, X-ray data alone are incapable to show the overall perspective. In a subsequent paper of this series (Li et al., in preparation), we will further explore the properties of highly obscured AGNs and their host galaxies by combining the wealth of multi-wavelength data in the CDFs, aiming at  testing the merger-triggered AGN-galaxy co-evolution scenario \citep{Sanders1988}.

\acknowledgments 
We thank the anonymous referee for valuable comments and suggestions. We thank David Ballantyne, Murray Brightman, Giorgio Lanzuisi and Yoshihiro Ueda for providing relevant data, and Tahir Yaqoob and Mislav Balokovi\'{c} for helpful discussions. J.Y.L., Y.Q.X., M.Y.S., and X.C.Z. acknowledge support from the 973 Program (2015CB857004), the National Natural Science Foundation of China (NSFC-11890693, 11473026, 11421303),  the CAS Frontier Science Key Research Program (QYZDJ-SSW-SLH006), and the K.C. Wong Education Foundation. 
M.Y.S and X.C.Z. acknowledge the support from the National Natural Science Foundation 
of China (NSFC-11603022) and the China Postdoctoral Science Foundation 
(2016M600485). 
X.C.Z also acknowledges the financial support from CSC(China Scholarship Council)-Leiden University joint scholarship program. 
FV acknowledges support from CONICYT and CASSACA through the Fourth call for tenders of the CAS-CONICYT Fund. 
T.M.H. acknowledges the support from the Chinese Academy of Sciences (CAS) and the National Commission for Scientific and Technological Research of Chile (CONICYT) through a CAS-CONICYT Joint Postdoctoral Fellowship administered by the CAS South America Center for Astronomy (CASSACA) in Santiago, Chile. 
B.L. acknowledges financial support from the National Key R\&D Program of China grant 2016YFA0400702 and National Natural Science Foundation of China grant 11673010.
CV acknowledges financial support from the Italian Space Agency (ASI) under the contracts ASI-INAF I/037/12/0 and ASI-INAF n.2017-14-H.0.
RG acknowledges support from the agreement ASI-INAF n.2017-14-H.O.
X.S. acknowledges support from the NSFC-11573001 and 11822301.  


\appendix 
\section{Justification of the adopted models and parameters}
\label{app:model}
In the MYTorus configuration, since the reproccessed components strongly depend on the inclination angle $\theta$, the equatorial \nh~and the assumed \fr, the best-fit results will also depend on the assumed geometries and parameters. However, due to the low S/N for most of the spectra, we have to make prior assumptions about the input parameters and simply fix them to values that their representativeness has not yet been physically validated. Here we justify that our spectral fitting results are not significantly influenced by the assumed parameters.

We first fit the spectra for the high count sources (counts \>~200) with $\theta$ and \fr~set as free parameters. The best-fitted $\theta$ peaks at $\theta < 70^\circ$, therefore this time we choose to fix it at $65^{\circ}$ for all sources. Then we re-fit the spectra by keeping \fr~free and obtain the best-fit results. For the low count sources, we fix \fr~at 2.0 for sources with detected reprocessed components, which is roughly the average value for high count sources. 

\begin{figure}
\centering
\includegraphics[width=0.8\linewidth]{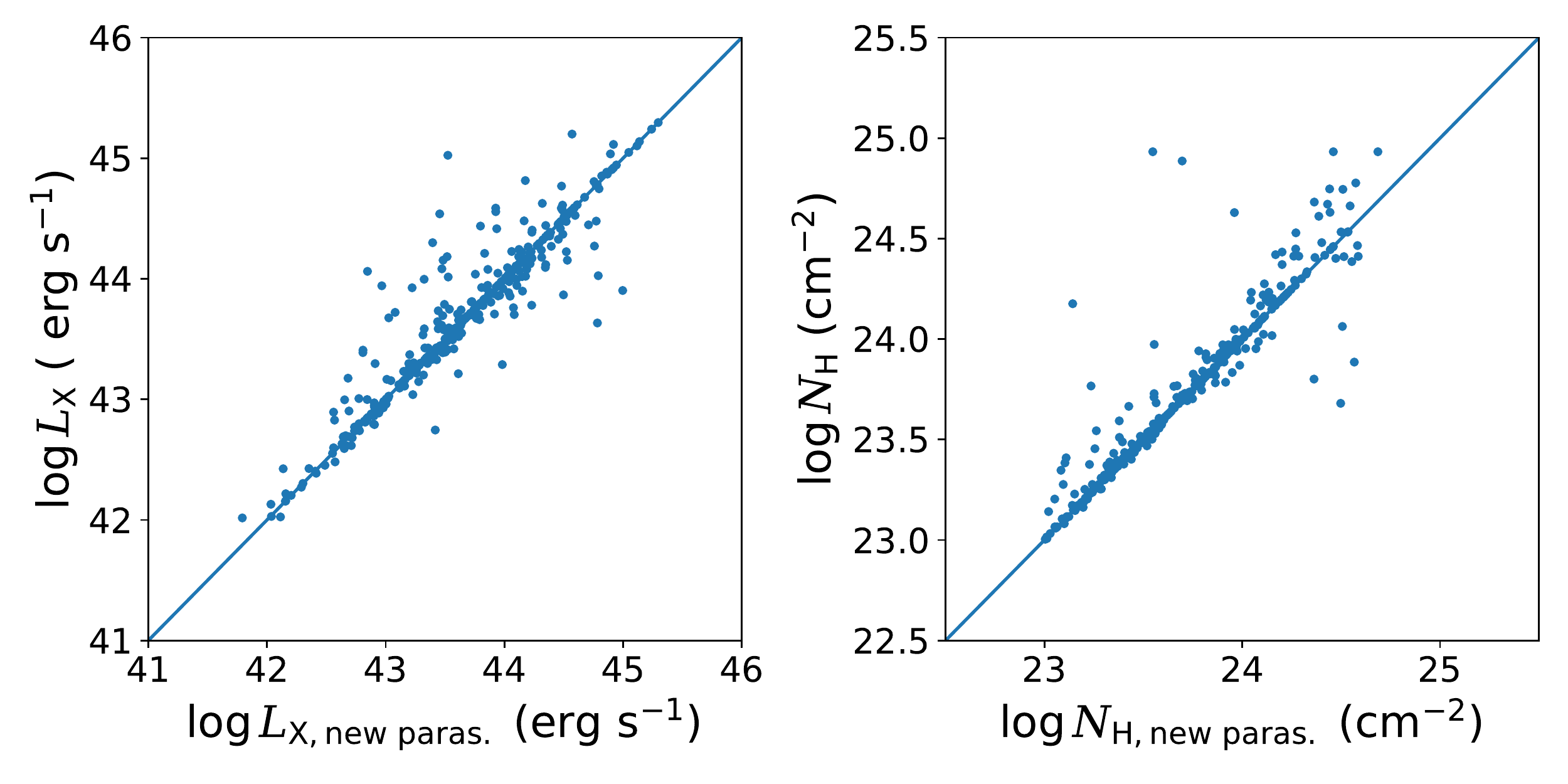}
\caption{Comparison between the spectral fitting results obtained in Section \ref{sec:spectra} and those obtained in Appendix~A using the new model parameters.}
\label{fig:cmp_previous}
\end{figure}

The comparison of the results with those obtained in Section  \ref{subsec:model} is shown in Figure \ref{fig:cmp_previous}. Despite of small scatters, the results are consistent. Therefore our simple choices of the fixed parameters are reasonable. The consistency between the MYTorus results and the Borus results obtained in Section  \ref{subsec:ref} also validates our spectral fitting strategy.

\end{document}